# Gli Strumenti Scientifici di Interesse Storico del «Lussana», del «Vittorio Emanuele», e del «Quarenghi» di Bergamo

**Erasmo RECAMI, Sergio PIZZIGALLI, Ettore PARIGI, Francesco DE VINCENTIS e Virgilio BORLOTTI**

**Volume Pubblicato col contributo della**
***«PRO UNIVERSITATE BERGOMENSI»***

---

**Dipartimento di Ingegneria**
***Università degli Studi di Bergamo***

***Volume a cura di:***

*Erasmo Recami* (Facoltà di Ingegneria, Università di Bergamo)
E-mail: recami@mi.infn.it

*Sergio Pizzigalli* (Liceo Scientifico “Lussana”, Bergamo)
E-mail: pizzi@uninetcom.it

*Ettore Parigi* (I.T.I.S. “Paleocapa”, Bergamo)
E-mail: parett@infinito.it

*Francesco De Vincentis* (I.T.G. “Quarenghi” e Università di Bergamo)
E-mail: devincent@libero.it

*Virgilio Borlotti* (I.T.C. “Vittorio Emanuele”, Bergamo)
E-mail: maieuticacmd@libero.it

# 0. PREFAZIONI

***Prefazione del Dirigente Scolastico del Liceo "Lussana" di Bergamo***

Dal lontano 1926, anno di fondazione, ai primi anni sessanta, il Liceo Lussana passava da circa 40 a poco meno di 500 studenti in un panorama scolastico, pur con le debite proporzioni e la relativa crescita quantitativa, ancora fortemente elitaria (il Lussana rimaneva infatti l'unico Liceo Scientifico Statale di Bergamo e provincia).

Siccome il periodo di acquisizione degli strumenti scientifici di cui tratta questo repertorio, così sagacemente predisposto dal prof. Recami, è concentrato nella parte iniziale di tale segmento temporale, non si può non rimanere positivamente colpiti da quanta importanza venisse riservata dall'allora Direzione Scolastica al ruolo di queste dotazioni, oltre che al loro uso nell'incremento delle acquisizioni scientifiche degli alunni.

A questa tradizione di scuola seria e rigorosa è nostro preciso dovere restare collegati, come è nostro dovere incrementarla accettando a pieno titolo il profondo rinnovamento tecnologico, così connesso all'informatica e alla multimedialità, e la sfida difficile ma certamente esaltante di una scuola di alta qualità non più per pochi ma per molti.

Per una comunità scolastica che si ispira a convinzioni del genere e si sforza di tradurle in pratica ogni giorno, il lavoro del prof. Recami e del suo staff è oltremodo prezioso perché non si configura come un repertorio ma come spia di una metodologia di apprendimento vissuto negli anni, fatta di passione scientifica e culturale.

*Prof. Cesare Quarenghi*

***Prefazione del Dirigente Scolastico dell'ITC "Vittorio Emanuele II" di Bergamo***

L'Istituto Tecnico "Vittorio Emanuele II" di Bergamo, essendo stato fondato nel lontano 1862 ed avendo dato origine a tutti gli Istituti a carattere tecnico-scientifico di Bergamo, ha avuto modo di accumulare nel tempo un certo numero di strumenti fisici antichi.

Le vicissitudini storiche, non ultima l'occupazione tedesca dell'edificio nel periodo bellico e i conseguenti traslochi del materiale didattico, hanno in parte disperso la ricca dotazione dell'Istituto.

Ciò che rimane, però, è di grande interesse e ben volentieri viene messo a disposizione dei visitatori durante le annuali Settimane della cultura scientifica e tecnologica.

Nell'attuale momento storico nel quale la Scienza e la Tecnologia hanno fatto passi giganteschi che non erano facilmente immaginabili, il mostrare, come fa questa pubblicazione, gli antichi strumenti scientifici serve a storicizzare questo nostro tumultuoso progresso e a spingerci ad alcune salutari riflessioni.

*Prof. Dario Frigerio*

***Prefazione del Dirigente Scolastico dell'I.S.I.S. "G. Quarenghi" di Bergamo***

La collezione di antichi strumenti topografici dell'Istituto "G. Quarenghi" costituisce parte di quel patrimonio storico presente sul territorio che è bene venga conosciuto e compreso.

Meritevole è pertanto l'iniziativa sorta per la passione del gruppo di cultori di Storiografia Scientifica che si è impegnato nella redazione del presente volume.

Credo e mi auguro che questo libro possa fornire una effettiva utilità agli studiosi della materia e agli studenti dei nostri corsi.

Si ringraziano gli autori ed in particolare il prof. F. De Vincentis che ha curato le schede degli strumenti topografici del nostro Istituto, la Pro Universitate Bergomensi che ha "sponsorizzato" l'opera e tutti coloro che hanno reso possibile la interessante pubblicazione.

*Prof. Francesco Fabio*

# 1. INTRODUZIONE

L'Università degli studi di Bergamo è di recente costituzione. In Bergamo e provincia gli strumenti scientifici di interesse storico sono da ricercarsi, quindi, nelle Scuole Medie Superiori di antica fondazione.

Una notevole eccezione è costituita dalla nota collezione di Giorgio Mirandola, che comprende circa duemila strumenti, includendo quelli tecnologici, e che probabilmente è la maggiore collezione privata del genere in Italia.

In base a una rapida indagine organizzata da G.Mirandola, già Assessore alla Cultura della Provincia (ed effettuata da L.Serra Perani e P.Serra Bailo), strumenti scientifici di interesse storico si trovano -oltre che nelle Scuole di cui qui ci occuperemo- soprattutto presso:

- Il Liceo Ginnasio «Sarpi» di Bergamo [alcuni strumenti si trovano presso il Museo di Scienze Naturali di Bergamo, a cura del Direttore di questo Museo]. Tutto questo materiale è stato studiato dalle citate L.Serra Perali e P.Serra Bailo, per conto di G.Mirandola. Un ottimo lavoro è stato effettuato indipendentemente da Giacomo Sechi, il quale ha prodotto tra l'altro un CD multimediale sul Museo degli strumeti scientifici antichi posseduti dalla Scuola: pure questo CD andrebbe senz'altro pubblicato; esso è comunque accessibile in rete, nel sito del «Sarpi», presso *http://www.liceosarpi.bg.it/risorse/FISICA_SITO/collezione.htm* .

- L'Istituto Magistrale «Secco Suardo» di Bergamo

- L' I.T.I.S. «Paleocapa» di Bergamo

- La Biblioteca di Treviglio, in cui è conservata ed esposta una grossa collezione di strumenti provenienti dalla Scuola Media «Grossi».

In questa pubblicazione ci limitiamo all'interessante materiale di carattere *fisico* esistente presso: 1) l' **I.T.C. «Vittorio Emanuele II»** (la prima Scuola media superiore nata a Bergamo: fondata nel 1862); 2) presso l'adiacente **Liceo Scientifico «Lussana»**, e 3) presso l'**I.T.G. «Quarenghi»**, tutti di Bergamo.

Il presente lavoro è stato possibile grazie alla collaborazione:

a) per la redazione delle schede, del Prof. Ettore Parigi, dell'«I.T.I.S. Paleocapa» di Bergamo;
b) per il Liceo Lussana, del suo Dirigente Scolastico Prof. Cesare Quarenghi, del

docente di Fisica Professor Sergio Pizzigalli e del Tecnico di laboratorio Sig. Adriano Gabuzzi;

c) per l'Istituto Vittorio Emanuele, del suo Dirigente Scolastico Prof. Dario Frigerio, e del Tecnico di laboratorio Sig. Virgilio Borlotti. Segnaliamo anche lo scritto «Selezione dell'antica strumentazione dei gabinetti scientifici», preparato da V. Borlotti e Rodolfo Vittori come guida per le visite organizzate durante le Settimane della Scienza, "Settimane" bandite annualmente dal MURST (ora MIUR);

d) per l'Istituto Quarenghi (che possiede soprattutto strumenti topografici) del suo Dirigente Scolastico Prof. Francesco Fabio e del Prof. Francesco De Vincentis. Cogliamo l'occasione per ricordare che l'ex-Assessorato alla Cultura di cui si è detto ha organizzato la catalogazione e schedatura -anche fotografica- di tutti gli strumenti topografici di interesse storico esistenti nelle province di Bergamo e di Milano, compresi quelli del Politecnico: la pubblicazione del risultato di tale rilevamento sarebbe di grande interesse culturale.

Le ricerche che hanno condotto a questo libretto sono state parzialmente finanziate dal **C.N.R.** (ex-Comitato 15), mentre le spese di pubblicazione sono state coperte dalla **Pro Universitate Bergomensi** (grazie al cortese e fattivo interessamento del Prof. Alberto Castoldi, M. Rettore, e dei Presidi di Facoltà, in particolare i Professori Antonio Perdichizzi e Maria I.Bertocchi, e più ancora del Presidente della P.U.B., Cav.d.Lav. Dottor Emilio Zanetti, e del Dottor Roberto Terranova).

Cominciamo con l'elencare i principali strumenti. Quelli del «Lussana» (cento strumenti) sono della prima metà del novecento; mentre quelli del «Vittorio Emanuele» sono: 52 strumenti della seconda metà dell'ottocento, e 9 strumenti della prima metà del novecento; e quelli del «Quarenghi» sono: 8 della seconda metà dell'ottocento e 18 della prima metà del novecento. La data limite superiore convenzionalmente scelta è quella del 1950 circa. Quando non diversamente indicato, essi si trovano presso i Laboratori di fisica (e, nel caso del «Quarenghi», presso il laboratorio di Topografia).

# 2. ELENCO DEGLI STRUMENTI

## *2.1 - **Elenco degli Strumenti del Liceo Scientifico «Lussana»***

Il nome dello strumento è seguito dal suo numero di inventario; dal nome della Ditta fabbricatrice o, in mancanza, della Ditta fornitrice; e quindi dalla data di acquisizione dello strumento da parte del Liceo, quale risulta dagli Inventari esistenti. Il numero di inventario di cui sopra è quello «della Provincia». Quando si tratta invece di quello «dello Stato», ciò è esplicitamente indicato con la sigla «i.s.». Gli strumenti autocostruiti più antichi sono spesso opera del Tecnico di laboratorio Signor Vogini. L'attuale tecnico di Laboratorio è il Sig. p.i. Adriano Gabuzzi.

1) Bilancia analitica
i.s.183 *(c/o Lab. di Chimica)*
Galileo Sartorius, Milano
???

2) Bilancia
Inv.204 *(c/o Lab. di Chimica)*
???
???

3) orologio con scappamento ad ancora
Inv.500; AP 1475/76
???
<1925

4) Ponte di Wheatstone, con cursore e cuffia telefonica
Inv.569
???
???

5) Rocchetto di Ruhmkorff
i.s.561 *(c/o Lab. di Chimica)*
???
??? *(VED. SCHEDA)*

6) Apparecchio per studio della caduta dei gravi
724
Martini, Trento
1925

7) Macchina per produrre rotazione
725
Phywe
1925 ***(VED. FOTOGRAFIA 1)***

8) Modello di pompa aspirante in vetro, con vaschetta
727
Martini, Trento
1925

9) Densimetro/Aerometro
729
Martini, Trento
1925

10) Tubo a vuoto di Newton
735
Trevisini, Milano
1925

11) Barometro a sifone, con 2 termometri
736
Tironi, Bergamo
1925

12) Tavolo con cinque tubi sonori, in legno
737--741
Martini, Trento
1925

13-14) Coppia di Diapason, con cassette di risonanza in legno
741 e 742
Trevisini, Milano
1925

15) Modello della distribuzione in macchina a vapore: con biella e manovella
744
Trevisini, Milano
1925 ***(VED. FOTOGRAFIA 2)***

16) Cassetta di materiale per elettrostatica composto da: pennacchio di carta, scampanio elettrico, mulinello elettrico, 6 lastre di vetro con applicazioni in stagnola per visualizzare linee di forza, catenelle, palline e punte metalliche

751
Phywe
1925

17) Elettrometro a foglia d'alluminio e condensatore
752
Trevisini, Milano
1925 (manca la foglia di Al)

18) Macchina elettrostatica di Whimshurst
753
Martini, Trento
1925 *(VED. SCHEDA)*

19) Galvanometro verticale a bobina mobile
756
Physikalishes Werkstätten, Gottingen [n.6345 del costruttore]
1925 ***(VED. FOTOGRAFIA 3)***

20) Martelletto di Wagner
757
Phywe
1925 (***VED. FOTOGRAFIA 4***, rappresentante anche lo strumento 73 [inv. n.1310])

21) Grande bussola magnetica (ad aghi multipli), in custodia metallica
759
W.Rudolph, Brema
1925

22) Stereoscopio, a doppia immagine
763
Tironi, Bergamo
1925

23) «Figura» di Plateau in ottone [per «bolle» di sapone (superfici minime)]
772
autocostruita
1925

24) Modello di elica sviluppabile, in legno
796
Martini, Trento
1926

25) Altre «figure» di Plateau in ottone [per «bolle» di sapone (superfici minime)]
797, 798
Martini, Trento
1926

26) Arco voltaico
803
Phywe
1926

27) Sei Tubi di vetro con diverse rarefazioni dell'aria
806
Martini, Trento
1926

28) Apparecchio per verifica della legge di Boyle (secondo Mueller), con termometro ad aria
811
Phyve
1926 *(VED. SCHEDA)* (***VED. FOTOGRAFIA 5***, che illustra anche lo strumento 29 [inv. n.1031])

29) Incastellatura per apparecchio di Boyle-Mariotte
1031
Limonta, Bergamo
1929 (***VED. FOTOGRAFIA 5***, che mostra anche lo strumento inv. n.811)

30-31) Stazione telefonica trasmittente, con microfono ad alta sensibilità. Idem, stazione ricevente.
847, 848
Phywe
1926 ***(VED. FOTOGRAFIA 6)***

32) Altoparlante a tromba
854
Trovati, Milano
1926

33) Sonòmetro
859
Rota, Bergamo
1926

34) Doppio cuneo in legno, con facce rivestite di velluto, per studi sull'attrito
888
Martini, Trento
1926

35) tubo a raggi catodici a croce di Malta
AP 495
???
1927

36) Doppio cilindro, per misura di conducibilità termica
918
autocostruito
1927

37) Apparecchio di Haldat
947
Resti, Milano
1927 *(VED. SCHEDA)*

38-39) Stazioni radiotelegrafiche trasmittente e ricevente, con 2 pile Grenet e 3 pile Leclanché
948, 949
Resti, Milano
1927 ***(VED. FOTOGRAFIA 7)***

40) Apparecchio universale per riflessione e rifrazione
954
Galileo (con *scheda* della Galileo)
1928 ***(VED. FOTOGRAFIA 8)***

41) Conduttore a pera per esperimenti di elettrostatica
1028
autocostruito
1929

42) Riproduzione della pila di Volta a colonna
1038 *(c/o Lab. di Chimica)*
autocostruita
1929 *(VED. SCHEDA)* ***(VED. FOTOGRAFIA 9)***

43) Pila di Volta a vaschette
1039 *(c/o Lab. di Chimica)*
autocostruita
1929 *(VED. SCHEDA)* ***(VED. FOTOGRAFIA 10)***

44) Elettròforo di Volta
1045
De Monti
1929 ***(VED. FOTOGRAFIA 11)***

45) Piano inclinato in legno, con carrucola e contrappeso
1047
Phywe
1929

46) Supporto regolabile in legno (forse per archetto di contrabbasso)
1048
De Monti
1929

47) Pompa premente
1050
De Monti
1929

48) Spettroscopio a treppiede [incompleto]
1072
Phywe
1929

49) Galvanòmetro astatico di Nobili
1085
autocostruito
1929 *(VED. SCHEDA)*

50) Alternatore
AP 965
Officine Galileo Firenze
1929 ??

51) Tubo di Crookes
2728 ??
???
1931

52) Spinteròmetro a due sferette
1087
autocostruito (Vogini)
1931

53) Tubo per raggi X, con boccia di raffreddamento e supporto in legno
1095
? [il supporto in legno è autocostruito: Vogini]
1931 ***(VED. FOTOGRAFIA 12)***

54) Fontana intermittente, in vetro
1115
Resti, Milano
1931

55-56) Due Tubi di Crookes
1146, 1147
Bernardi, Trento
1931/33

57) Galvanometro a riflessione
1168
Physikalishes Werkstätten, Gottingen [n.6345 del costruttore Bernardi di Trento, intermediario]
1934 ? ***(VED. FOTOGRAFIA 13)***

58) Pallone a vuoto, per pesare l'aria
1176
Caironi, Bergamo
1934

59) Accumulatore a Piombo
1179 *(c/o Lab. di Chimica)*
???
1934

60) Apparecchio per onde stazionarie
1184
autocostruito
1934

61) Campana a vuoto con supporto
???
???
1934 ?

62) Apparecchio per confronto della densità di due liquidi
1191
autocostruito
1935/36

63) Tubo per raggi X
1205
Cacciari, Milano
1936

64) Barometro di Torricelli, con supporto in legno
1215
autocostruiti (Vogini)
1936

65) Dinamo più anello di Pacinotti
1216
autocostruito
1936 *(VED. SCHEDA)* ***(VED. FOTOGRAFIA 14)***

66) Oscillografo di Lemoin (galvanometro per correnti rapidamente variabili)
1221
autocostruito (Vogini)
1933/39 ***(VED. FOTOGRAFIA 15)***

67) Cassetta di resistori
AP. 2996
Allocchio - Bacchini
Collaudata nel 1939

68) Cassetta di resistori
2834
Delft (Olanda)
???

69) Fenditura con vite micrometrica
1236
S.I.A.S. (Società Italiana Apparecchi Scientifici), Milano
1940

70) Tubo di Crookes a pera
1237
?
1940 ***(VED. FOTOGRAFIA 16)***

71) Microfono a laringe
(non inventariato)
Esercito americano
?

72) Modelli (in legno) per facilitare il riconoscimento di bombe
n. 54 *(c/o Lab. di Chimica)*
???
~1945

73) Macchina a vapore
1310
Ottica Gentili, Bergamo
1944 (***VED. FOTOGRAFIA 4***, che mostra anche lo strumento inv. n.757)

74) Telefono sfilabile
1339
S.I.A.S., Milano
1947

75-77) Voltmetri: da 15, 80 e 400 volt di fondo scala
1341, 1342, 1343
S.I.A.S., Milano
1947 ***(VED. FOTOGRAFIA 17)***

78) Igròmetro a capello
1365
Tironi, Bergamo
1947

79) Contagiri
1385
(dono della Caproni)
1947

80) Anemòmetro sezionato
1389
(Dono della Caproni)
1947

81) Dinamo/Alternatore, per c.c. e per c.a.
1422
?
1948

82) Condensatore di Epino
1429
ITAS ??
1948

83) Tubo di Braun
1432
S.I.A.S., Milano
1948 (***VED. FOTOGRAFIA 18***, che mostra pure l0 strumento n.84 [inv. n.2024])

84) Alimentatore per Tubo di Braun
2024
S.I.A.S., Milano
1948 (***VED. FOTOGRAFIA 18***, che mostra pure lo strumento inv. n.1432)

85) Motorino elettromagnetico rettilineo, con piccola pompa (rotti)
1441
Secretan, Parigi
1948

86) Pila Leclanché, piccola
1453
?
1948 *(VED. SCHEDA)*

87) Microscopio ottico, con spettrometro a due viste appaiate (per confronto diretto di due spettri)
1690 *(c/o Lab. di Chimica)*
Koristka, Milano
1948 ? ***(VED. FOTOGRAFIA 19)***

88) Motorino con elettromagnete a U, e volano
1942
Secretan, Parigi [dono dell'alunno Marinoni]
1948 ***(VED. FOTOGRAFIA 20)***

89) Apparecchio rotante, per studio di fototropismo
2089 *(c/o Lab. di Chimica)*
???
???

90) Cassetta di derivatori
2834
Kipp, Delft (Olanda)
1953

91) Reostati
2839

Belotti (Bergamo?)
1953

92) Termoscopio di Looser, con numerosi accessori (per misure di Effetto Joule, in resistenze e in gas)
2841
???
1953 *(VED. SCHEDA)* ***(VED. FOTOGRAFIA 21)***

93) Raddrizzatore termoionico
2959
Philips
195...

94) Macchina per proiezioni
???
???
???

95) Bobina di induzione
3000
???
1954

96) Macchina per rotazione
3002
Marelli
???

97) Contatore Geiger con altoparlante, più accessori (contacolpi, tubo di Braun)
3082
S.I.A.S., Milano
1958

98) Bilancia
P3106 (massiera: P3105) *(c/o Lab. di Chimica)*
???
? ***(VED. FOTOGRAFIA 22)***

99) Camera (a nebbia) di Wilson
3120
Chima, Milano
1959

100) Ebulliometro di Vidal-Malligand
P. 3108 ? (c/o Lab. di Chimica)
Kleemann, Vienna
1930 ???

101) Apparecchio per termologia (??)
(non inventariato)
Galileo
???

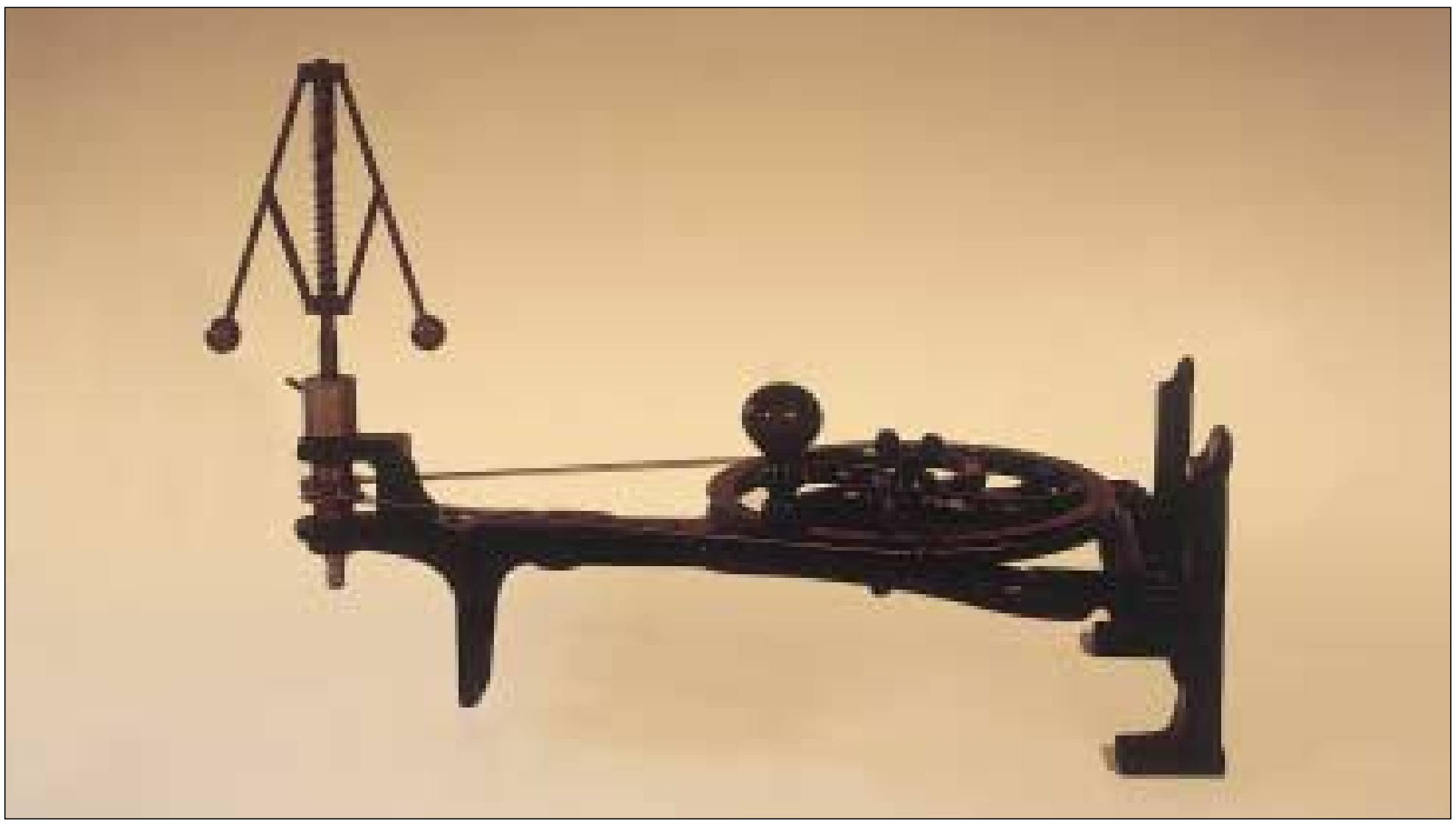

*Foto 1 - Macchina per produrre rotazione.*

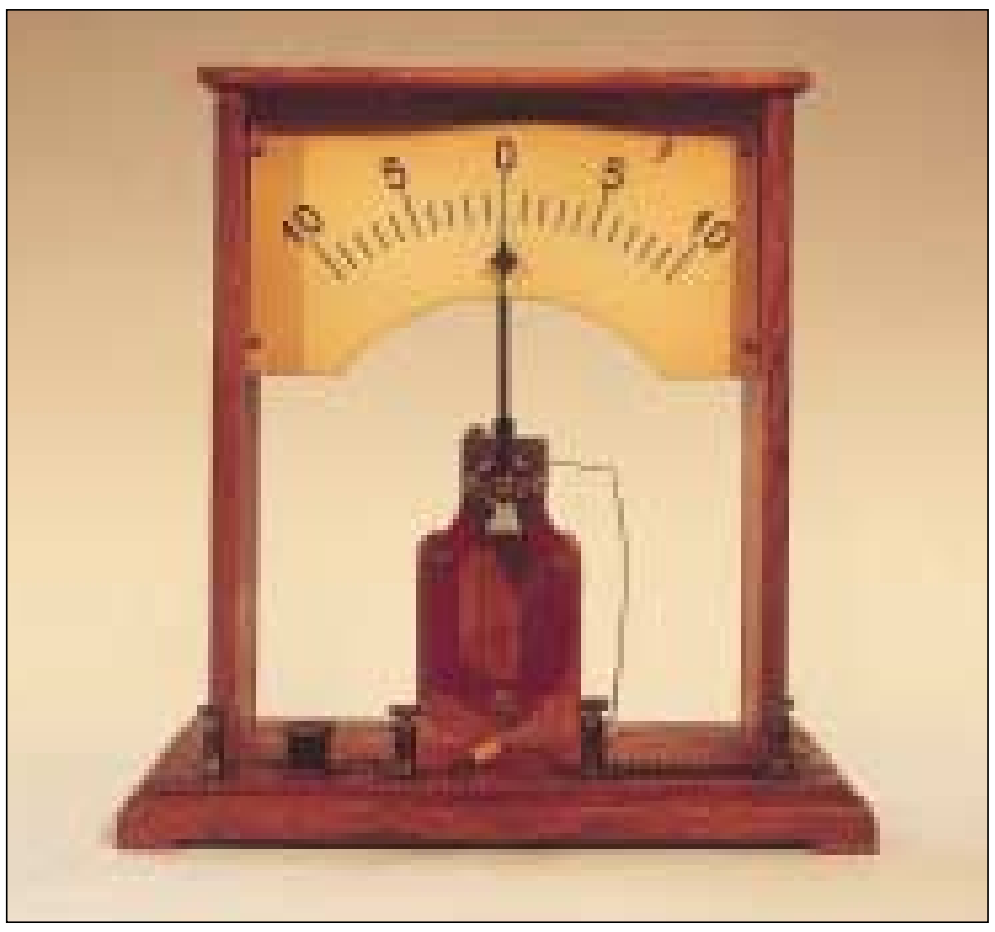

*Foto 3 - Galvanometro verticale a bobina mobile.*

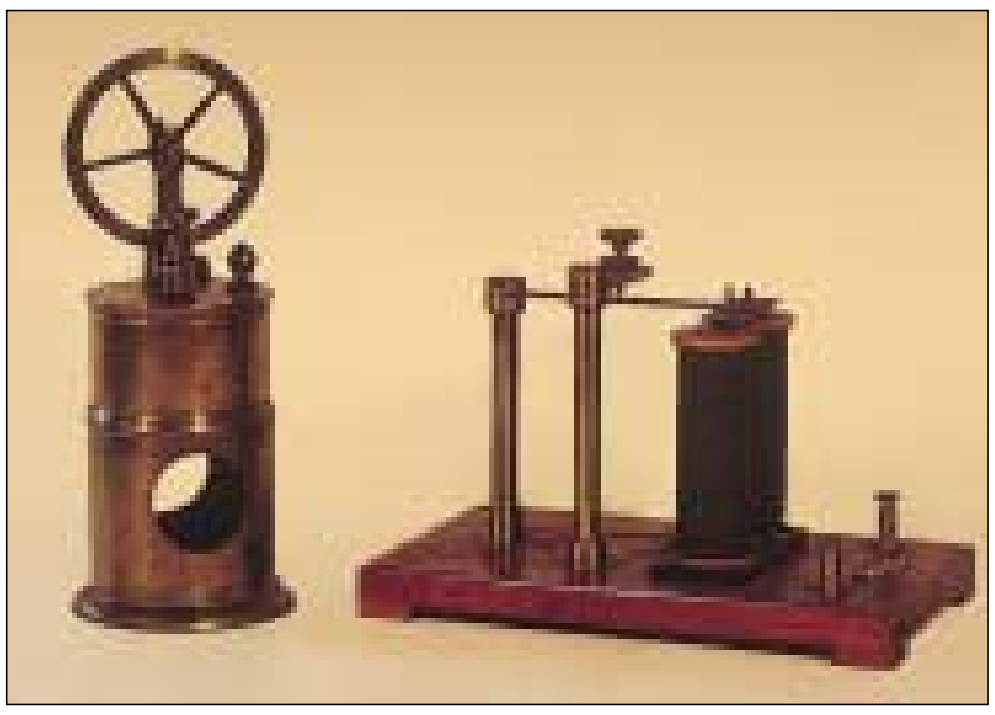

*Foto 4 - Macchina a vapore e Martelletto di Wagner.*

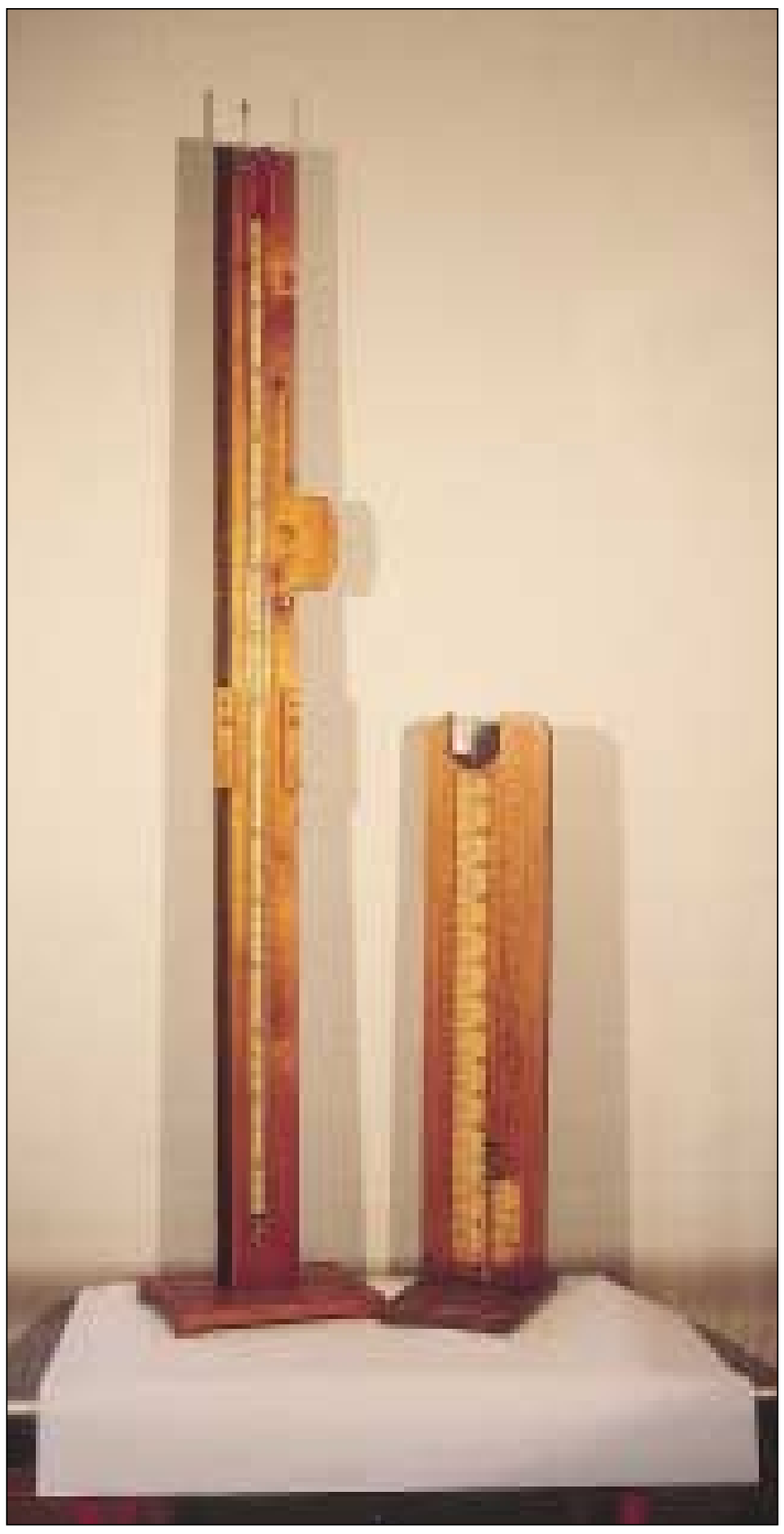

*Foto 5 - Apparecchio per verifica della legge di Boyle (secondo Mueller), con termometro ad aria.*

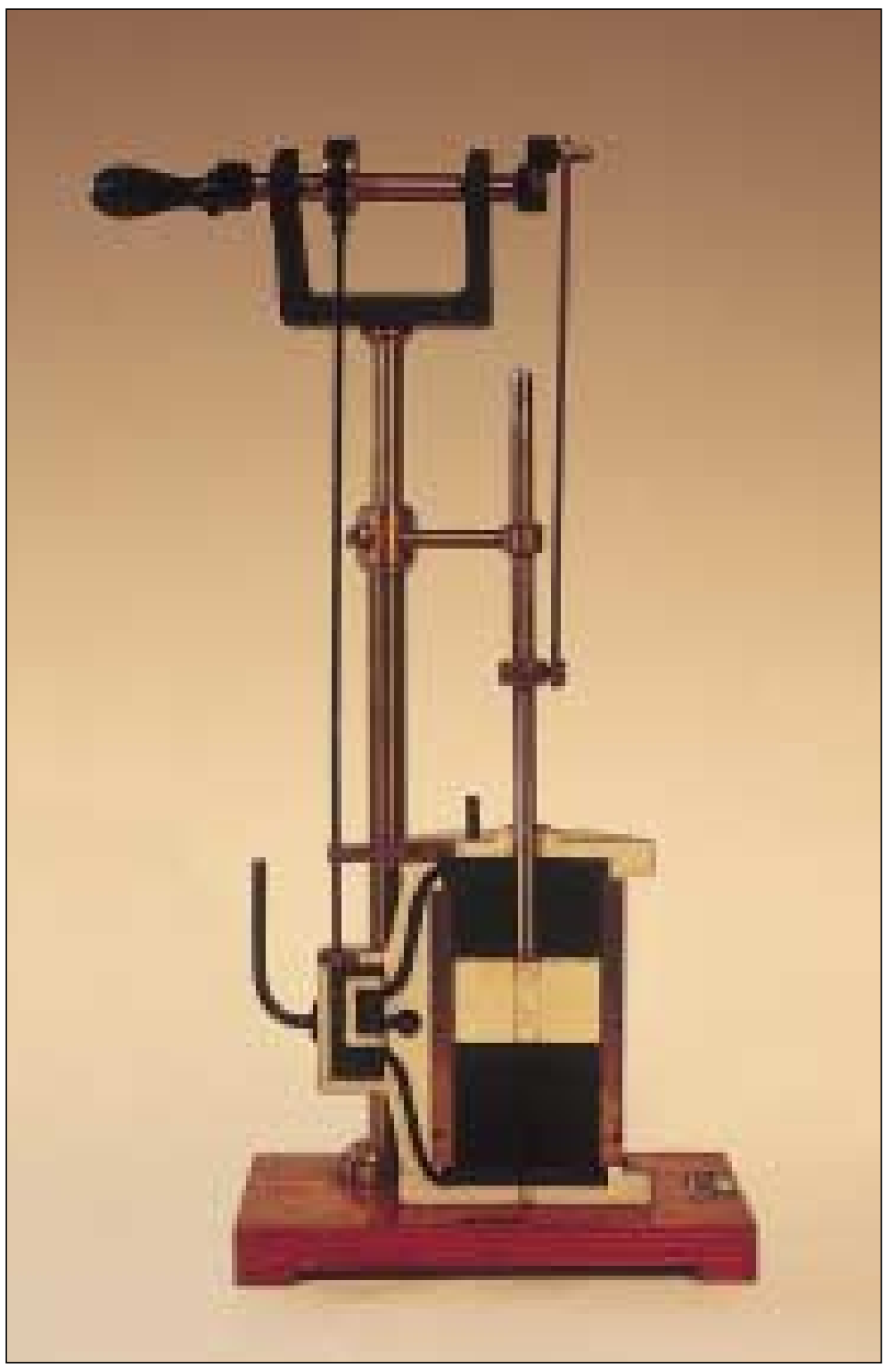

*Foto 2 - Modello della distribuzione in macchina a vapore: con biella e manovella.*

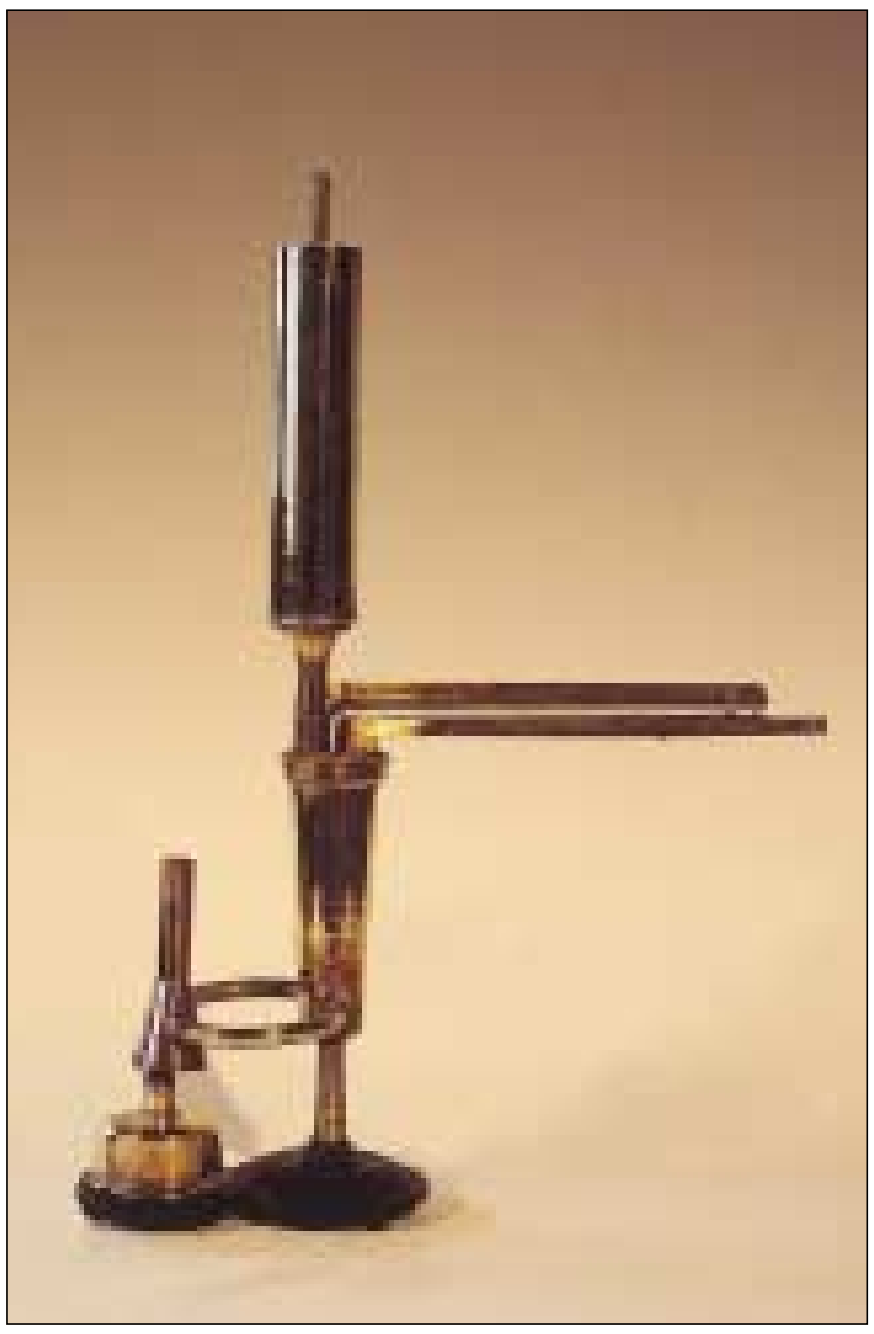

*Foto 23 - Ebulliometro di Vidal-Malligand.*

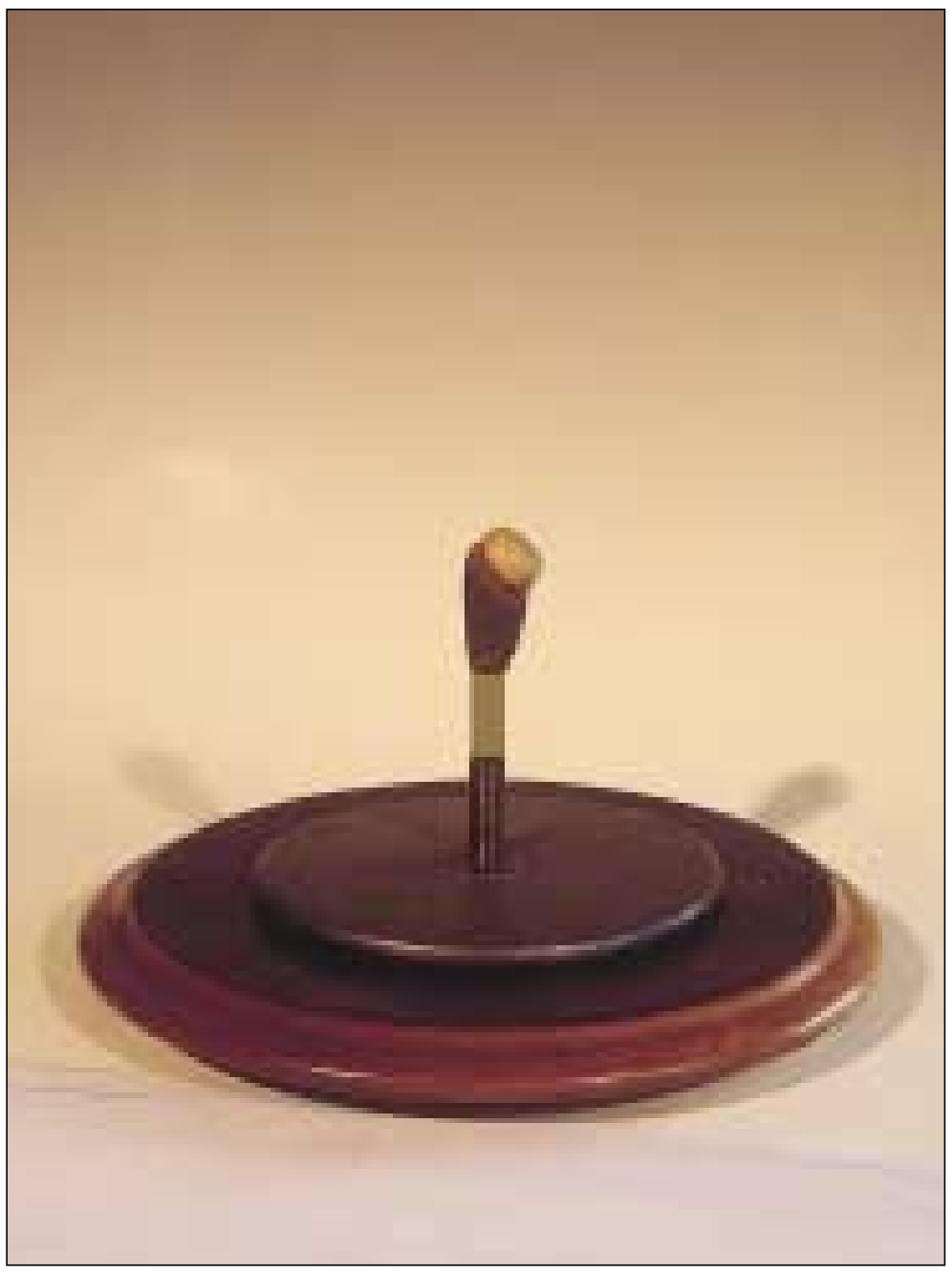

*Foto 11 - Elettròforo di Volta.*

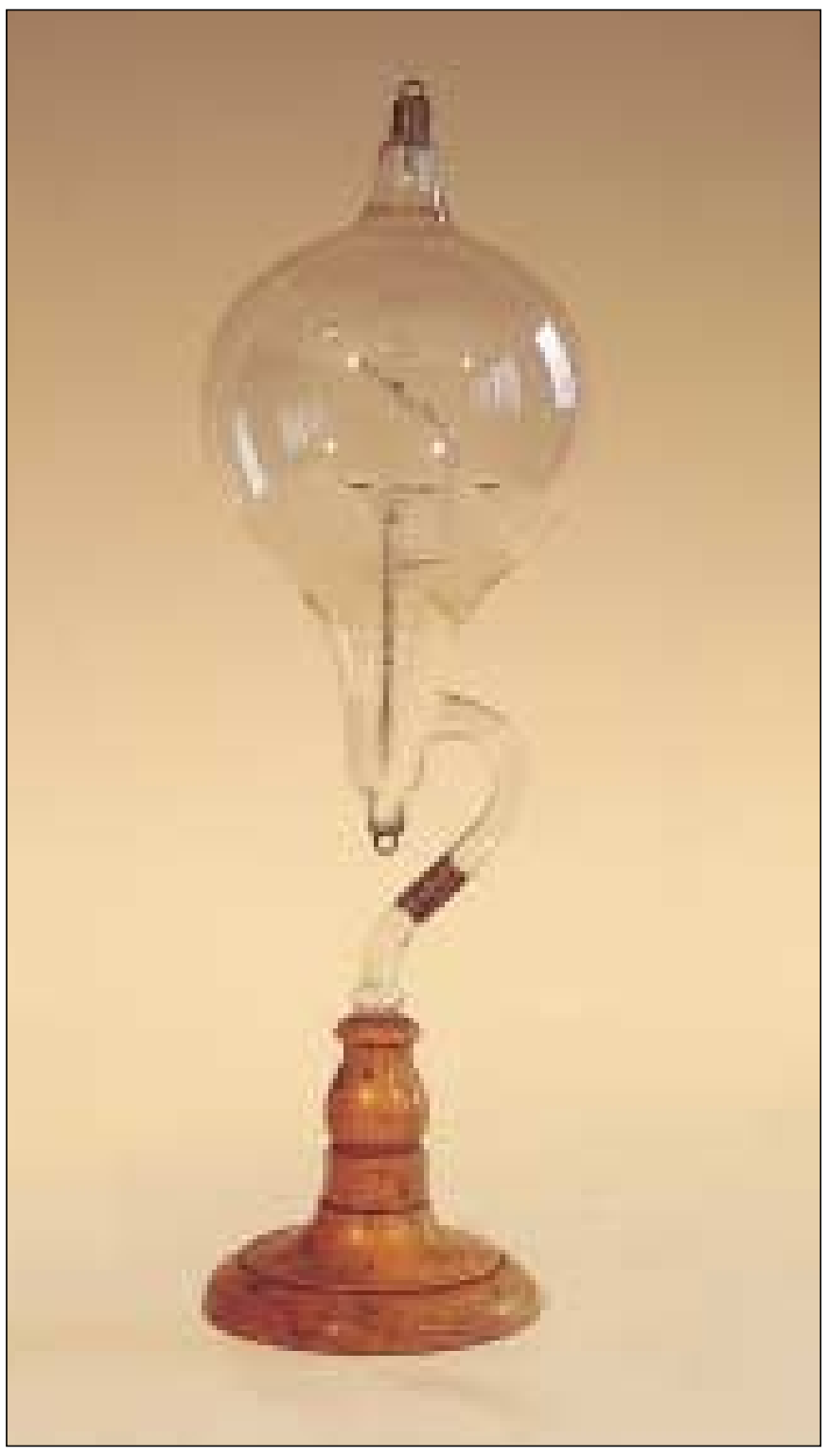

*Foto 16 - Tubo di Crookes a pera.*

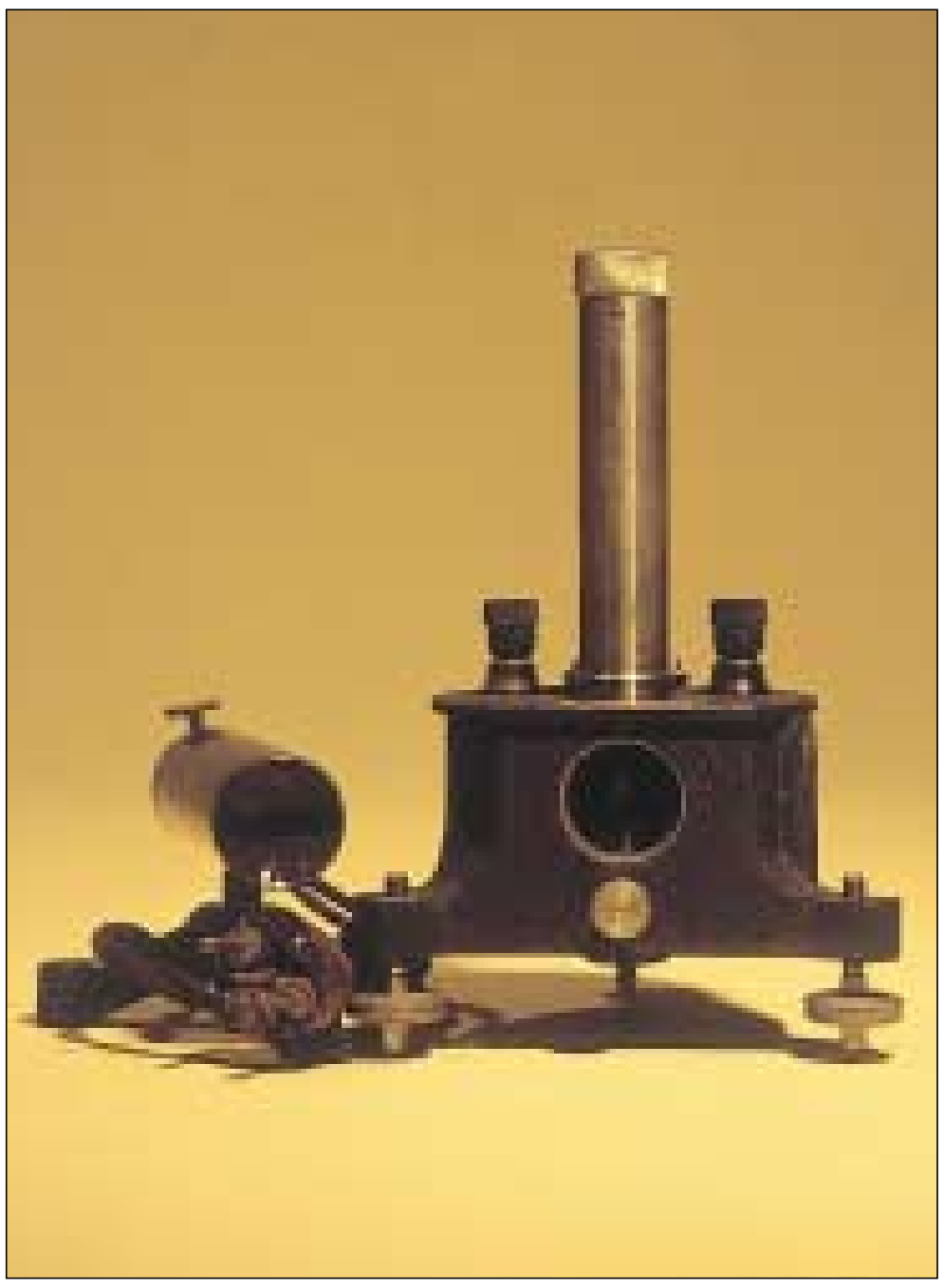

*Foto 13 - Galvanometro a riflessione.*

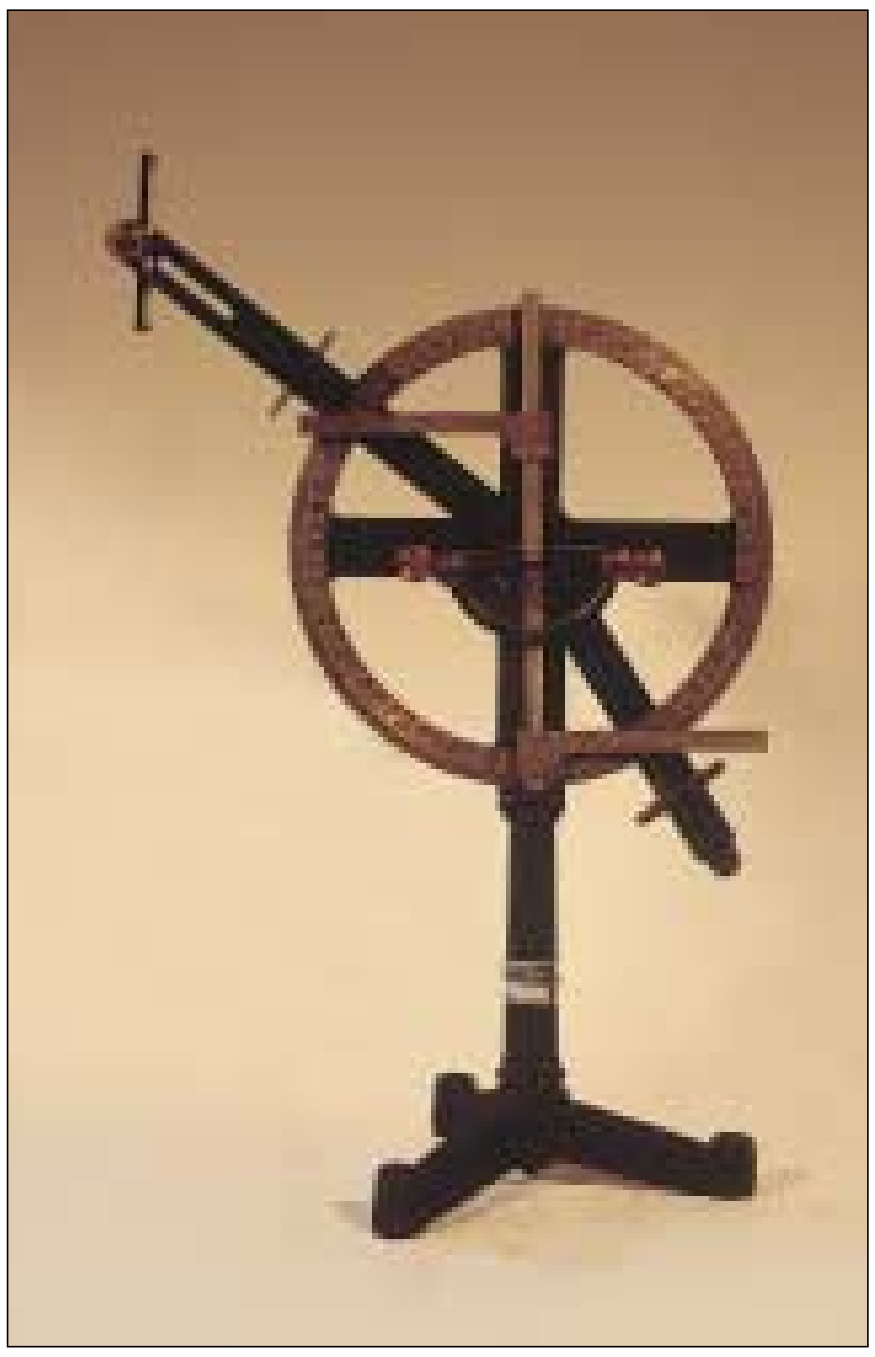

*Foto 8 - Apparecchio universale per riflessioni e rifrazioni.*

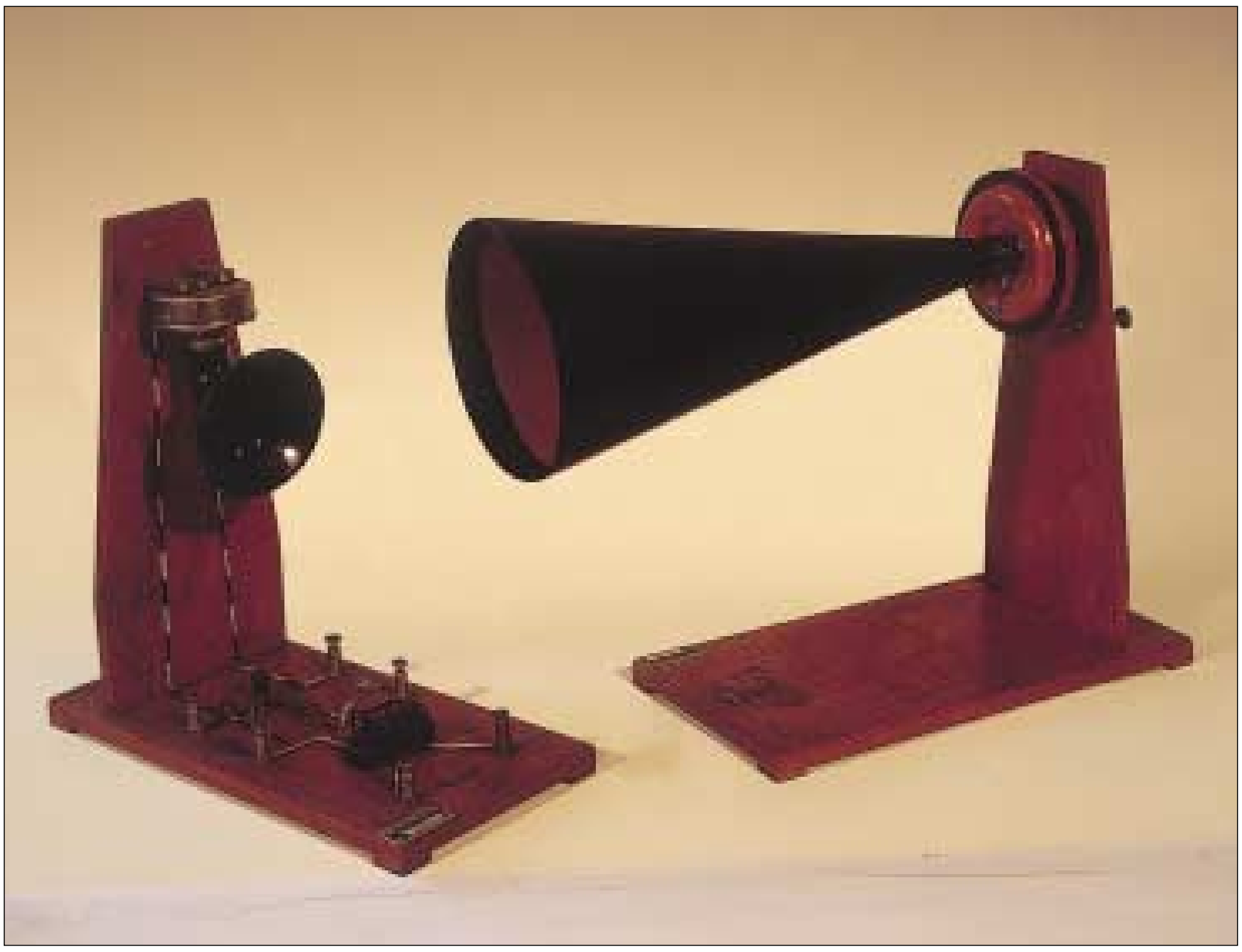

*Foto 6 - Stazione telefonica trasmittente, con microfono ad alta sensibilità.*

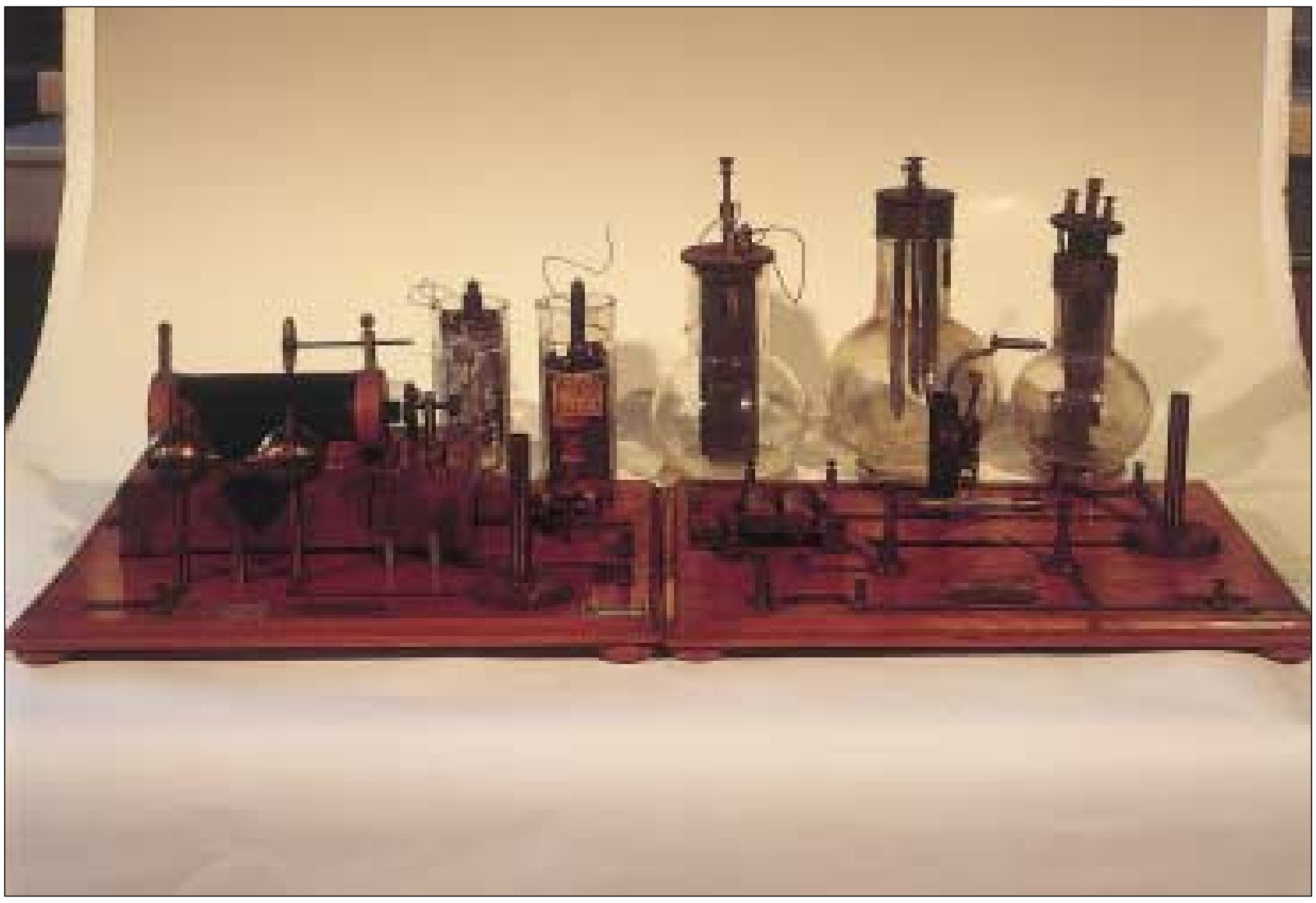

*Foto 7 - Stazioni radiotelegrafiche trasmittente e ricevute, con due pile Grenet e tre pile Leclanché.*

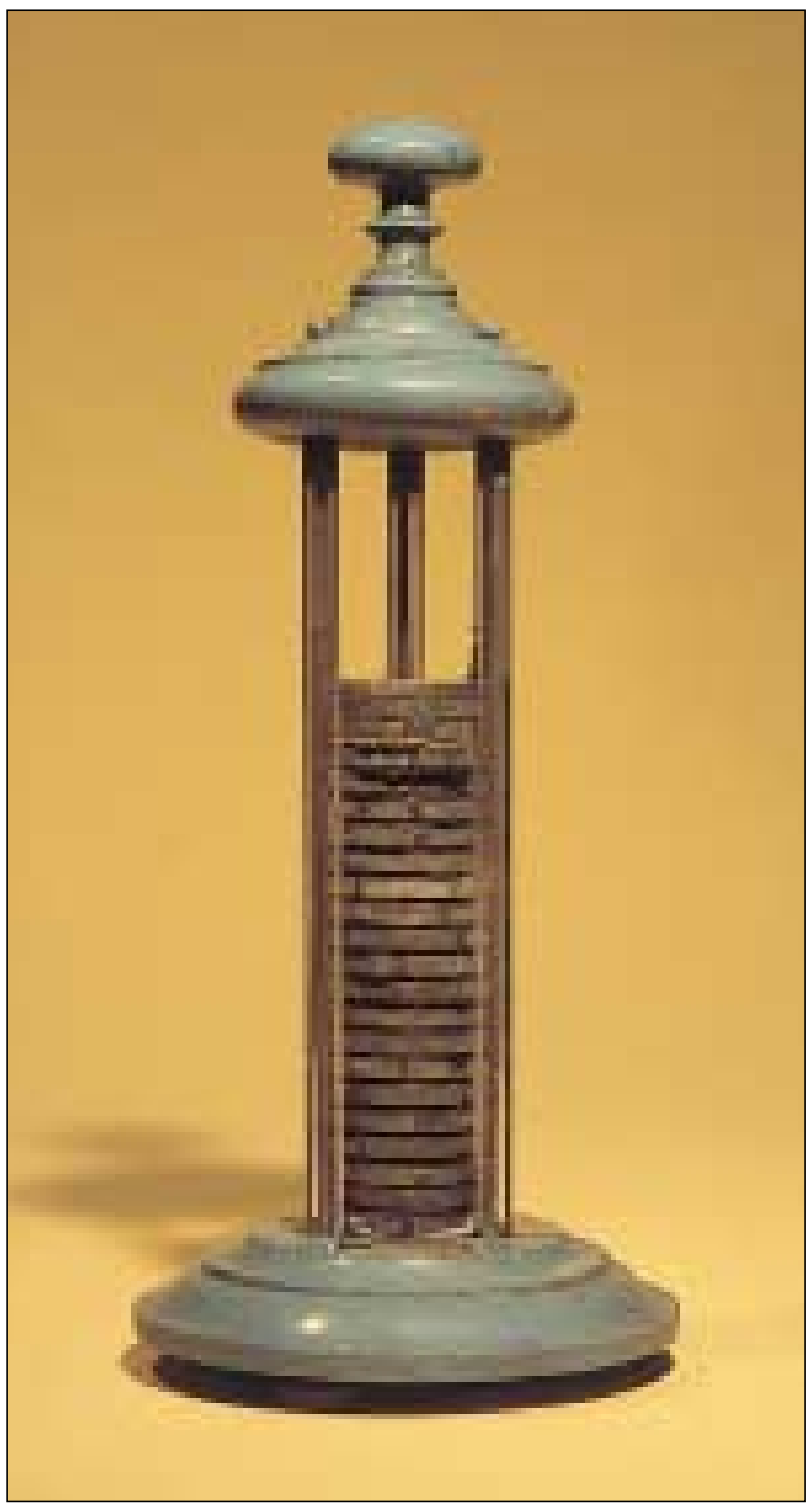

*Foto 9 - Riproduzione della pila di Volta a colonna.*

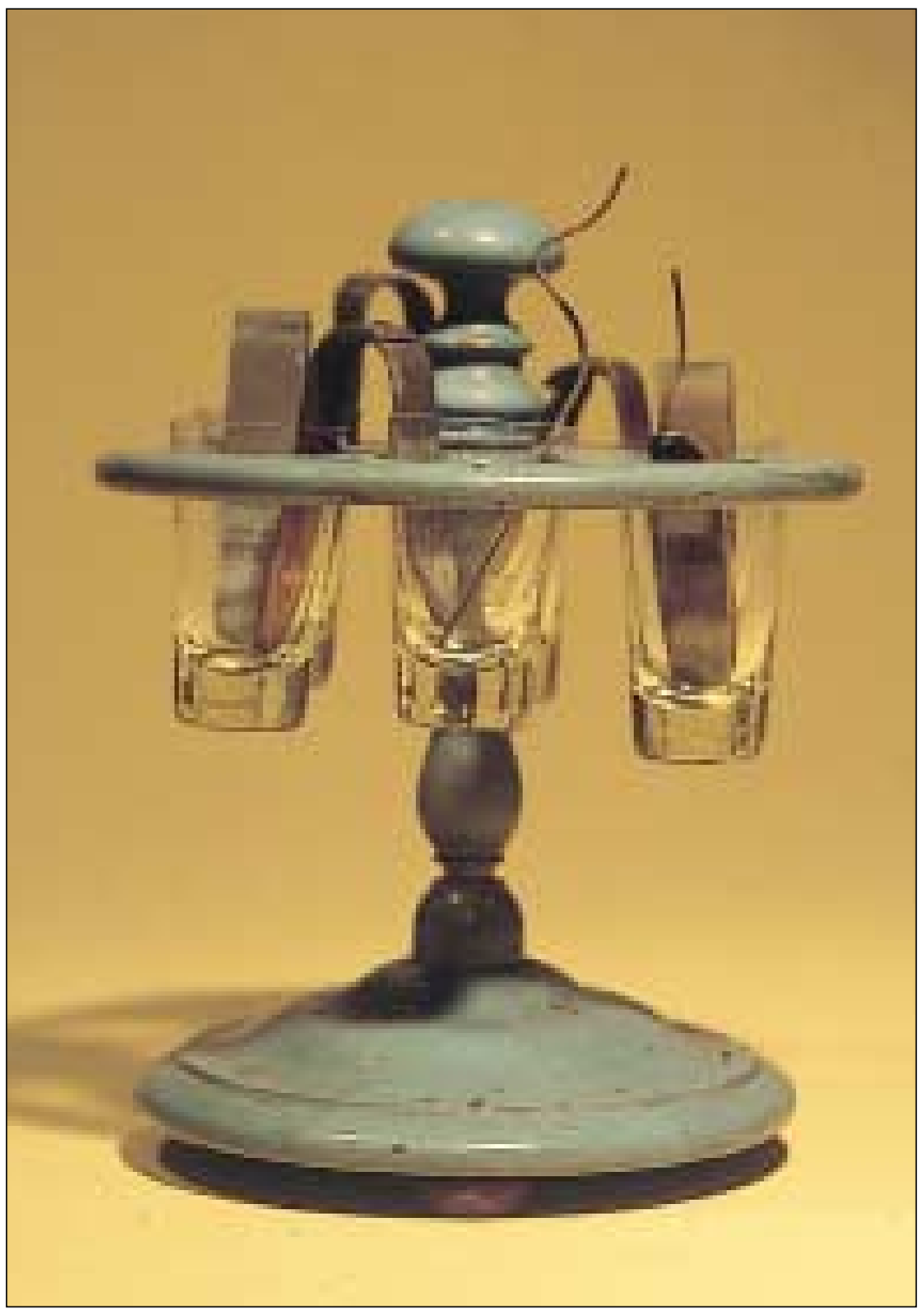

*Foto 10 - Pila di Volta a vaschette.*

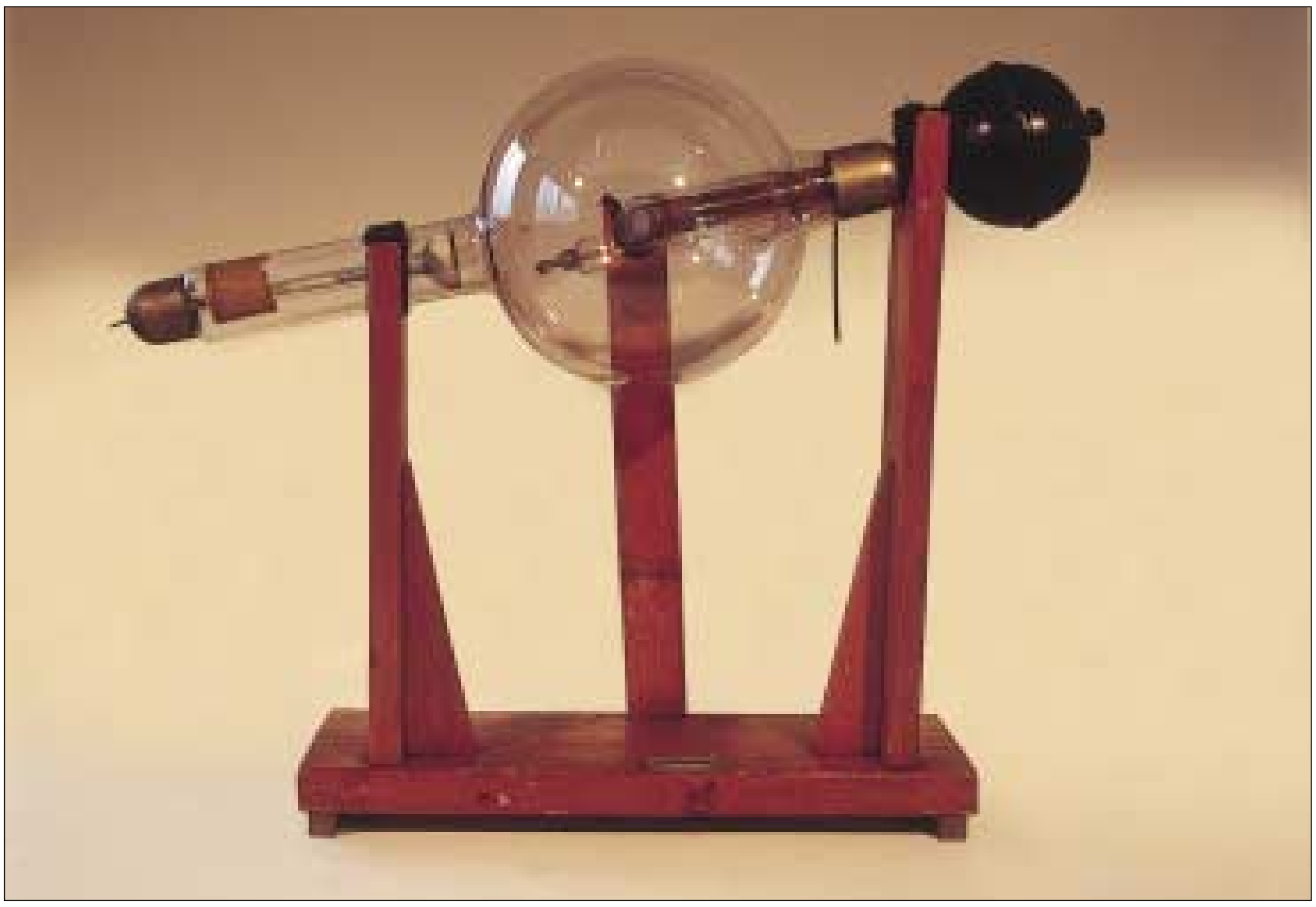

*Foto 12 - Tubo per raggi X, con boccia di raffreddamento.*

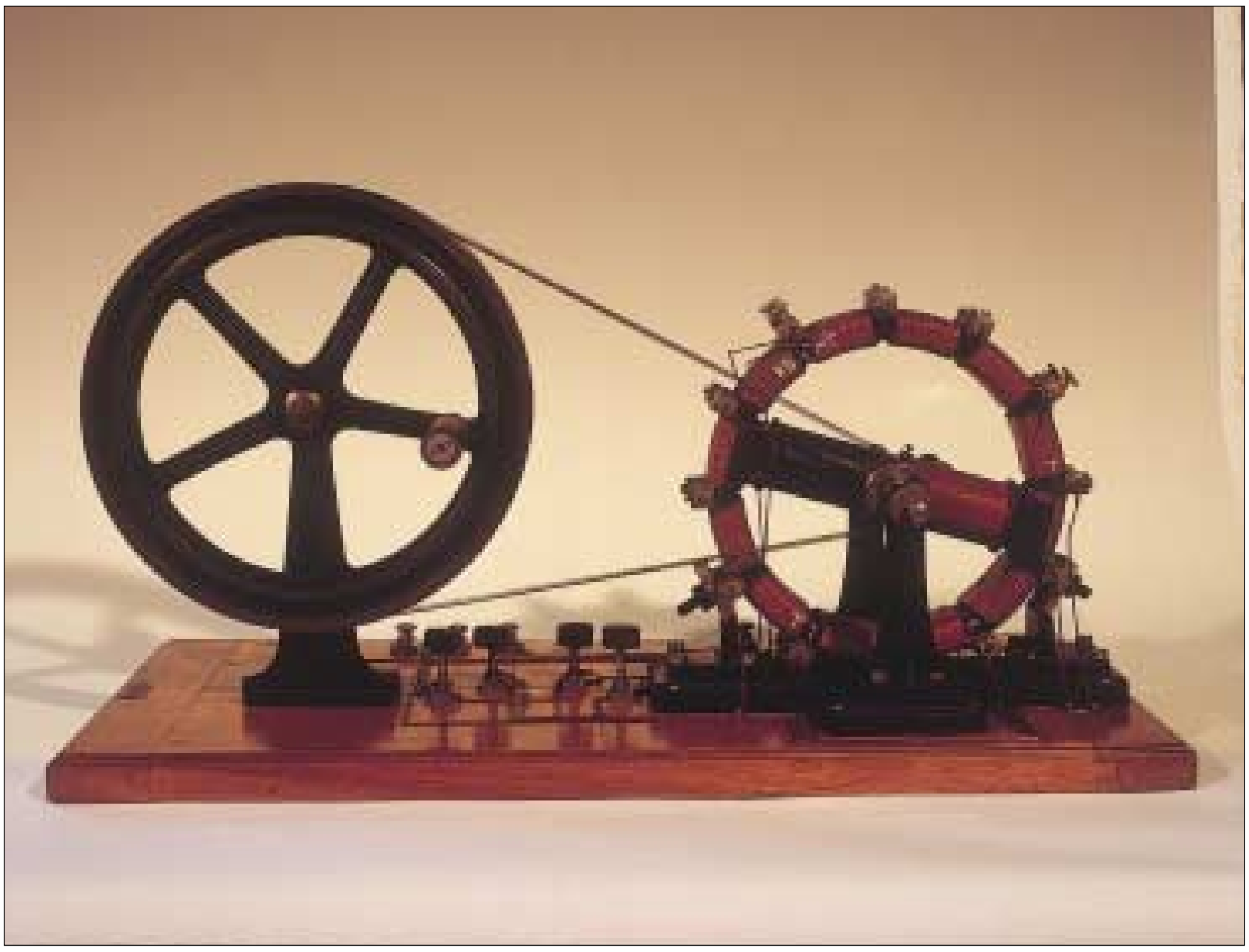

*Foto 14 - Anello di Pacinotti.*

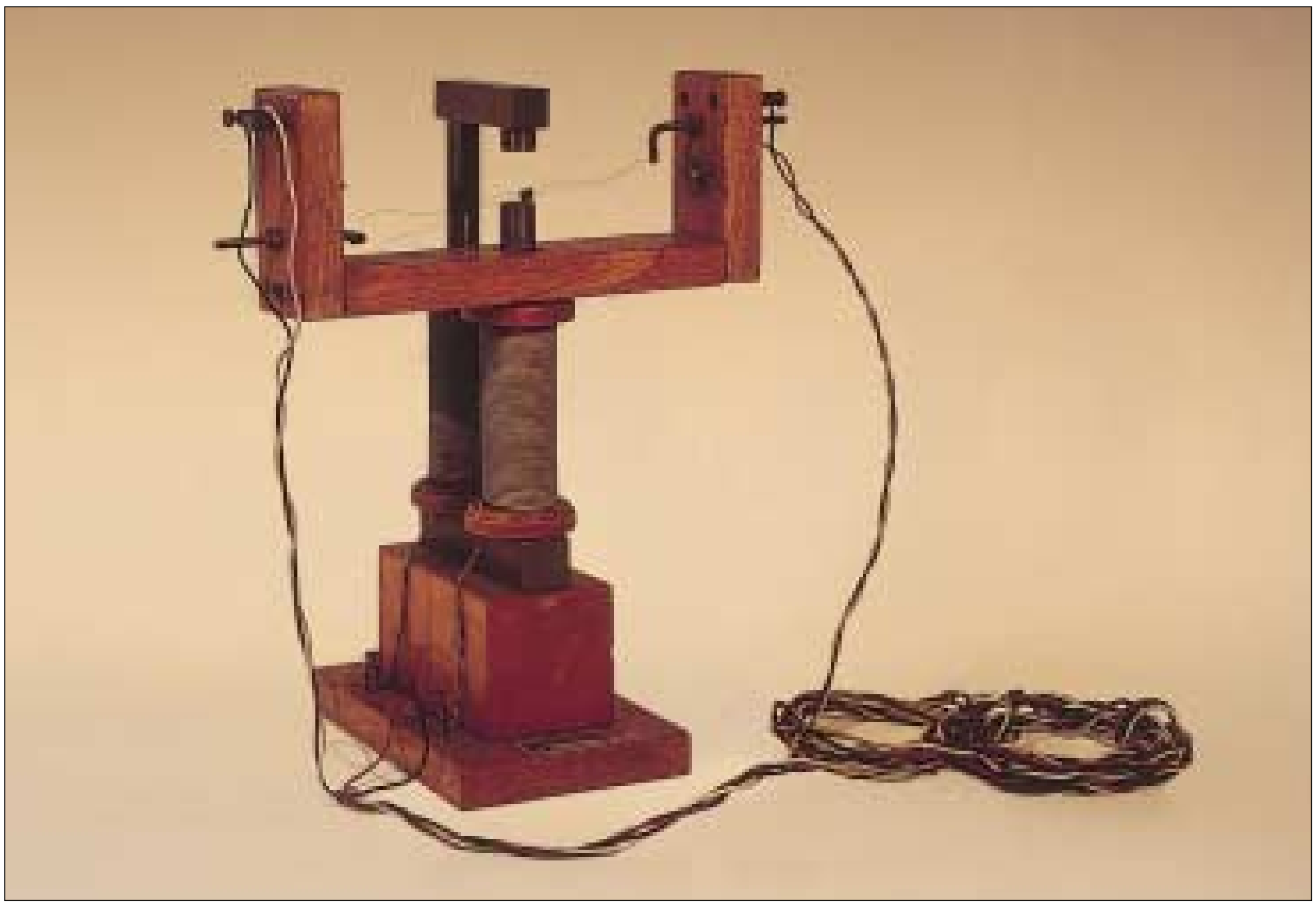

***Foto 15 - Oscillografo di Lemoin (galvanometro per correnti rapidamente variabili).***

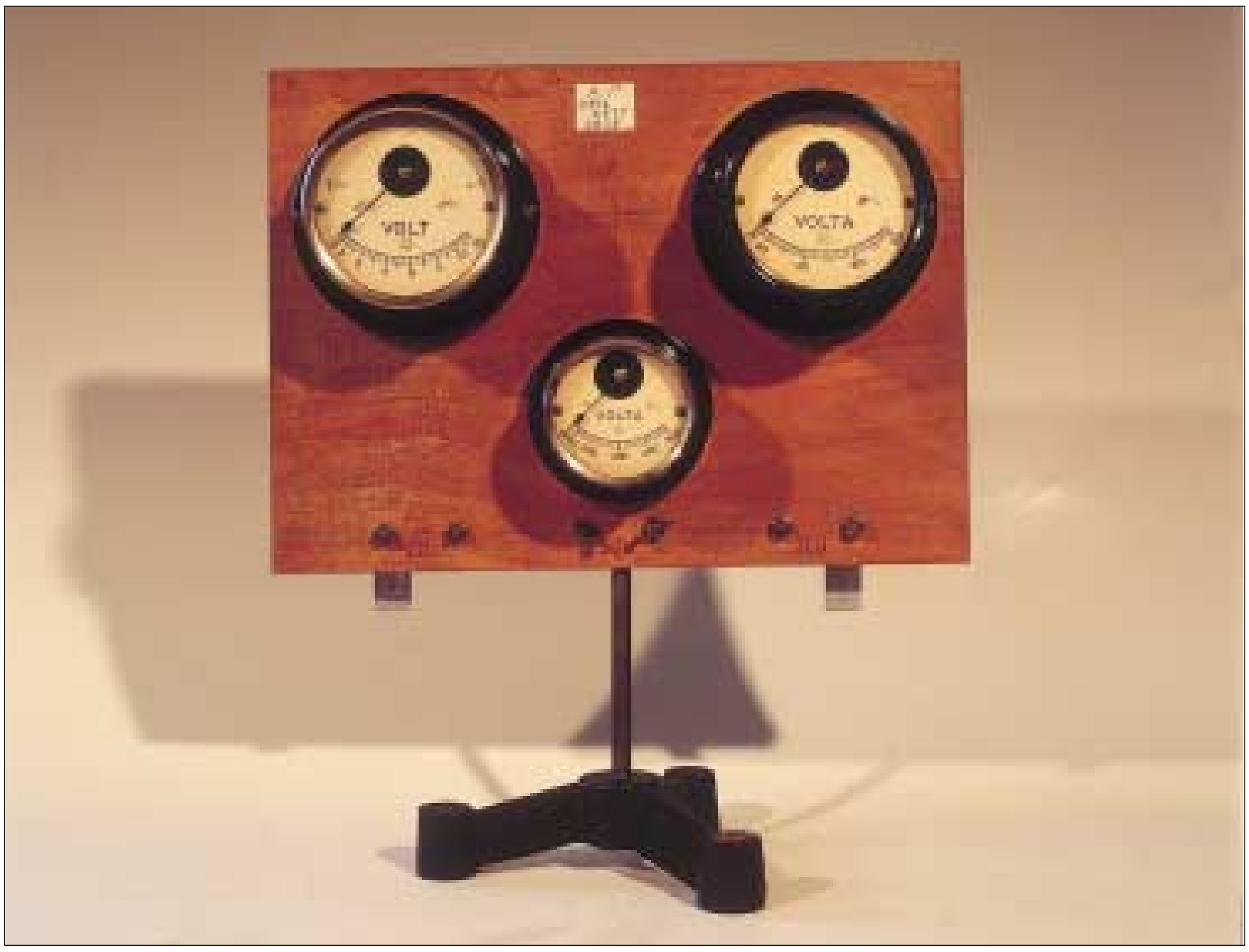

***Foto 17 - Voltmetri: da 15, 80 e 400 volt di fondo scala.***

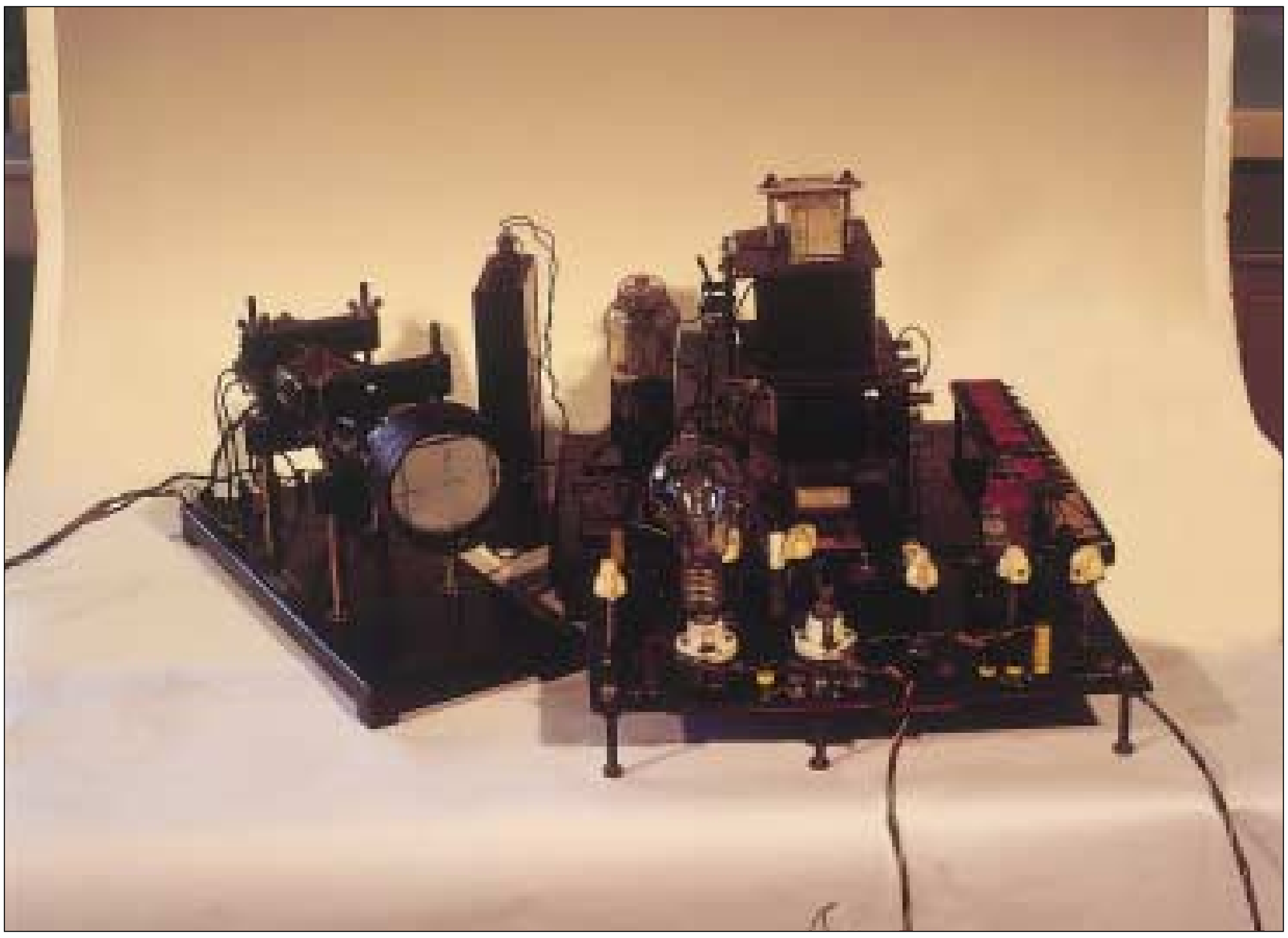

*Foto 18 - Tubo di Brown, con alimentatore.*

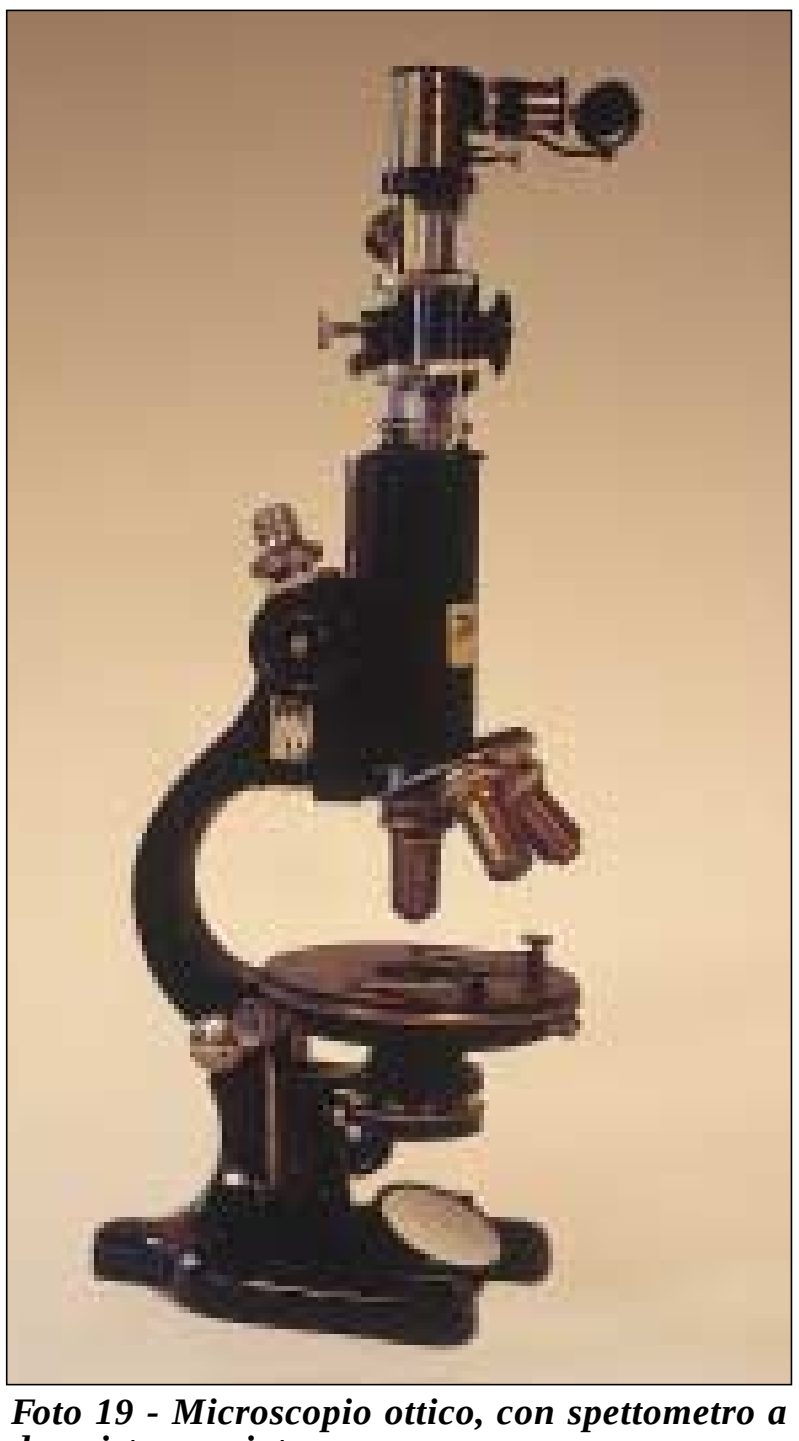

*Foto 19 - Microscopio ottico, con spettometro a due viste appaiate.*

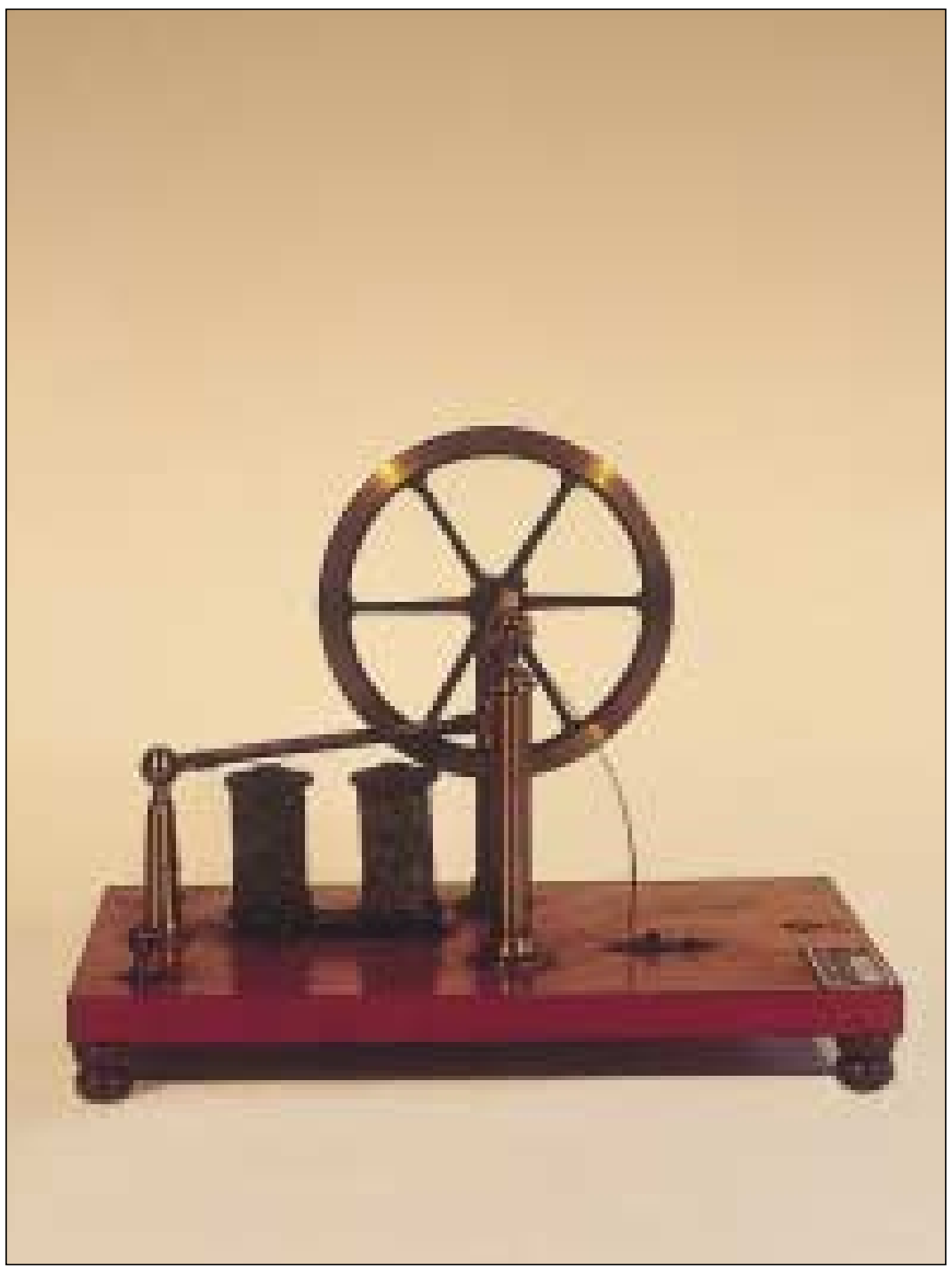

*Foto 20 - Motorino con elettromagnete a U, e volano.*

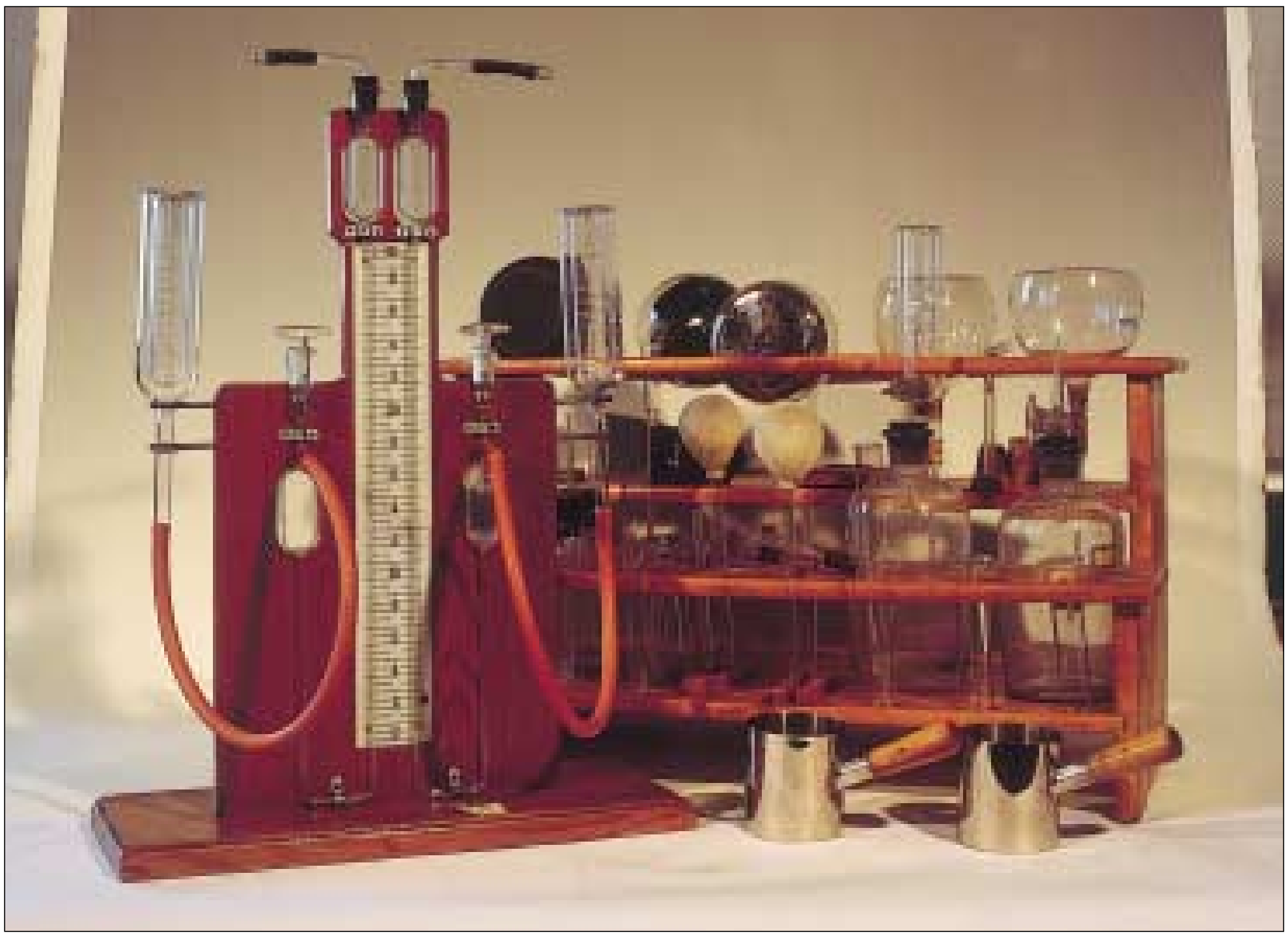

*Foto 21 - Termoscopio di Looser, con numerosi accessori*

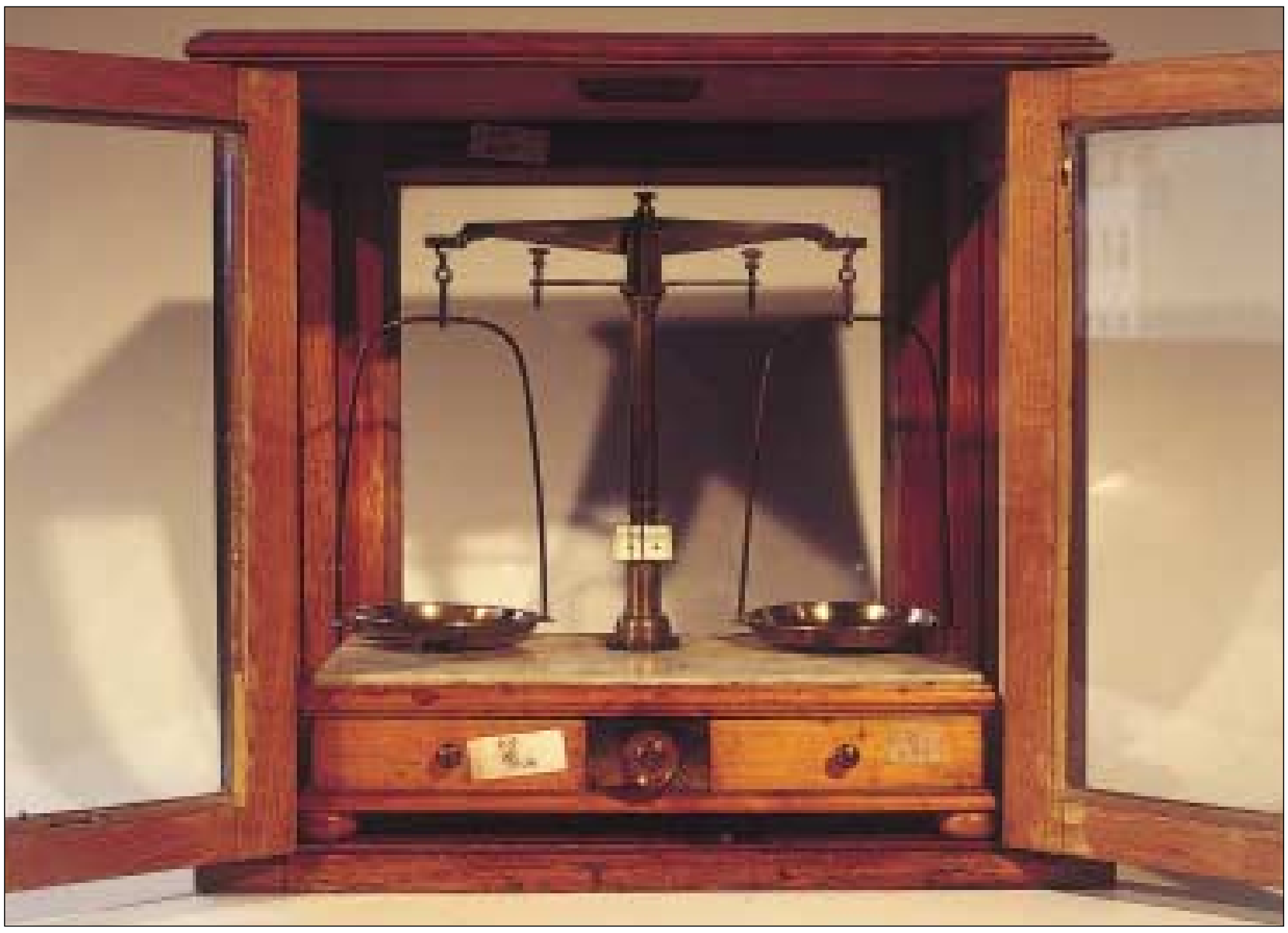

*Foto 22 - Bilancia analitica*

## *2.2 - **Elenco degli Strumenti dell'I.T.C. «Vittorio Emanuele»***

Elenco degli Strumenti dell'Istituto Tecnico Commerciale «Vittorio Emanuele II» di Bergamo. In questo elenco non appare il numero di catalogo degli strumenti.

1) Modello di organo
?
1870

2) Macchina di Ramsden
?
? *(VED. SCHEDA)*

3) Metro campione
Istituto Metrico e Metrologico di Roma
1878

4) Galvanometro di Nobili
S.I.A.S. (Società Italiana Apparecchi Scientifici), Milano
1880 *(VED. SCHEDA)*

5) Pendolo di Galileo Galilei
S.I.A.S., Milano
1885

6) Regolatore di Watt
?
1890

7) Pompa aspirante in metallo
?
1893

8) Sonòmetro
?
1896

9) Apparecchio di Hoope
Emilio Resti, Milano
1897

10) «Paradosso meccanico»
S.I.A.S., Milano
1900

11) Dilatometro
Emilio Resti, Milano
1900

12) Anello di Gravesande
?
1900

13) Pile Leclanché
S.I.A.S., Milano
1900 *(VED. SCHEDA)* ***(VED. FOTOGRAFIE 24 e 25)***

14) Tubi di Crookes (cinque)
S.I.A.S., Milano
1900 ***(VED. FOTOGRAFIE 26 e 27)***

15) Densimetro
?
1900

16) Coppia di Diapason
Emilio Resti, Milano
1900

17) Apparecchio di Pellat
S.I.A.S., Milano
1907 *(VED. SCHEDA)* ***(VED. FOTOGRAFIA 28)***

18) Barometro di Torricelli
Officine Galileo
1910

19) Pila di Volta
S.I.A.S., Milano
1910 *(VED. SCHEDA)* ***(VED. FOTOGRAFIA 29)***

20) Elettrocalamita
Emilio Resti, Milano
1910

21) Sferòmetro
Istituto Metrico e Metrologico di Roma
1910

22) Barògrafo Registratore Richard
S.I.A.S., Milano
1920 *(VED. SCHEDA)* ***(VEDERE FOTOGRAFIA 30)***

23) Rotaia di Laplace
S.I.A.S., Milano
1920

24) Rocchetto di Ruhmkorff
S.I.A.S., Milano
1920 *(VED. SCHEDA)* ***(VED. FOTOGRAFIA 31)***

25) Cassetta di Ingenhousz
Emilio Resti, Milano
1920

26) Pendolo di Foucault
Leybold
1920

27) Torchio idraulico
Angelo Cattaneo, Milano
1920

28) Macchina di Whimshurst
S.I.A.S., Milano
1925 *(VED. SCHEDA)* ***(VED. FOTOGRAFIA 32)***

29) Modello di Manometro
Angelo Cattaneo, Milano
1929

30) Telegrafo di Morse
S.I.A.S., Milano
1930 *(VED. SCHEDA)* ***(VED. FOTOGRAFIA 33)***

31) Voltamperometro
S.I.A.S., Milano
1930

32) Galvanometro dimostrativo
S.I.A.S., Milano
1930 ***(VED. FOTOGRAFIA 34 )***

33) Galvanometro di D’Arsonval
S.I.A.S., Milano
1930 *(VED. SCHEDA)*

34) Modelli di valvole: Triodo; Pentodo
Angelo Cattaneo, Milano
1930 ***(VED. FOTOGRAFIA 35)***

35) Condensatore di Epino
S.I.A.S., Milano
1930 ***(VED. FOTOGRAFIA 36)***

36) Anello di Pacinotti
Emilio Resti, Milano
1930 *(VED. SCHEDA)* ***(VED. FOTOGRAFIA 37)***

37) Alternatore polifase
Officine Galileo, Firenze
1935

38) Modello di Distillatore
Carlo Erba
1940

39) Ruota di Barlow
S.I.A.S., Milano
1940 ***(VED. FOTOGRAFIA 38)***

40) Pompa a vuoto
Leybold
1950

41) Disco di Weinhold
Angelo Cattaneo, Milano
1950

42) Raddrizzatore a lamina
Leybold
1950

43) Apparecchio di Joule
Officine Galileo, Firenze
1960 *(VED. SCHEDA)*

44) Bilancia di torsione
Welch Scientific Company, Chicago (Usa)
1960

45) Pila di Grenet
Officine Galileo, Firenze
1965 *(VED. SCHEDA)* ***(VED. FOTOGRAFIA 39)***

Della prima metà del ‘900 esistono inoltre i seguenti strumenti: 46) Coppia di Diapason; 47) Vaschetta di ottone; 48) Termoscopio di Looser; 49) Bilancia elettrostatica (1904); 50) Arganetto idraulico (1920); 51) Torchio idraulico (1920); 52) Gabbia di Faraday (1920); 53) Quadro oscillante (1925); 54) Condensatore variabile (1930); 55) Contatore di Geiger-Mueller; 56) Termoscopio (1950). Con data 1955 esistono poi: 57) Generatore di van de Graf; 58) Conduttori elettrostatici. Infine, del 1960 circa, si segnalano: 59) Vaschetta ondoscopica; 60) Apparecchio per la dinamica dei fluidi; 61) Giroscopio.

### *2.3 -* ***Elenco degli Strumenti dell’I.T.G. «Quarenghi».***

Il catalogo da noi redatto degli strumenti topografici di interesse storico del Quarenghi coincide con l’elenco del materiale schedato: si rimanda quindi al relativo Capitolo

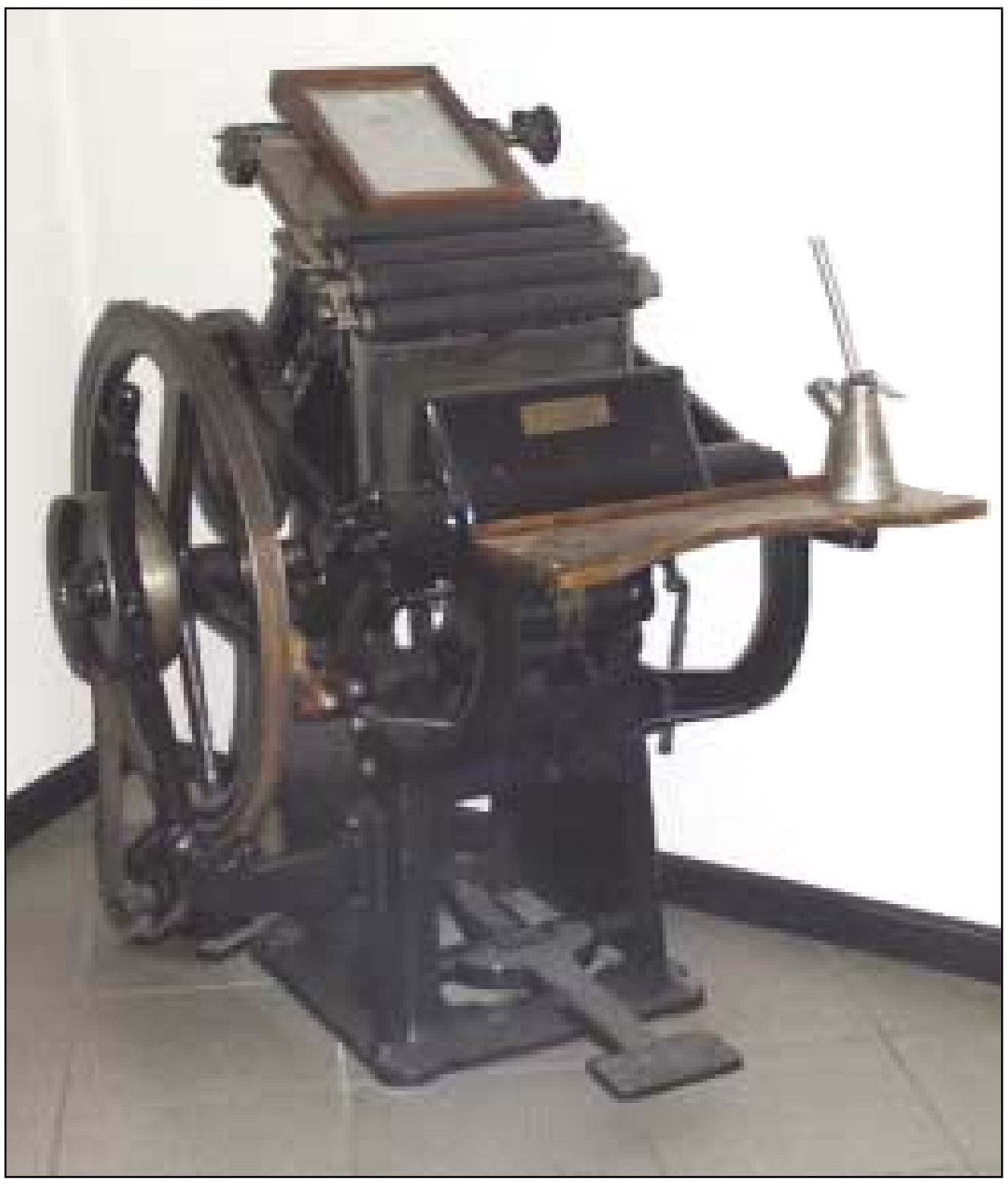

***Foto 23***
***Macchina da stampa tipografica “E. Saroglia”, Torino, del 1922. Essa è proprietà della Tipolito Castel snc di BG, che ha realizzato la produzione di questo volume.***

# 3. «SCHEDE» DEGLI STRUMENTI PIU` SIGNIFICATIVI DI «LUSSANA» E «VITTORIO EMANUELE».

## *GALVANOMETRO DI DEPREZ E D'ARSONVAL*

Nel 1820 Hans Oersted scoprì che un filo percorso da corrente agisce su di un ago magnetico posto nelle sue vicinanze facendolo ruotare in direzione perpendicolare al filo.
La scoperta ebbe immediate ripercussioni all' interno della comunità scientifica e fu seguita da una vera esplosione di ricerche. In particolare emerse la possibilità di misure di correnti attraverso il galvanometro, strumento il cui principio di funzionamento è basato sull' interazione che si esercita tra una bobina percorsa da una corrente ed un magnete fisso. La corrente viene fatta passare in un filo conduttore, che si avvolge formando alcune spire, su un leggero telaietto libero di ruotare attorno ai fili di sospensione (attraverso cui passa la corrente). Il telaietto è sospeso tra i poli nord e sud di una forte calamita permanente.

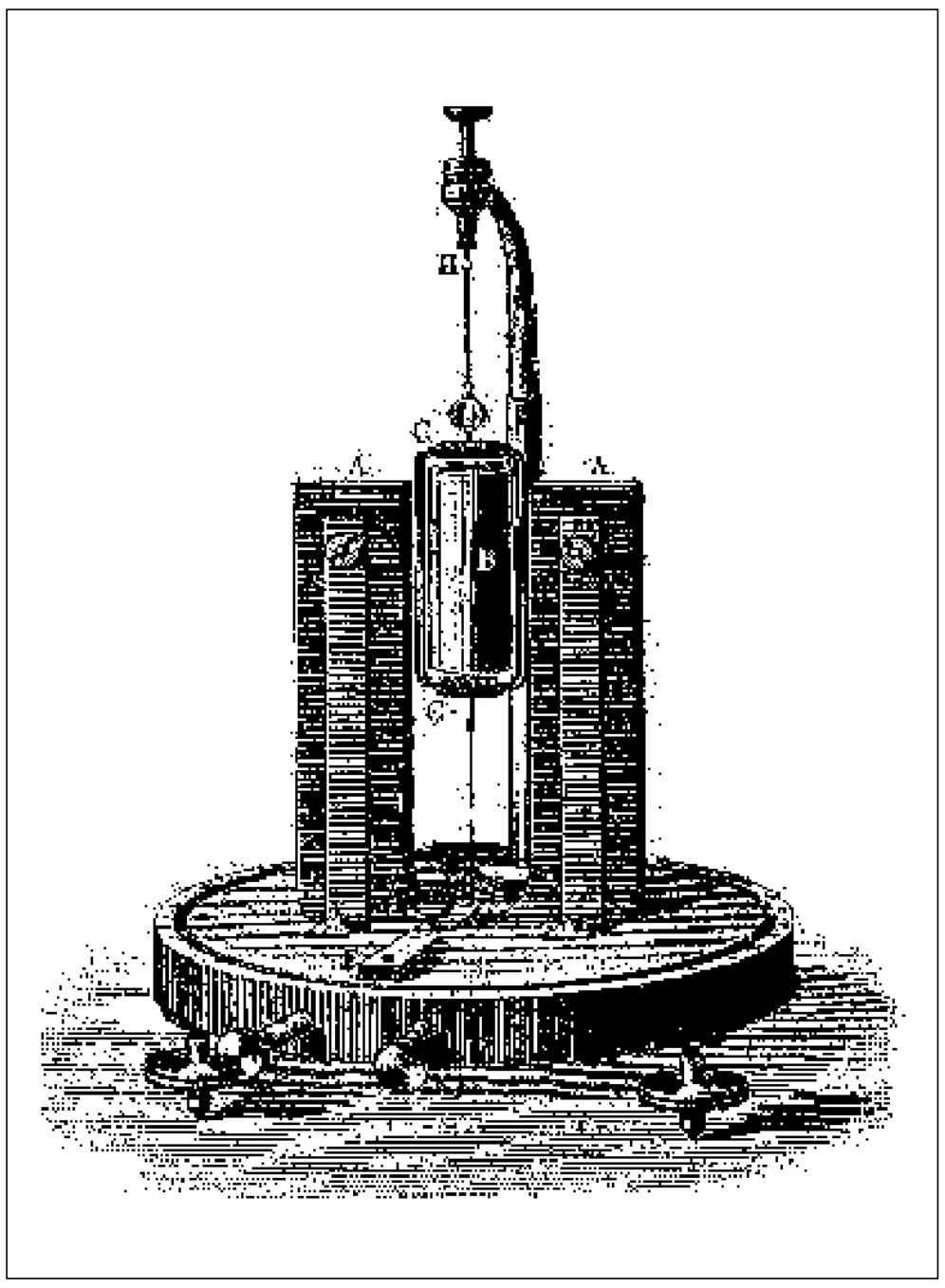

### *DESCRIZIONE*

In questo apparecchio il campo magnetico è generato da un magnete a ferro di cavallo (A-A') in cui i poli sono rivolti verso l' alto e la base è incastrata in uno zoccolo di ebanite. Nella base, sorretta da tre viti calanti, sono incastrate a loro volta in due piccole bacinelle, chiuse da vetri, aventi la funzione di segnalare la posizione orizzontale della base: funzionano in pratica da livella.
Un sottile filo isolato è avvolto attorno ad un quadro (o equipaggio) mobile leggerissimo, sospeso tra due fili metallici, i quali servono sia a portare corrente all' avvolgimento sia a fornire la coppia elastica che risulterà quale reazione alla loro torsione quando passa corrente (fili C-C'). Al centro del quadro vi è un cilindro (B) che rafforza l' effetto magnetico sulla corrente per induzione elettromagnetica.
In alto il filo (H) è sospeso ad un robusto gambo di ottone con un sistema a due movimenti, uno di rotazione della bobina (per disporla nella posizione giusta), l' altro in verticale per regolarne l'altezza. Il secondo filo (E) arriva all'estremità di una lamina che è fissata alla periferia dello zoccolo; questa ha una vite al centro che permette di regolare la tensione del filo, e quindi l' intensità del momento di richiamo, per ottenere un funzionamento idoneo dello strumento al passaggio della corrente. Un piccolissimo specchio incollato al filo permette di leggere l'angolo di torsione con grande precisione; infatti esso riflette (su una scala in celluloide disposta ad un metro di distanza dal filo) l' immagine di un crocifilo illuminato posto sotto la scala stessa. Con questo dispositivo il galvanometro ha un indice senza peso, lungo due metri.
Questo accorgimento può essere usato con tutti i galvanometri in cui lo specchio è fissato al filo di torsione. Il cilindro metallico (B) rinforza l' intensità del campo magnetico e quindi anche lo smorzamento del galvanometro, che è completamente aperiodico.

***FUNZIONAMENTO***

Quando passa la corrente, il piano del quadro tende a porsi perpendicolarmente alla retta che congiunge i poli della calamita. L'azione equilibrante all' effetto di rotazione del campo magnetico sulla corrente è svolta dalla torsione elastica del filo.
La bobina è sostenuta da un filo a torsione (H-C); il momento elastico di torsione $m_\alpha = k_m\alpha$ equilibra il momento magnetico $m_i = k_i i$. Per cui all'equilibrio si ha: $k_m\alpha + k_i i = 0$, ovvero: $i = \alpha(k_m/k_i)$.
Lo specchio mostrato in figura e il metodo di Poggendorf servono per la misura di $\alpha$.
Si definiscono $A = i/\alpha$ il fattore di riduzione del galvanometro, e $R = \delta/i$ la costante dello strumento: dove $\delta$ è il numero di divisioni della scala, e $i$ è l'intensità di corrente misurata.
La sensibilità di uno strumento del genere può essere valutata sapendo che, con un circuito di 200 ohm di resistenza e con una durata di oscillazione del sistema mobile di 10 secondi, la deviazione di un millimetro sulla scala distante due metri corrisponde ad una corrente di intensità $i = 22{,}267 \times 10^{-9}$ A.

***USO***

Gli apparecchi a quadro mobile possiedono il vantaggio di non essere influenzati in modo apprezzabile dalle correnti o calamite eventualmente vicine. Inoltre, il movimento del quadro può riuscire smorzato grazie alle correnti d'induzione che si sviluppano quando esso si sposta nel campo della calamita.
Tale smorzamento si può variare, modificando la resistenza del circuito esterno all'apparecchio.

## *BAROMETRO REGISTRATORE RICHARD*

L' apparecchio è costituito da otto capsule, simili ai tamburi di un barometro aneroide, sovrapposte in modo che le loro deformazioni vengano a sommarsi.

I movimenti prodotti da tali deformazioni sono trasmessi, per mezzo di un sistema di leve, ad un indice, portante un pennino che lascia una traccia sopra un foglio di carta millimetrata.

Le capsule cilindriche, a pareti metalliche sottilissime e assai flessibili, di poca altezza, hanno basi larghe e scanalate concentricamente, per aumentarne l' elasticità.

Ogni capsula, fatto il vuoto, viene chiusa ermeticamente: al suo interno contiene una molla a spirale che, avvolgendosi attorno ad un asse mediano, resiste alla pressione esterna.

Le dimensioni delle diverse parti descritte sono tali che l'escursione del pennino rimane amplificata in modo da corrispondere a quella del barometro a mercurio: cioè, una traccia sulla carta ampia un millimetro equivale alla pressione di un millimetro di colonna di mercurio.

La carta diagrammata viene avvolta attorno al cilindro di ottone ed è fissata ad esso con una molla. Essa è rigata orizzontalmente e verticalmente: le linee orizzontali corrispondono al tempo espresso in ore, le linee verticali alla pressione espressa in millimetri di mercurio.

Quando la posizione del pennino non corrisponde alla reale pressione atmosferica, è possibile rettificare lo strumento introducendo l' imboccatura più piccola della chiave di taratura nel foro collocato al di sotto dell' apparecchio: la rettifica viene ottenuta girando a destra o a sinistra la chiave in modo da riportare il pennino nella posizione corretta ottenuta per confronto con un barometro a mercurio di precisione.

La riproducibilità delle misurazioni è influenzata anche dal giusto fissaggio del pennino alla leva di trasmissione; essa sarà ben regolata quando, facendo oscillare l'indice, questo si muove liberamente e la punta ritorna esattamente al suo posto. L'apparecchio è fissato su di una base di legno ed è ricoperto da una custodia a vetri. Un bottone che sporge in basso, muovendosi a destra permette di far avanzare un'asta metallica verticale, con cui si allontana o si avvicina il pennino, quando si vuole cambiare la carta millimetrata o spostare l' apparecchio.

Prima di utilizzare lo strumento, occorre controllare se l'orologio che determina il movimento del tamburo rotante è ben regolato e se l'indicazione della pressione è esatta.

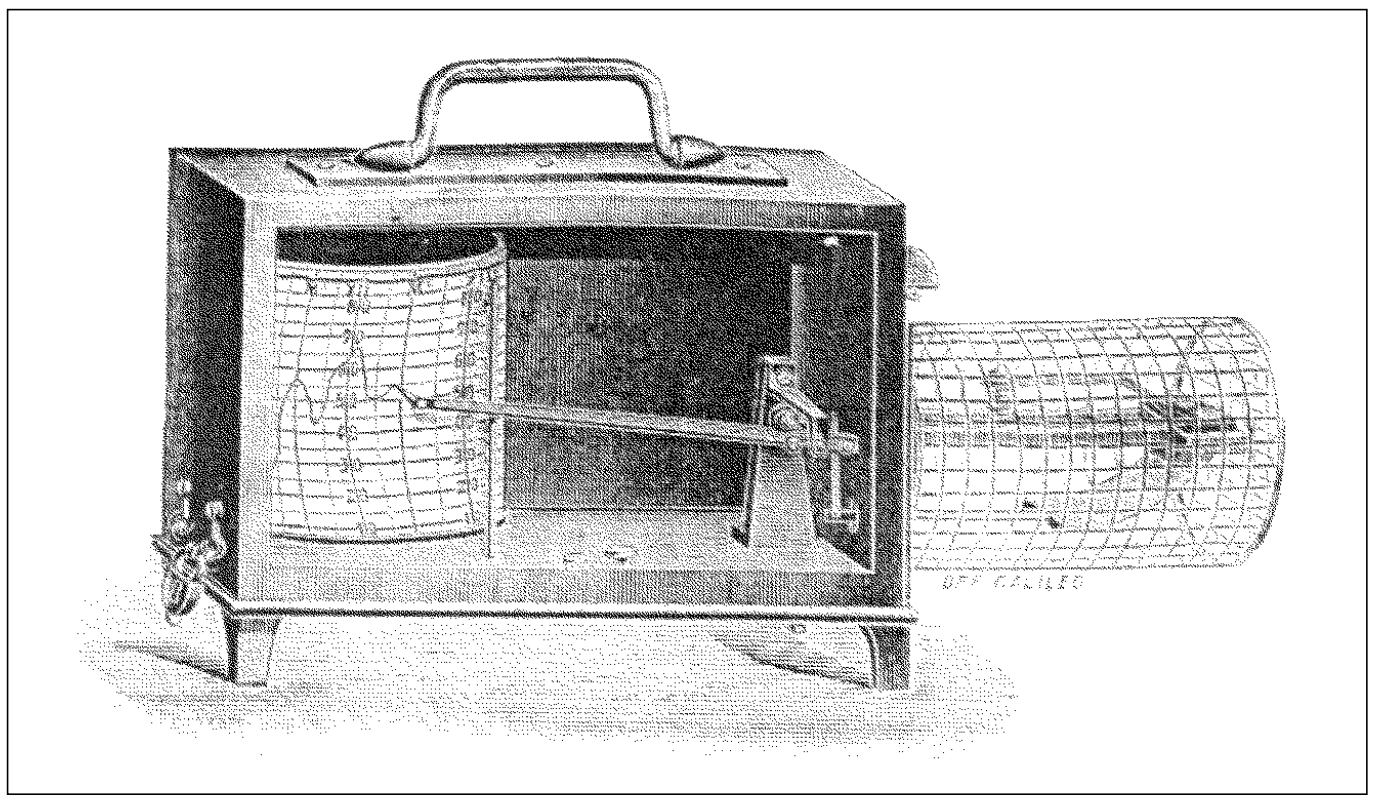

## *APPARECCHIO DI BOYLE - MARIOTTE*

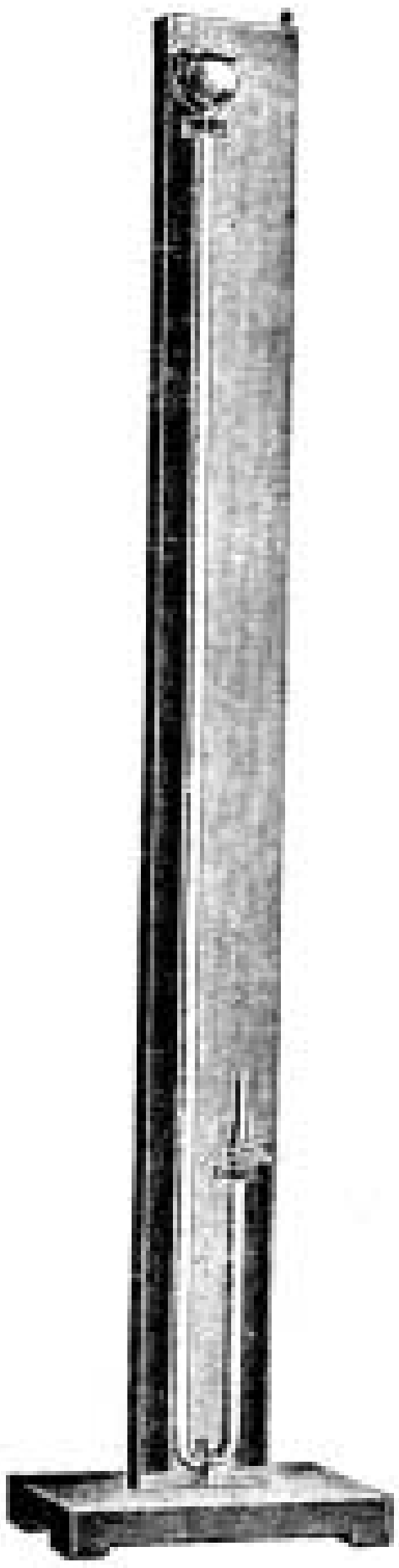

***PREMESSA***

I gas furono sottoposti a trasformazioni varie per misurare la variazione delle loro grandezze macro-scopiche e proprietà tipiche. Uno dei primi studi effettuati storicamente è stato quello relativo a trasformazioni isotermiche del gas, cioè a variazioni della pressione del gas con il volume, a temperatura costante. La legge così scoperta venne enunciata da Boyle e poi verificata da Mariotte. Nel 1684 l' abate Mariotte fu il primo a stabilire la legge della compressibilità del gas:
«Rimanendo la temperatura invariata, i volumi che assume una data massa di gas sono inversamente proporzionali alle pressioni che essa sostiene».

***DESCRIZIONE DELL'APPARECCHIO DI MARIOTTE***

Per l'aria la legge suddetta si verifica con l'apparato detto tubo di Mariotte.

Su una tavoletta di legno, mantenuta in direzione verticale, è fissato un tubo di vetro curvato a sifone i cui due rami hanno diversa lunghezza. Vicino al ramo più breve, che è chiuso, si ha una scala che indica capacità uguali; parallelamente, nel ramo più lungo, si trova pure una scala graduata in centimetri.
Gli zeri di ambedue le scale cadono sopra la medesima linea orizzontale.
Inizialmente si introduce mercurio nei due rami, in modo tale che esso tocchi il livello zero nei due rami. L'aria rinchiusa nel ramo più corto si trova allora alla pressione atmosferica che gravita sulla superficie del mercurio nel ramo più lungo. Si versa del mercurio nel tubo finché la pressione che ne risulta riduce della metà il volume dell' aria imprigionata nel tubo più corto, cioè finché questo volume, inizialmente 10, si riduca al valore 5.
Misurando ora la differenza di livello del mercurio nei due tubi, si trova che essa eguaglia esattamente l'altezza del mercurio di un barometro esterno, al momento in cui si svolge l' esperienza. Quindi la presenza della colonna di mercurio e dell'aria causa una pressione di 2 atm, e il volume dell'aria si trova ridotto alla metà: ovvero, al raddoppiare della pressione esterna, il volume si riduce alla metà. Quindi, se la pressione esterna diventa n volte più grande, il volume diventa 1/n del volume iniziale. La legge di Mariotte è stata verificata per l'aria, sino a 27 atmosfere, da Dulong e Arago, mediante apparecchio analogo a quello qui illustrato.
Esiste la possibilità di usare altri strumenti per verificare la legge di Boyle - Mariotte, quali quello di Pizzarello nelle sue più recenti versioni.

***OSSERVAZIONI***

Per molto tempo la legge di Boyle è stata ammessa come esatta, non solo per l' aria, ma anche per tutti gli altri gas.
Numerose esperienze, eseguite a questo proposito da Regnault, portarono però a concludere che la legge di Boyle - Mariotte vale con precisione tanto minore quanto più il gas si trova vicino alle condizioni di liquefazione. Ma quando un gas è molto lontano dal suo punto di liquefazione (come nel caso di ossigeno, idrogeno e azoto alle temperature ordinarie) la legge di Boyle si può applicare senza sensibili errori.

## *PILA DI VOLTA (1800)*

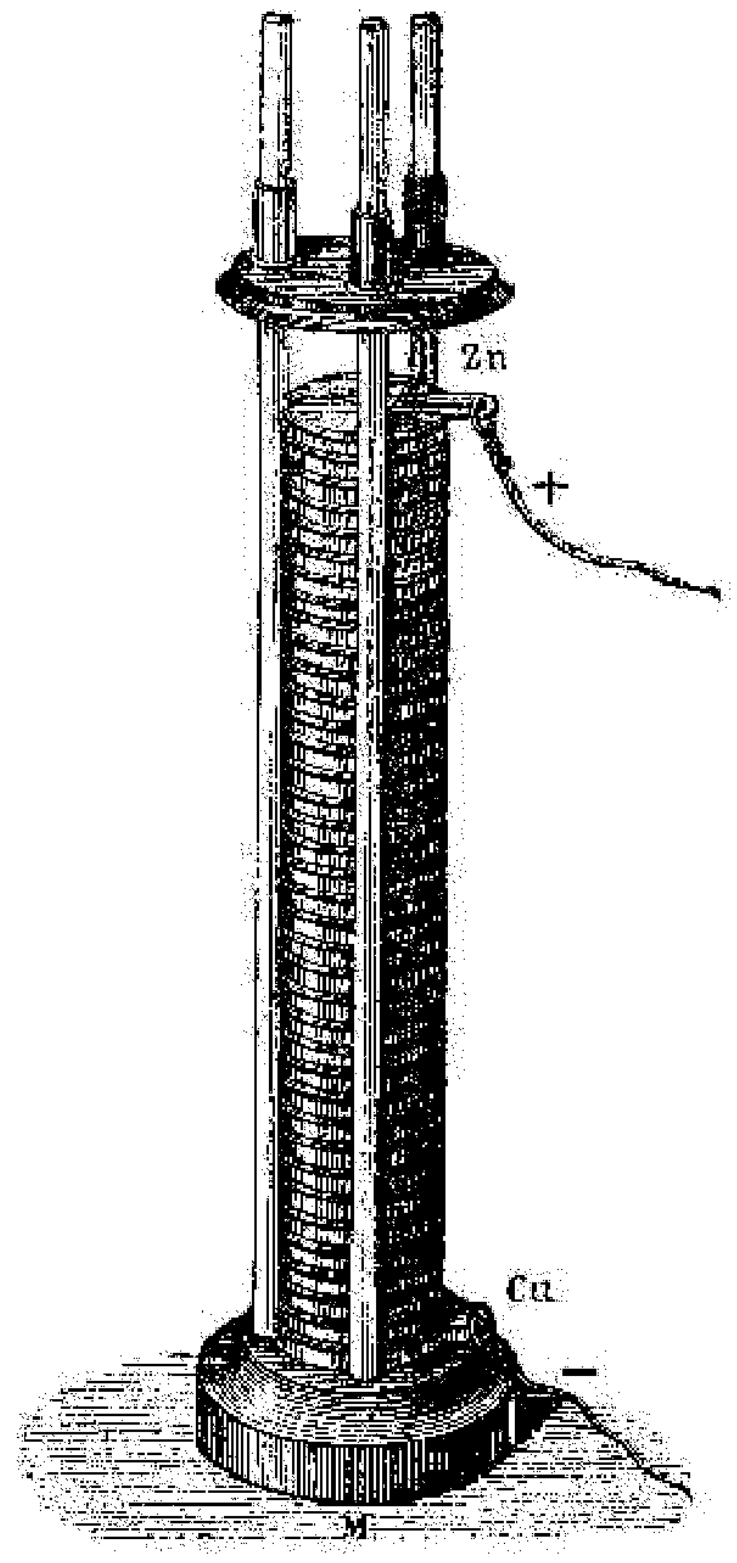

Fig. 384.

La scoperta della pila risolse il problema della costruzione di un generatore di corrente, ossia di un apparecchio in grado di far muovere cariche elettriche lungo un circuito chiuso. Questo generatore era costituito da una pila cilindrica, formata da coppie di dischi di rame e di zinco separate da un disco di panno imbevuto di acqua acidulata; gli estremi della pila erano collegati da due morsetti. Volta diede notizia della sua scoperta in una lettera dell'anno 1800 alla Royal Society di Londra. Leggiamone parte del testo:
«Le mie esperienze sono sull'elettricità eccitata dal semplice mutuo contatto di metalli di differenti specie e dal contatto di altri conduttori, differenti anche essi tra loro, sia liquidi, sia contenenti qualche umore al quale essi debbono il loro potere conduttivo...
«Ora se questa colonna arriva a contenere circa 40 di questi strati o coppie di metalli, essa sarà capace di far dare segni all' elettrometro, e non solo…».
L' apparecchio, presentato poi –come noto– a Napoleone, prese il nome di pila (dal francese «appareil a pile») nello stesso anno 1800. Si dà ora il nome generico di pila a tutti gli apparati che servono a fornire elettricità dinamica.

### *DESCRIZIONE*

Una base cilindrica in legno, posta su un treppiede, sorregge molte piastre cilindriche di rame e zinco, alternate. La prima e l'ultima piastra sono rispettivamente di rame e di zinco, e sono connesse a due morsetti ai quali collegare il circuito esterno. Più precisamente, la pila è composta da una serie di coppie di dischi di rame e di zinco, sovrapposti nel seguente ordine: un disco di rame, un disco di zinco, ed una rotella di panno bagnato con acqua, acidulata mediante acido solforico (10 parti di acqua, una di acido). Ogni coppia di dischi di rame e zinco dicesi elemento. Ogni pila è costituita da una quarantina di elementi (vedi figura). Ogni coppia genera sulle facce esterne una d.d.p. (differenza di potenziale) che si somma a quella delle altre coppie, poste in serie; la complessiva d.d.p. è di circa 1,1 volt. Connettendo i due poli estremi con fili metallici, detti reofori, si produce in essi un moto di cariche (dal polo negativo verso il polo positivo) che tende a ristabilire l'equilibrio elettrico.

***FUNZIONAMENTO***

Ciascun elemento della pila di Volta è composto da due elettrodi, uno di rame e uno di zinco, collegati tra loro ed a contatto con un panno intriso di una soluzione in acqua di acido solforico ($H_2SO_4$ ). I metalli rame e zinco per la diversa tendenza a perdere elettroni periferici danno luogo a ioni positivi secondo la loro affinità o elettronegatività (posizione nella serie elettrochimica degli elementi). Lo zinco tende a perdere elettroni più facilmente del rame. Quando mettiamo in contatto una lastra di rame con una di zinco, quest' ultima invierà elettroni al rame, cosicché la lastra di zinco rimane carica positivamente e quella di rame si carica negativamente. Quando con un filo conduttore si congiungono i due elettrodi, gli elettroni passano attraverso il filo ristabilendo una condizione di equilibrio elettrico. Per elettrolisi, però, l' $H_2SO_4$ viene scisso in ioni $H^+$ e ioni $SO_4^{--}$. Gli ioni $H^+$ si depositano sull' elettrodo di rame dando origine alla reazione $Cu + 2H^+ + 2e^- \rightarrow Cu + H_2$ . All'elettrodo di zinco avviene (con gli ioni $SO_4^{--}$) la reazione $Zn + SO_4^{--} \rightarrow ZnSO_4 + 2e^-$ . Lo zinco, che ha perso elettroni attraverso il filo, si trasforma in ioni $Zn^{++}$ in soluzione, e a sua volta il rame acquista elettroni. A circuito chiuso ha luogo il passaggio di una corrente continua. La f.e.m. della pila è circa 1,09 volt. Più elementi in successione danno luogo ad una forza elettromotrice (f.e.m.) somma delle f.e.m. di ciascun elemento. La pila di Volta ha il fondamentale difetto della polarizzazione: l'idrogeno, secondo il verso della corrente, va sulla lastra di rame e la ricopre con un velo gassoso che ostacola il passaggio di corrente; e la pila comincia a “scaricarsi”.

## *PILA LECLANCHE'*

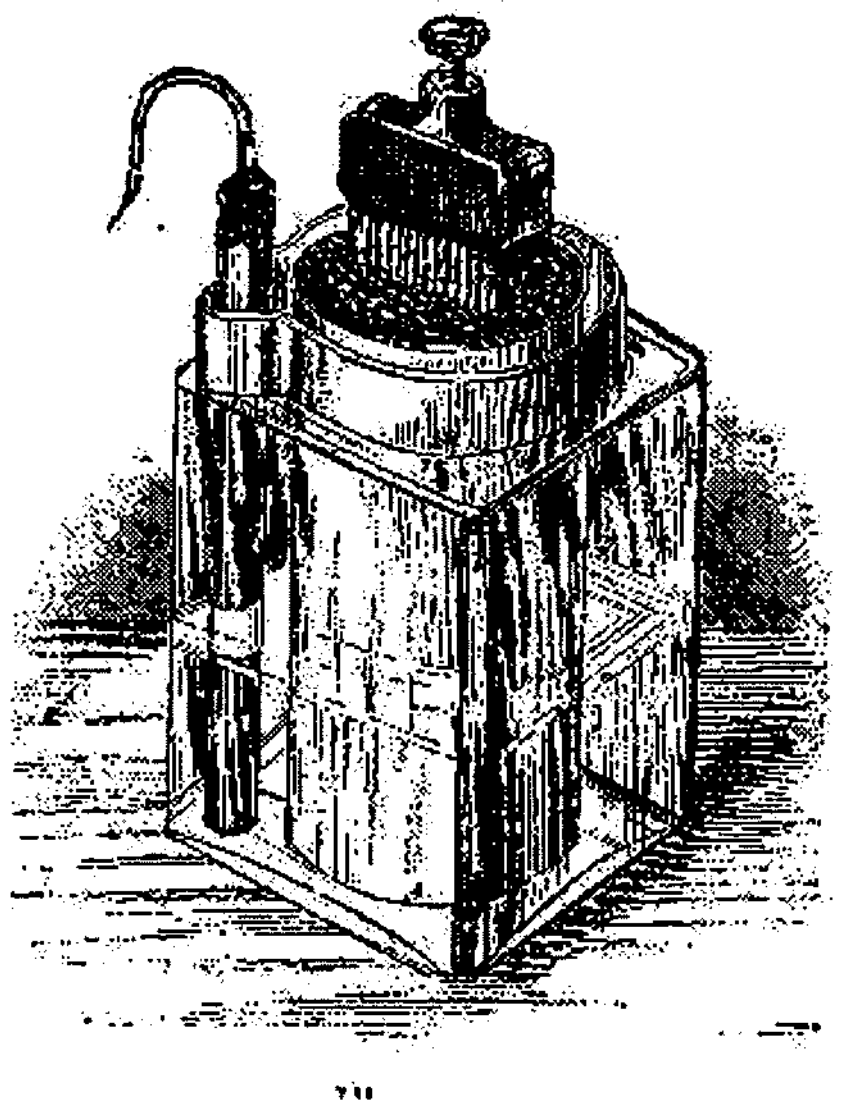

### *DESCRIZIONE*

Il primo elettrodo è costituito da un bastone di zinco amalgamato, immerso a metà in una soluzione di cloruro di ammonio, che funziona da elettrodo positivo. Il secondo elemento è un cilindro di carbone immerso in un piccolo vaso poroso circondato da una mescolanza di coke e di biossido di manganese, che serve da sostanza depolarizzatrice. E' una pila molto usata, specialmente nel caso in cui il circuito debba rimanere aperto per lunghi intervalli di tempo, come succedeva nei campanelli elettrici. La forza elettromotrice è di 1,48 volt.
Abbiamo anche la versione della pila Leclanché a secco, che può essere utile in molti casi risultando essa molto più comoda in svariate applicazioni.

### *FUNZIONAMENTO*

La pila Leclanché era costituita originariamente da un bacchetta di zinco (elettrodo negativo), immersa in una soluzione di $NH_4Cl$, e da una bacchetta di carbone (elettrodo positivo), inserita in un impasto di $MnO_2$. L' elemento negativo e quello positivo sono separati da un setto poroso, che impedisce che gli ioni vengano a contatto e si neutralizzino.
Le reazioni che avvengono al polo negativo sono $Zn^{2+} + 2NH_4Cl \rightarrow ZnCl_2 + 2NH_3 + H_2$.
Al polo positivo si ha l' antipolarizzazione, con la riduzione dello ione idrogeno: $2MnO_2 + 2H^+ \rightarrow Mn_2O_3 + H_2O$ . La f.e.m prodotta è di 1,5 - 1,6 volt. Questa pila ha l' inconveniente dell' arricchimento in $ZnCl_2$, mentre si impoverisce di $NH_4Cl$ .
Le successive pile Leclanché sono state a secco; in esse l' elettrolita non è liquido, ma è costituito da un impasto.
La pila Leclanché ha le seguenti caratteristiche: f.e.m. di 1,5 V; corrente di 1 A; «capacità» di 225 A h .

## *PILA DI GRENET*

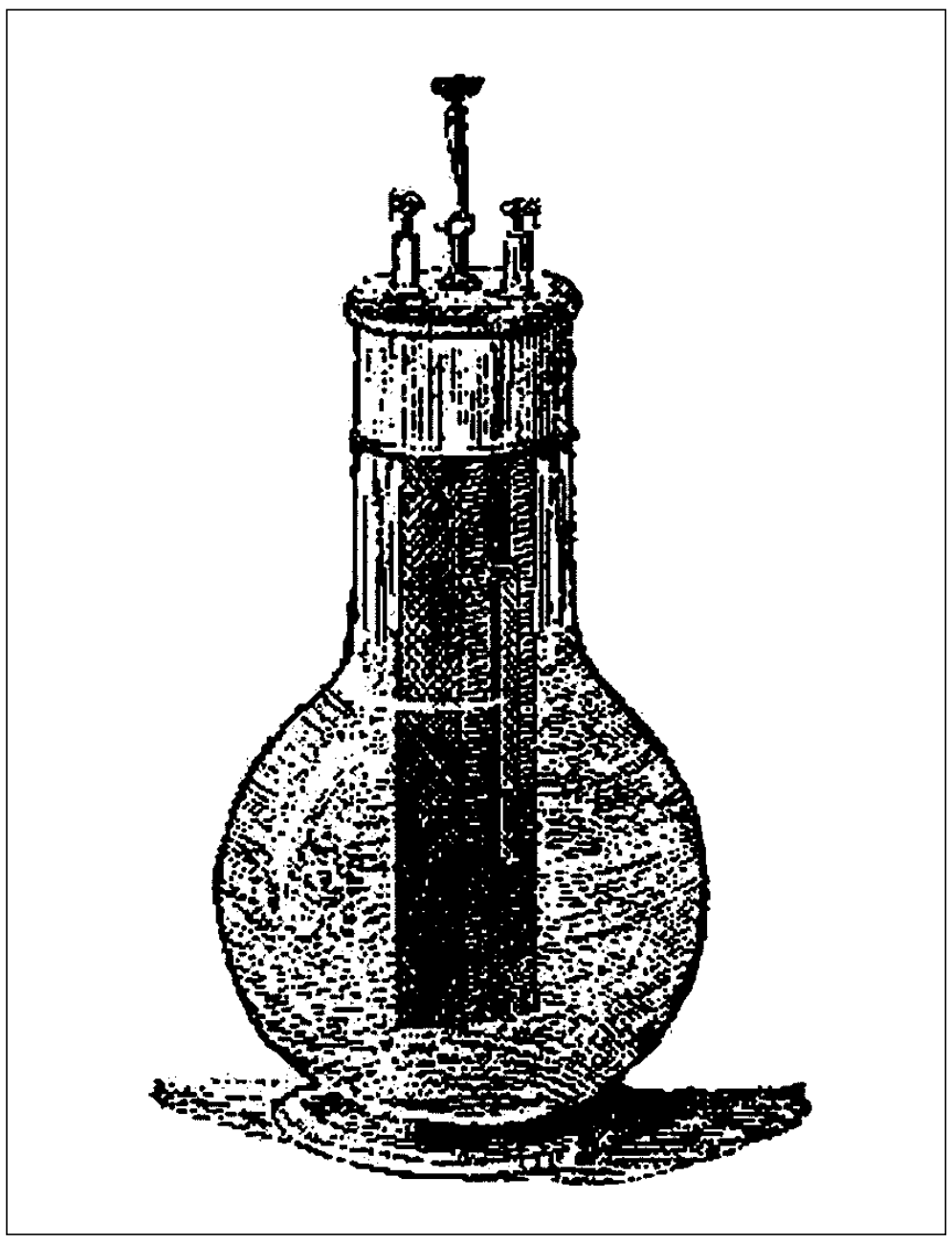

### *DESCRIZIONE*

Lo strumento è costituito da una boccia di vetro che contiene per 2/3 una soluzione di acido solforico, secondo la proporzione di Poggendorf (22 parti di acido solforico, 100 di acqua e 17 di bicromato). I bastoncini paralleli di carbone di storta, immersi nel liquido, sono fissati in un tappo di ottone, collocato in un coperchio di ebanite. Tra i due bastoncini di carbone e ad esse parallela è fissata una lastra di zinco amalgamato, unita ad un'asta di ottone pure fissata nel tappo di ottone e immobilizzata da una vite (posta sul coperchio di ebanite ed isolata dai cilindri di carbone).
Il serrafilo munito della lastra di zinco è il polo negativo; il polo positivo è il serrafilo dell'elettrodo di carbone. Il bicromato potassico ha la funzione di fissare l' idrogeno svolgentesi per la reazione dello zinco con acido.
La pila di Grenet è assai comoda, non dà esalazioni, e produce una f.e.m. di 2,12 volt. Ha inoltre come buona caratteristica un elettrodo di zinco estraibile, ad evitare reazioni secondarie.

### *FUNZIONAMENTO*

La pila Grenet è ad un solo liquido: è costituita da una bottiglia contenente bicromato potassico, $K_2Cr_2O_7$, e acido solforico $H_2SO_4$. La reazione chimica principale è analoga a quella della pila di Volta; la reazione di antipolarizzazione è la seguente: $Cr_2O_7^{--} + 3H_2 + 8H^+ \rightarrow 2Cr^{3+} + 7H_2O$ .
Fra composti la reazione invece è la seguente:

$$K_2Cr_2O_7 + 3H_2 + 4H_2SO_4 \rightarrow K_2SO_4 + Cr_2(SO_4)_3 + 7H_2O \ .$$

## *APPARECCHI DI DE HALDAT E DI PELLAT*

Sono Apparecchi molto simili, e aventi la medesima finalità. Non riteniamo essenziale, qui, distinguerli.

L'apparecchio fu costruito (1910) per misurare la pressione prodotta dalla gravità di un liquido: essa è una funzione della accelerazione di gravità, della densità e della altezza h della colonna di liquido secondo la legge di Stevino:

$$P = \rho g h$$

dove:
$\rho$ = densità del liquido alla temperatura della misura
$g$ = accelerazione di gravità nel luogo della misura
$h$ = altezza della colonna di liquido.

Dal Trattato elementare di fisica di Oreste Murani (1923) riportiamo quanto segue:
« Esso consiste in un tubo di vetro ricurvo verticalmente alle sue estremità, e riunito ad un tubo di ferro, sul quale si avvitano dei vasi di forma diversa, aventi però tutti una base con la medesima area. Si versa del mercurio nel tubo ricurvo: la superficie terminale di questo liquido forma il fondo del vaso, ed è essa che sopporta la pressione idrostatica dell'acqua che riempie il vaso superiore fino alla cima. Qualunque siano la forma dei vasi e le loro dimensioni, facendo arrivare l'acqua fino alla detta cima, si vedrà nell' altro ramo salire il livello del mercurio allo stesso punto».

L'apparecchio serve anche a dimostrare che la pressione è indipendente dalla massa d'acqua, mentre dipende solo dall' altezza della colonna di liquido; pertanto, si possono produrre pressioni considerevoli con piccole quantità di liquido.

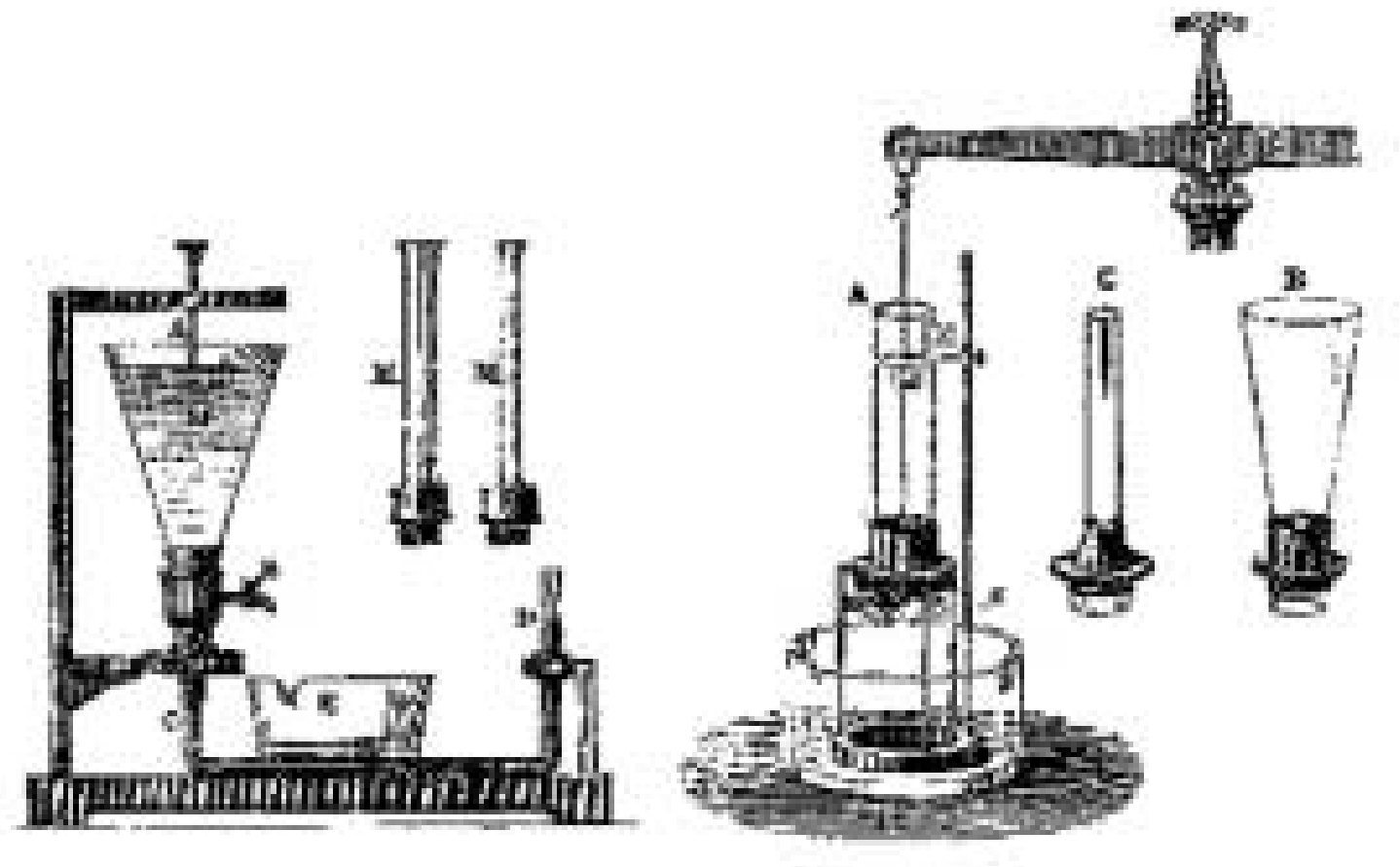

## *APPARECCHIO DI JOULE*

La «forza viva» impiegata nel vincere l'attrito non si distrugge, ma semplicemente si trasforma in energia termica (calore). In realtà lavoro e calore possono trasformarsi l'uno nell' altro; come è noto, si può ottenere del calore spendendo del lavoro, e, viceversa, con la spesa di calore si può (sotto certe condizioni) ottenere del lavoro.
Le esperienze dimostranti la trasformazione reciproca di lavoro in calore sono varie. Ricordiamo ad esempio quelle di Rumford (1758), di Davy (1802), e di Tyndall, che qui brevemente descriviamo. L'apparecchio ideato da Tyndall consiste in un tubo di ottone riempito di alcool o di etere, a cui si imprime un rapido movimento di rotazione per mezzo di due pulegge di diverso diametro, mentre lo si stringe fra due assicelle di legno. Il liquido, rimescolandosi, si riscalda fino a bollire; il suo vapore, in virtù della forze elastiche, fa saltar via il tappo.
F.R.Mayer spiegò questo risultato con il fatto che lo scambio «occulto» fra calore $Q$ e lavoro $L$ deriva dal fatto che il calore è un'energia cinetica (microscopica), della stessa natura di quella meccanica (macroscopica). Se le due energie vengono misurate con le stesse unità di misura, si ha che $JQ = L$, con $J$ = equivalente meccanico del calore.
Verso la metà del diciannovesimo secolo lo studio delle macchine termiche a ciclo periodico condusse a considerarle come sistemi meccanici capaci di ritornare allo stato iniziale, e di presentare sempre lo stesso funzionamento. Si era notato che ad ogni ciclo si cedeva al sistema una certa quantità di calore $Q$ mentre il sistema erogava una quantità di lavoro $L$, con un rapporto pressoché costante tra $Q$ e $L$ , almeno in una stessa macchina.

### *DESCRIZIONE DELLO STRUMENTO*

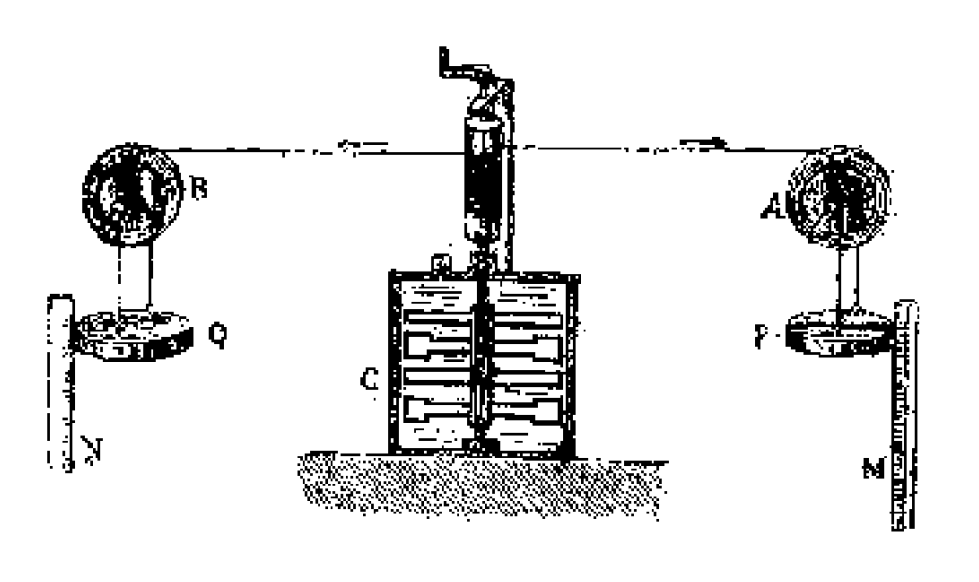

Joule fece i suoi primi esperimenti nel 1844. La rappresentazione del «mulinello di Joule» è data in figura.
L'apparecchio è costituito da un sistema di palette, portate da un asse che può girare all'interno di un calorimetro C contenente dell'acqua. Questo asse è messo in rotazione mediante due funicelle avvolte su di un cilindro S e passanti nelle gole di due carrucole A e B, che vengono a loro volta poste in rotazione per mezzo di pesi (P e Q). Per il movimento delle palette rotanti, si sviluppa fra queste e l'acqua un forte attrito, che viene reso maggiore mediante un sistema di palette fisse all'asse C, le quali impediscono all'acqua di assumere un movimento di rotazione. I pesi P e Q sono portati nella posizione più alta e quindi vengono lasciati cadere, con il ché l'asse si mette in rotazione. Le scale M e N sono regoli misuratori di quota. Durante la discesa dei pesi, viene compiuto un lavoro il quale è impiegato per:

a) vincere gli attriti delle parti dell'apparecchio esterne al calorimetro;
b) comunicare una certa energia cinetica ai pesi
c) riscaldare l' acqua in virtù degli attriti dovuti al moto delle palette.

***FUNZIONAMENTO***

Il lavoro fatto dal campo gravitazionale nella caduta di due pesi uguali P e Q, indicando con P il valore in chilogrammi del peso di ciascuno di essi e con h l'altezza in metri, è dato da $L = 2Ph$ .
Per ottenere una quantità apprezzabile di lavoro meccanico, si ripete *n* volte la caduta, cosicché $L = 2nPh$. Valutiamo ora il lavoro necessario a vincere gli attriti esterni al calorimetro, lavoro indicato con $l$ .
Determiniamo il peso additivo necessario, *p*, perché, mentre il calorimetro è vuoto, l'asse ruoti con la stessa velocità costante di quando è messo in rotazione dai pesi *P* e *Q* (in presenza di acqua generante attrito per il movimento delle palette): $l = n\,p\,h$.
Il valore dell'energia cinetica impressa all' insieme delle due masse durante la caduta è data in kgm da

$$E_{\text{cinetica}} = N_{\text{pesi}} \left( \frac{1}{2}\ \text{v}^2\ \frac{P}{g} \right) N_{\text{cadute}} = N_{\text{cadute}}\ \text{v}^2\ \frac{P}{g}$$

Si ha quindi l'equazione di bilancio energetico

$$L_{\text{tot}} = L_{\text{motore}} - l_{\text{attrito}} - E_{\text{cinetica}} = \text{calore sviluppato, in kgm}$$

$$L_{\text{tot}} = L_{\text{motore}} - l_{\text{attrito}} - E_{\text{cinetica}} = JQ$$

dove: $Q$ = calore in Calorie; $J$ = equivalente meccanico della Caloria. La quantità di calore $Q$ viene valutata mediante un metodo calorimetrico, misurando la massa dell'acqua contenuta e l'innalzamento della temperatura della stessa, con le correzioni necessarie dovute alle dispersioni di calore nel recipiente. Quindi:

$$J = \frac{L_{\text{totale}}}{Q}$$

***ESPERIENZE EFFETTUATE***

Joule ottenne, da quaranta misure, il valore $J$ = 423,9 kgm/Cal, ove Cal indica la chilocaloria, kcal. Usando il mercurio al posto dell'acqua, da 50 misure ottenne il valore $J$ = 427,7 kgm/Cal; finalmente, facendo ruotare due dischi di ferro nell' interno di un calorimetro a mercurio, ottenne per $J$ il valore di 425,7 kgm/Cal .

## *TERMOSCOPIO DI LOOSER*

### *PREMESSA*

Il termoscopio di Looser è un termoscopio a liquido. Esso sfrutta il fatto che il volume di un corpo (solido, liquido, gassoso), a pressione costante (nel caso nostro la pressione atmosferica), varia al variare della temperatura.
La particolare utilità del termoscopio doppio di Looser sta nel fatto che esso permette anche una serie di esperienze di confronto delle proprietà termometriche e calorimetriche di diverse sostanze.

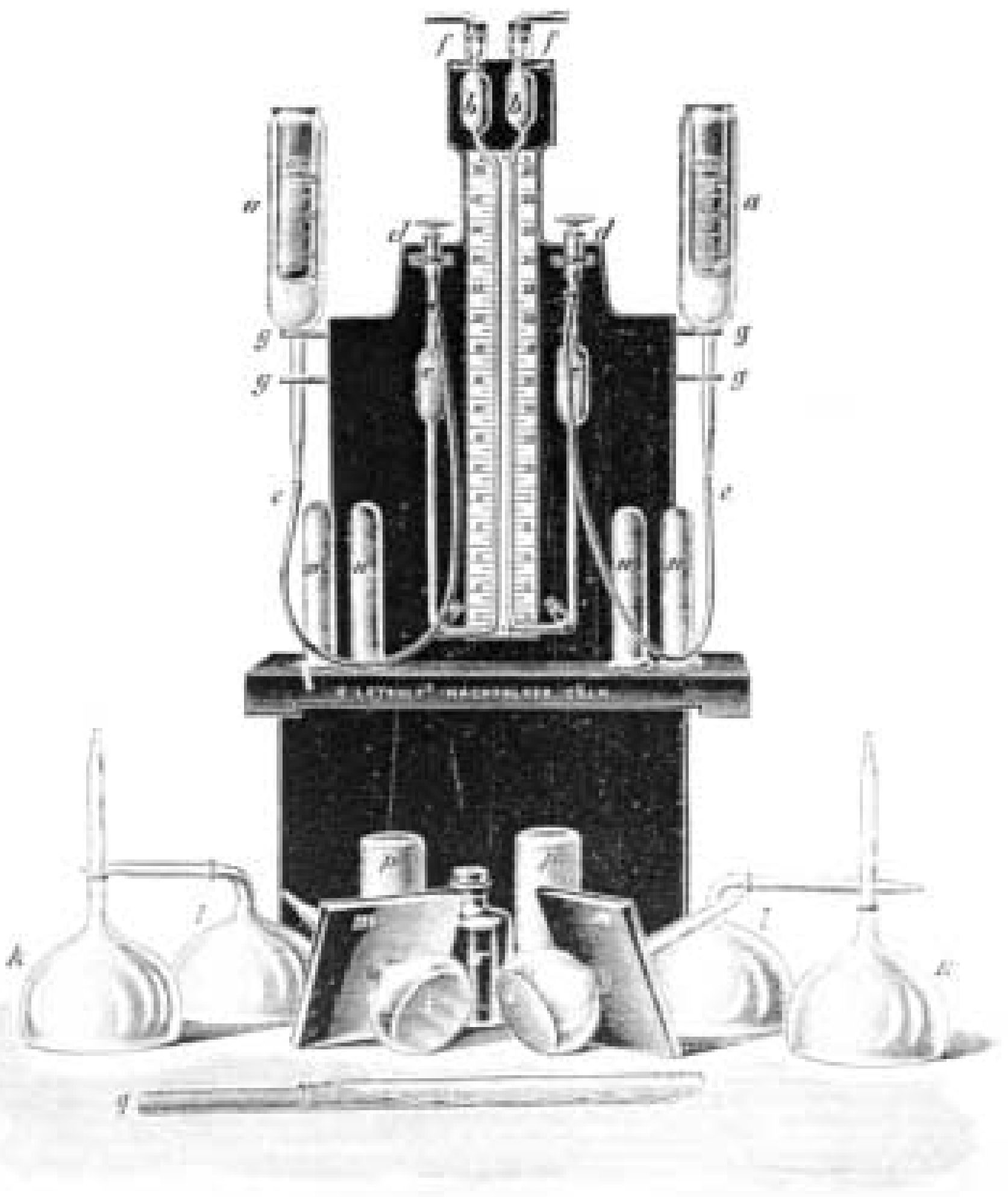

### *DESCRIZIONE*

In questo termoscopio le indicazioni dello strumento sono visibili sia di fronte, sia di lato, il che rende possibile a docente e allievi di seguire insieme l'innalzamento e l' abbassamento delle due colonne liquide e di leggere la differenza dei livelli sulla scala graduata.

L' apparecchio è composto essenzialmente da:
- il termoscopio propriamente detto
- due piccoli contenitori smerigliati (i)
- due grandi contenitori orizzontali (k)
- due grandi contenitori verticali (l)
- due raccoglitori a doppia parete, graduati in cc (centimetri cubici) (a)
- due tavolette in legno (m)
- quattro provette in vetro (n) da disporre nei provettoni (a)
- due contenitori tipo becker (p)
- apparecchio per la pulizia (q)
- una boccetta di liquido per il riempimento.

Si riempiono i due tubi manometrici con del liquido blu attraverso i due imbuti superiori (b), in modo che, essendo aperti i rubinetti, la colonna liquida salga a 15 cm di altezza nei quattro tubi. Se si formano bolle d'aria, si costringe il liquido nella parte allargata, in (b), soffiando nel tubo di gomma (c). Se si è versato troppo liquido, si apre il rubinetto (d) e si elimina l'eccesso del liquido usando una pipetta capillare e un pezzetto di carta da filtro. Si rimettono poi i rubinetti (f), così come i beccucci a gomito superiori. L' attacco di ciascun rubinetto porta un segno nero; quando questo segno è volto verso l'alto, il manometro comunica in condizioni normali con l'aria libera.
Nel monitoraggio qui rappresentato, i contenitori a provetta a doppia parete (a) sono collegati al termoscopio con i tubi di gomma (c). Al posto di essi, si possono mettere i contenitori (i), (k), (l) con gli anelli (g) [se necessario, ci sono due tavolette in legno (m) che permettono di non toccare le parti in vetro con le mani], e si fissano le loro estremità tubolari ai tubi di gomma (c). Ciò permette di procedere successivamente con rapidità a differenti esperienze, senza dimenticare di pulire sempre i diversi contenitori. L' apparecchio di pulitura (q), accluso allo strumento e avvolto di carta da filtro, serve appunto alla pulizia e all' asciugamento dei contenitori.

### *FUNZIONAMENTO, ED ESPERIENZE*

a) *Calore specifico:*
Si collegano al termoscopio i due raccoglitori graduati (a) e si mette in ciascuno acqua alla temperatura ambiente fino al livello di 20 cm.
In un recipiente pieno di acqua si riscaldano alla medesima temperatura un blocchetto di rame e uno di piombo di uguale volume, entrambi sospesi ad un filo. Si estraggono i blocchetti riscaldati, li si asciuga accuratamente con carta da filtro, e li si introduce nelle provette. Le altezze del liquido blu nelle colonne del termoscopio sono nel rapporto 1:3 corrispondente al rapporto dei calori specifici del rame e del piombo (0,093 e 0,032).

b) *Conducibilità termica dei solidi:*
Si mettono nelle provette, riempite di acqua, le due sbarrette di rame e di ferro piegate ad angolo retto (coprendo le loro estremità con gomma per proteggere le provette), e li si incrocia ad una distanza di circa 4 cm dalle altre estremità.
Si scalda il punto di incrocio con un becco Bunsen. Uno dei manometri indica rapidamente che il rame è il miglior conduttore.

c) *Conducibilità termica dei liquidi:*
Per dimostrare la diversa conducibilità dei liquidi, ci si serve di due provettoni doppi (figura in basso (C)). Si riempie di acqua o di alcool il vaso in vetro più lungo di uno dei due raccoglitori, e di glicerina o di mercurio il vaso più largo dell' altro raccoglitore. Poi si uniscono i due vasi interni più stretti con il termoscopio doppio, e si immergono simultaneamente i due raccoglitori in un vaso pieno di acqua calda. Dalla differente altezza delle colonne del termoscopio rileveremo la differenza di conducibilità dei due liquidi.

d) *Calore e Lavoro:*
Si riempie a metà con alcool il raccoglitore (a). Si martella su di una pietra, almeno per un minuto, il lingotto di stagno unito al filo metallico; poi si immerge lo stagno nell'alcool del raccoglitore (a) unito al termoscopio.
Si osserverà un innalzamento della colonna del termoscopio, che dimostra come il lavoro di martellamento si sia tradotto in calore.

e) *Calore di combinazione:*
Nelle provette (n) poste nei contenitori (a), in cui si è versato qualche ml di $HNO_3$ (per il rame) e di $H_2SO_4$ diluito (per lo zinco), si immerge con precauzione del Cu (nell' $HNO_3$) e dello Zn in lamina o in filo (nell' $H_2SO_4$). Il calore sviluppato dalla reazione viene rivelato dal termoscopio.

f) *Calore prodotto dalla corrente elettrica:*
(1) Il riscaldamento è proporzionale alla lunghezza del filo.
Ci si serve di due spirali di Pt, una di 15 e l'altra di 30 mm di lunghezza. Si riempiono d'alcool le due provette in modo che le spirali vi rimangano immerse fino ai punti di introduzione delle provette. Si pongono allora le provette nei contenitori, e si dispongono le spirali in serie in un circuito percorso da una corrente di circa 1 ampere. Gli spostamenti che si verificano nel termoscopio stanno allora tra di loro come 1 sta a 2.

(2) Ci si serve di una spirale di Pt (Platino) e di una di Ag (Argento) della stessa lunghezza come nell' esperienza vista in precedenza. Le resistenze specifiche dell' Ag e del Pt sono 0,94 e 4,1. Perciò la colonna liquida salirà circa 4 volte più velocemente dalla parte del Pt che dalla parte dell' Ag.

## *GALVANOMETRO DI NOBILI*

### *PREMESSA*

I galvanometri sono strumenti che permettono di misurare l'intensità di una corrente tramite i suoi effetti magnetici.
Il primo galvanometro, inventato dal tedesco Schweiger, si basava sull'azione della corrente, secondo la legge di Ampère, su di un ago magnetico. Esso era composto da un ago calamitato mobile posto all' interno di un telaio, attorno al quale era avvolto un filo di rame rivestito di seta.
Lo strumento di Nobili, inventato attorno al 1825, detto ad ago mobile, è fondato su di un principio simile.

### *DESCRIZIONE*

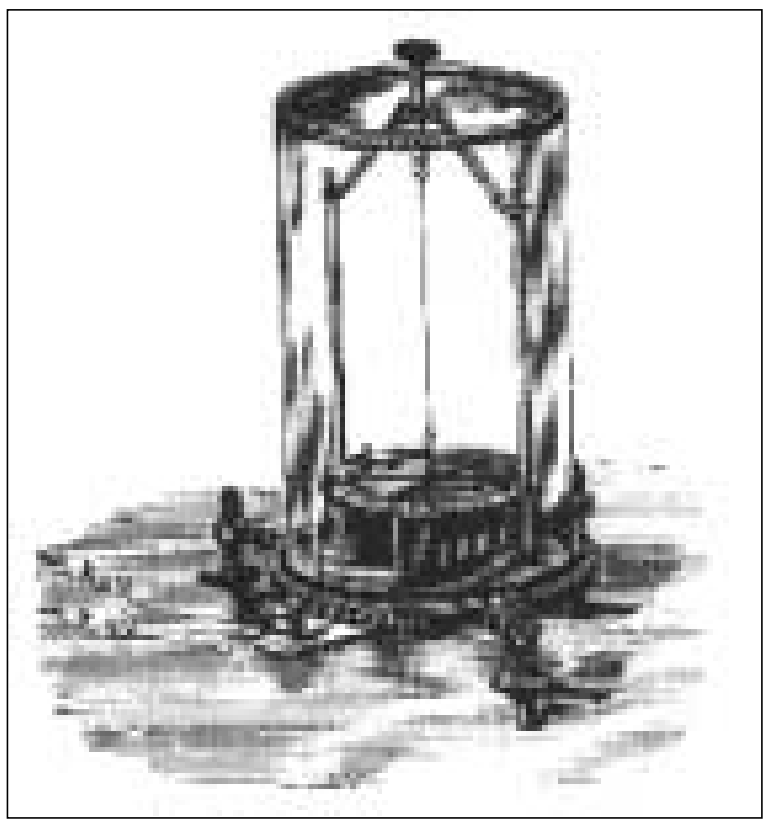

Il Galvanometro di Nobili è rappresentato in figura Vi è una base cilindrica in legno con due morsetti metallici, retta da tre piedini in ottone, regolabili. Al centro della base c'è una spirale di rame, avvolta in un rocchetto di legno, e contenente un primo ago disposto orizzontalmente. Sopra il rocchetto c'è un disco in ottone avente incisa una scala semicircolare con 180 divisioni. Al centro del disco vi è stato praticato un foro, al quale corrisponde un secondo ago metallico (S) parallelo al precedente, con poli orientati al contrario, e sollevato rispetto al disco. I due aghi connessi formano il cosiddetto sistema astatico. Sono sostenuti da un filo di seta (L) sorretto da un gancio orientabile con vite (K).
Per aumentare la sensibilità dello strumento il rocchetto è costituito da molte spire, che generano un intenso campo magnetico prodotto dal passaggio di corrente. Per ridurre l'effetto del campo magnetico terrestre, si utilizza il sistema astatico, costituito da due aghi magnetici identici, rigidamente collegati, paralleli e con polarità opposte: in modo che il momento magnetico del sistema sia praticamente uguale a zero. Le linee di forza del campo magnetico, generato dalla corrente, avendo all'interno e all'esterno del moltiplicatore senso opposto, agiscono concordemente su entrambi gli aghi. Per l'affidabilità delle misure è necessario che la durata dell' oscillazione dell' ago sia breve; come smorzatore si usa il disco di rame graduato. Una campana di vetro protegge lo strumento da correnti d'aria.

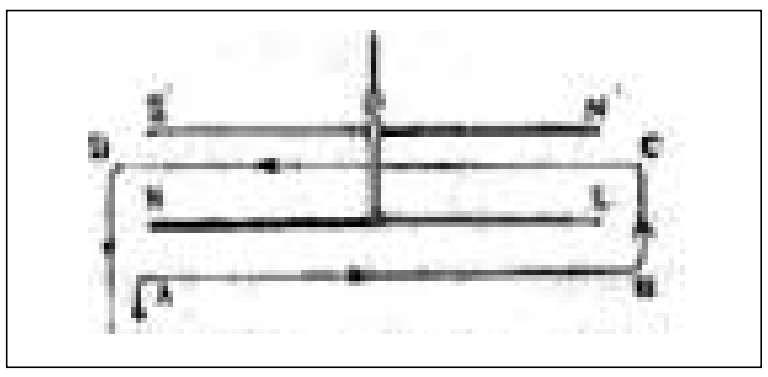

### *FUNZIONAMENTO*

Il funzionamento del galvanometro di Nobili è legato al sistema astatico rappresentato nel disegno. L' ago inferiore NS si trova tra le spire del rocchetto; il superiore S'N' ne resta invece fuori. Osservando la figura, si vede che il lato CD della spirale esercita sui due aghi un'azione concordante nel far deviare il sistema nel medesimo senso; e che il lato AB esercita sull'ago inferiore un'azione assai più rilevante di quella esercitata in senso contrario su quello superiore. Il principio di fun-zionamento è molto semplice: la corrente che attraversa il moltiplicatore genera un campo magnetico, che fa ruotare il sistema astatico. Dalla misura dell'angolo segnalato sul quadrante di rame (la sua tangente) otteniamo la misurazione della corrente, con la sua intensità e verso.

## *ANELLO DI PACINOTTI (1861)*

E’ un dispositivo elettro-meccanico: probabilmente il primo motore elettrico ideato al mondo. La parte essenziale della macchina è l’indotto, costituito dall’anello inventato da Pacinotti. Nel 1861 Pacinotti, con l’aiuto del meccanico Poggiali, realizzò una macchina elettromagnetica ad anello che fungeva ad un tempo da dinamo generatrice di corrente continua e da motore.
Pacinotti comprese immediatamente le enormi possibilità applicative del suo dispositivo, e ritardò volutamente la pubblicazione del suo lavoro per esplorare le nuove possibilità di utilizzazione di esso.
Ma nel 1869 un elettrologo belga, Zanobio Gramme, sfruttando l’invenzione di Pacinotti, brevettò il modello di dinamo e motore a corrente continua e lo lanciò su piano industriale. Per questo nel mondo viene ricordato erroneamente Gramme, al posto di Pacinotti. (Analogamente, non vengono riconosciute le priorità di Galileo Ferraris nella ideazione del primo motore elettrico a corrente alternata; e così via.)

### *DESCRIZIONE*

Una calamita orizzontale serve da induttore. L’ indotto (rotore) è l’anello di Pacinotti, costituito da un nucleo di ferro laminato a forma di anello e solidale con un asse di rotazione.
Attorno al nucleo è avvolto il circuito indotto, formato da numerosi gruppi di spire, collegati in serie fra loro. Il collettore è diviso in numerosi settori di rame, isolati fra loro, detti lamelle: ciascuno dei quali, mediante fili, è collegato alle spire. Due spazzole sfregano sul collettore e sono collegate con morsetti di uscita. L’indotto viene fatto ruotare per mezzo di un volante con manovella. La macchina è reversibile: ponendo in rotazione il rotore si genera corrente continua; mentre, alimentando l’anello con una sorgente di corrente continua, esso funziona come motore.

### *FUNZIONAMENTO DELLA MACCHINA*

Per spiegare il funzionamento della macchina è indifferente che l’anello di ferro dolce resti immobile e si faccia scorrere tutto all’intorno l’insieme delle spire, ovvero che il nucleo giri trascinando nel suo movimento le spire fissate su di esso.
Consideriamo dapprima (figura 1) una spira unica, e supponiamo che essa giri con l’ anello compiendo un giro completo. Consideriamo la variazione del flusso del campo magnetico concatenato con la spira che ruota da M ad M’ passando per S.. Tale variazione sarà uguale ma di segno contrario a quella che si produce nella medesima spira nella seconda parte del suo percorso, ovvero quando ruota da M’ ad M passando per N.
A tali variazioni del flusso corrisponderanno in ciascuna spira delle forze elettromotrici che saranno di segno opposto da una parte e dall’altra del diametro MM’: nel primo quarto di giro, da M a S, il flusso concatenato con la spira va diminuendo, mentre da S a M’, nel secondo quarto di giro, detto flusso concatenato va aumentando. La forza elettromotrice nella spira che da M si sposta fino ad M’ passando per S ha segno costante. Nell’altro mezzo giro, da M’ ad M passando per N, il flusso concatenato con la spira diminuisce da M’ sino ad N ed aumenta da N sino ad M; la f.e.m. conserva di nuovo lo stesso segno, che però è contrario a quello del primo mezzo giro. Tornata la spira nella sua posizione iniziale (M), la forza elettromotrice indotta riprende il segno e il valore primitivo. I valori della forza elettromotrice in ciascun punto del percorso, supposto il moto uniforme, possono essere rappresentati da una sinusoide, con valore massimo in corrispondenza di M e di M’, e valore 0 in corrispondenza di S e N..
Consideriamo ora l’anello di Pacinotti in figura 2, e l’insieme delle spire in movimento: in tutte le spire, che si muovono da un lato del diametro ab, le forze elettromotrici hanno uno stesso senso, contrario a quello delle forze elettromotrici che si producono dall’altro lato.
Il risultato è analogo a quello di una pila che abbia il polo positivo in a e il negativo in b. Il sistema equivale anzi ad una combinazione di due pile P e P’ (ved. figura 3) aventi i poli positivi riuniti in A, e i poli negativi in B. In questi punti devono collocarsi le spazzole destinate a portare la corrente nel circuito esterno AB. Durante la rotazione, ciascuna lamina del collettore prende successivamente il posto di quelle che la precedono. Appoggiando le spazzole alle due sezioni del collettore che hanno la massima differenza di potenziale, il circuito esterno sarà percorso da corrente elettrica ogni volta che due lamine del collettore vengono a contatto con le spazzole alle quali sono collegati i suoi capi.

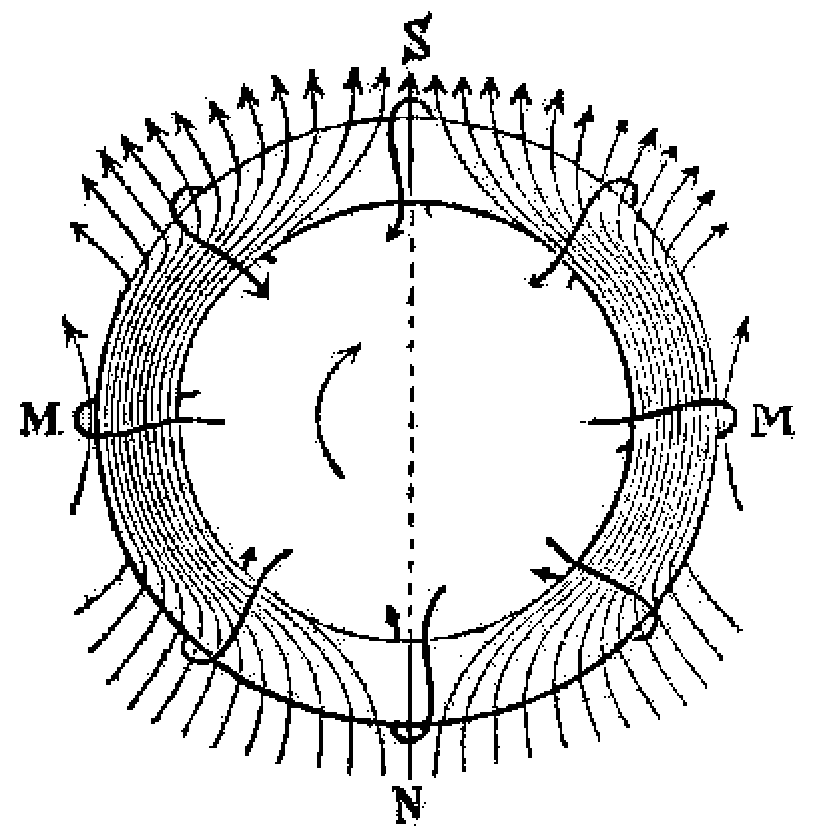

Figura 1

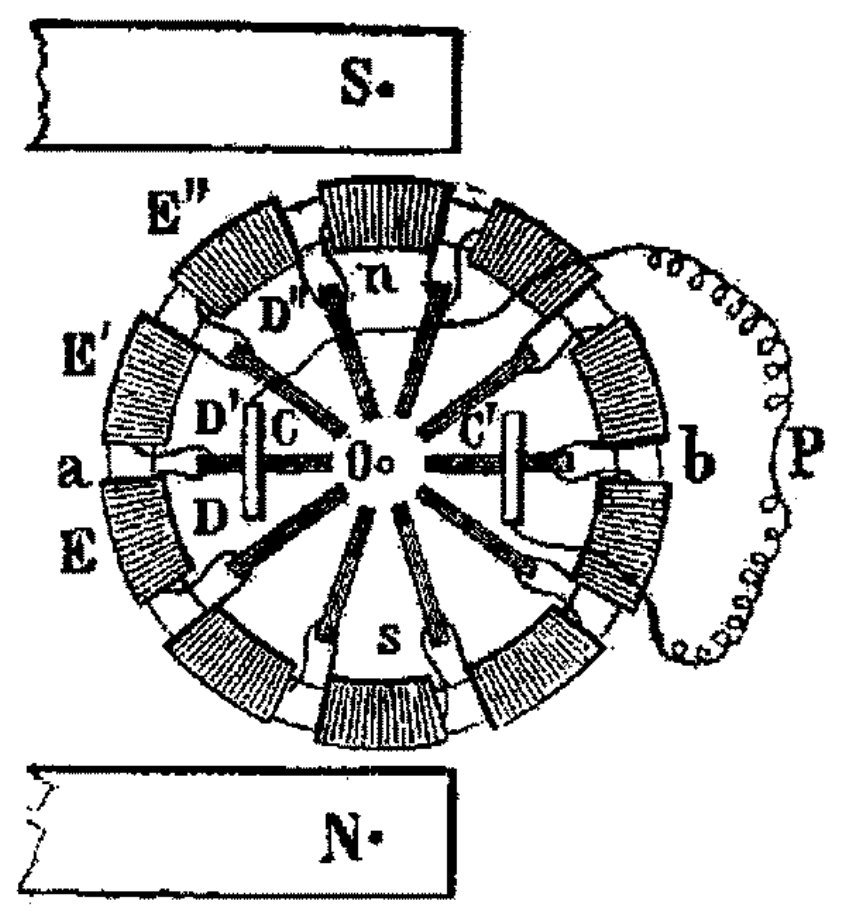

Figura 2

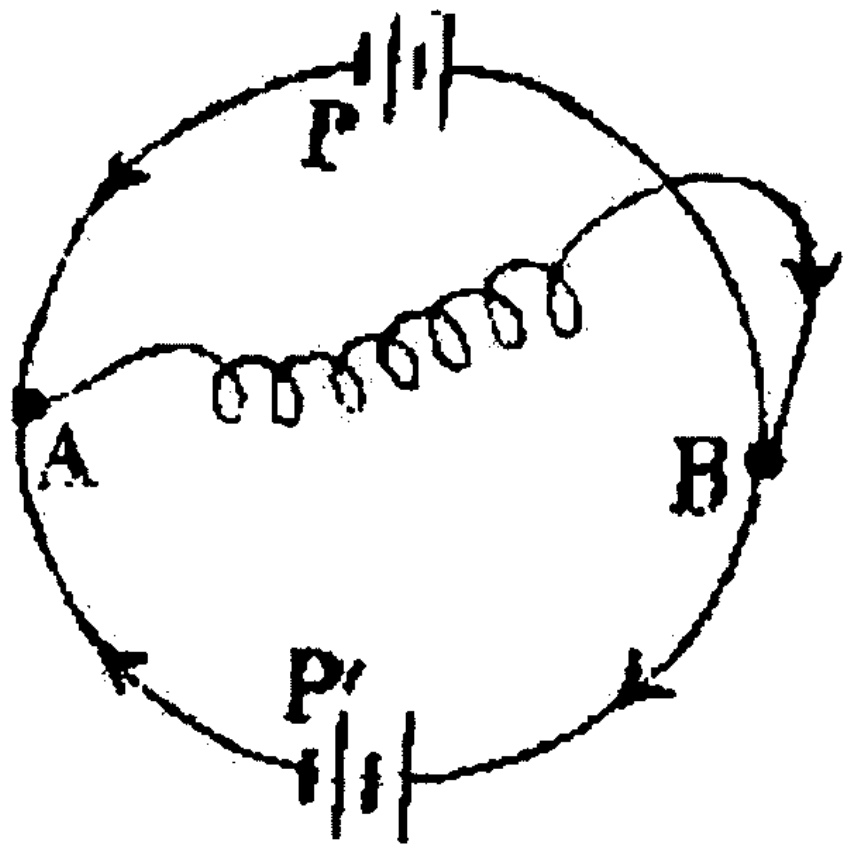

Figura 3

## *MACCHINA ELETTROSTATICA DI RAMSDEN*

La prima macchina elettrica fu ideata da Ottone di Guericke; essa consisteva in una sfera di zolfo fissata ad un'asse, che si faceva ruotare con una mano mentre l'altra mano, appoggiata sulla sfera, serviva da strofinatore. In seguito, alla sfera di zolfo si sostituì un cilindro di rame o di vetro, che veniva sempre strofinato da una mano. Nel 1740 Winkler, fisico tedesco, adottò per primo come strofinatore un cuscino di crini coperto di seta. Finalmente, nel 1766, Ramsden sostituì al cilindro di vetro un disco della stessa sostanza strofinato da quattro cuscinetti. Da allora la macchina di Ramsden rimase in voga per circa un secolo, perfezionata da John Cuthbertson, e fu la prima macchina elettrica largamente conosciuta ed utilizzata.

### *DESCRIZIONE*

Fra due sostegni di legno, si trova un disco, P, di vetro (diametro 0,5 m), fissato per il suo centro ad un asse che si fa ruotare per mezzo di una manovella M. Questo disco è compreso fra quattro strofinatori o cuscinetti di stoffa, fissi, impregnati di amalgama di Zinco e Solfuro di stagno. Nella direzione del suo diametro orizzontale il disco passa fra due tubi di ottone, curvati a ferro di cavallo, detti pettini, dotati di punte collocate ai lati e dirimpetto al disco. Questi pettini sono connessi a voluminosi tubi di ottone, C , di elevata capacità, detti conduttori, collegati per mezzo di un tubo di ottone di diametro minore.

### *FUNZIONAMENTO*

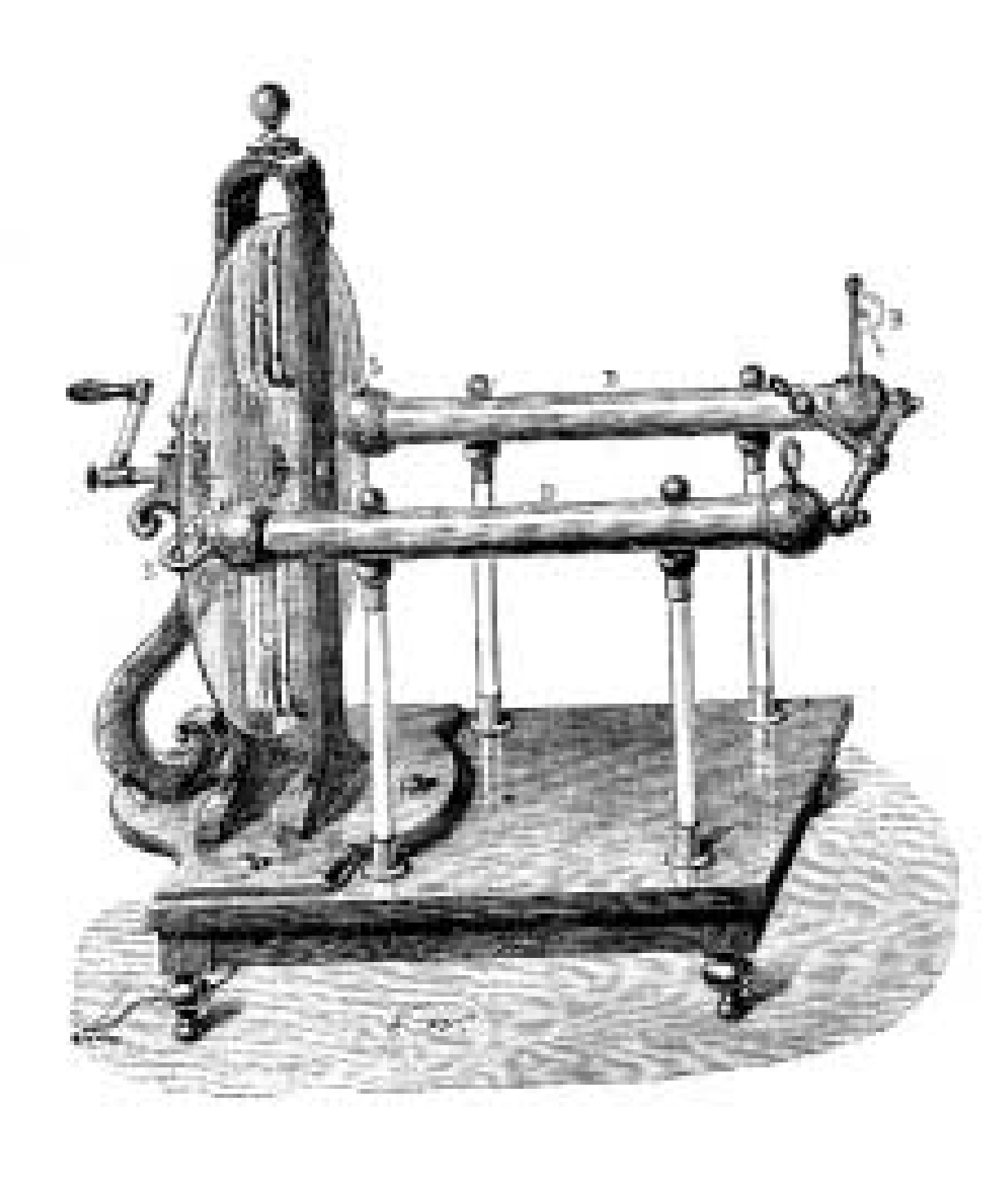

Il disco di vetro, messo in rotazione, si elettrizza positivamente per strofinio nella zona di contatto delle due coppie di cuscinetti. Questi ultimi, collegati con il suolo, D, perdono immediatamente la loro carica negativa.

L' elettricità positiva del disco agisce per induzione sui conduttori ed attrae le cariche negative che, attraverso i pettini, vanno a combinarsi con l'elettricità positiva del vetro e la neutralizzano.

I conduttori, nelle loro parti più lontane, restano carichi positivamente e possono trasmettere la propria carica ad altri conduttori per mezzo di eccitatori o collegamenti metallici.

## *ROCCHETTO DI RUHMKORFF*

### *PREMESSA*

Con l' avvento nelle prime quattro decadi dell'ottocento di sempre nuove e più potenti sorgenti di energia elettrica, quali le pile a partire dal 1800, le macchine magnetoelettriche a partire dal 1833 e le bobine ad induzione dal 1836, si passò a più significative applicazioni, soprattutto in medicina: con l'elettroterapia.
Nel 1851, H. Ruhmkorff, un costruttore di strumenti, di origine tedesca, stabilitosi a Parigi dal 1840, apportò consistenti miglioramenti alla bobina d'induzione, rendendo più grandi le f.e.m. indotte ed incrementando la «distanza esplosiva» dello spinterometro. Il suo modello, esposto a Parigi nel 1855, si affermò per tutta la seconda metà del secolo col nome di rocchetto di Ruhmkorff.

### *DESCRIZIONE*

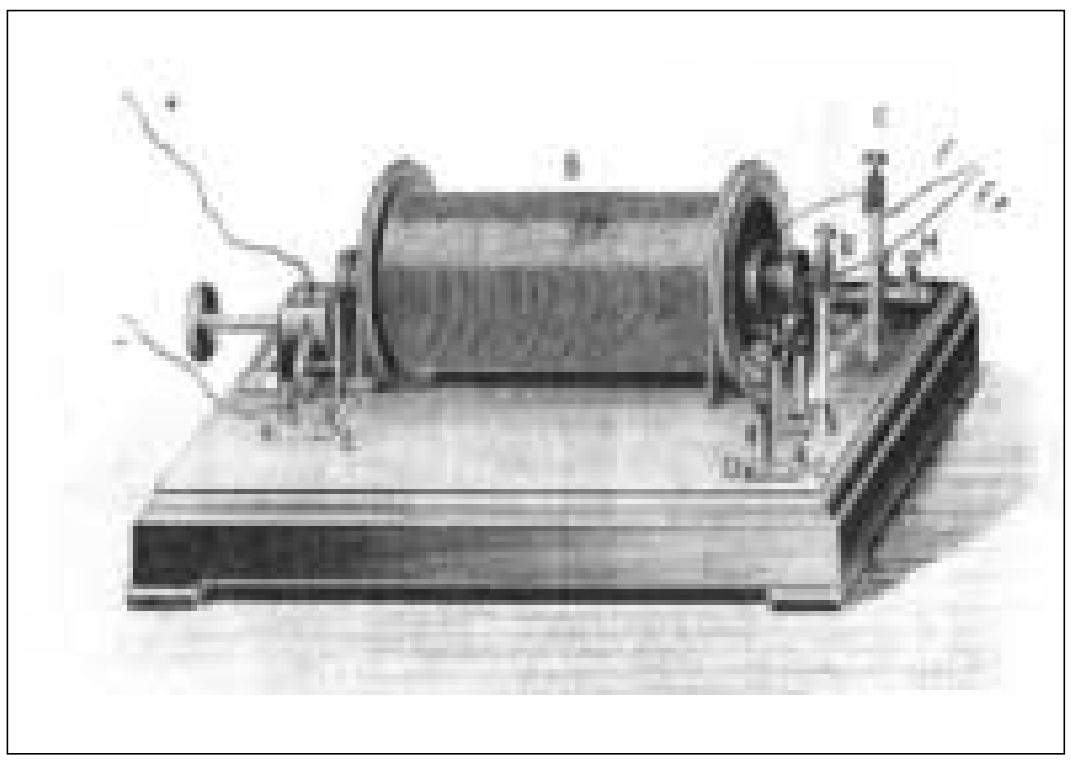

L'apparato di Ruhmkorff è composto da un cilindro cavo di legno, entro il quale vi è un fascio di filo di ferro, che sporge alle due estremità. Sul cilindro sono avvolti a bobina, l'uno sull' altro, due fili di rame, quello più grosso all'interno, l'altro, lunghissimo (parecchie migliaia di giri) e sottile, all' esterno.
Il filo induttore ha le estremità fissate alle colonnine metalliche D, F: l' indotto le ha nei serrafili B, C, retti dai sostegni isolanti. I reofori della pila, alimentatrice del rocchetto, vanno in R, R' ad un commutatore, che permette di invertire a piacere la corrente nel rocchetto primario.
Supponiamo che il commutatore sia girato in maniera che E' comunichi col polo positivo, così che la corrente passa da E' in F, entra nel rocchetto primario induttore, ed arriva alla colonna metallica D ed all' interruttore oscillante (formato da una molla recante un martelletto di ferro dolce). Così otteniamo una successione rapida di aperture e chiusure, ciascuna delle quali crea nel rocchetto secondario la corrispondente corrente indotta, che si potrà far passare per un conduttore qualunque inserito tra i due serrafili B, C.
La grande f.e.m. indotta nel secondario provoca una scarica nello spinterometro esterno al rocchetto; tale scarica può essere anche lunga un metro.

### *FUNZIONAMENTO*

Una rapida ed automatica apertura e chiusura del circuito è ottenuta con un interruttore a martello, costituito da una lamella di acciaio che poggia su una punta conduttrice, a cui è fissato un martelletto di ferro dolce affacciato al nucleo di ferro. Al passaggio della corrente nel circuito primario, il martello viene attratto dal nucleo del rocchetto e la lamella si stacca dalla punta causando l'apertura del circuito primario; per cui, non passando in esso più corrente, l'attrazione cessa e il martello richiude il circuito primario, ritornando a contatto con la punta conduttrice, e il ciclo ricomincia. Questo determina una forza elettromotrice indotta *V”* legata alla f.e.m. *V’* secondo il rapporto tra il numero delle spire $n''$ del secondario e quello $n'$ del primario, secondo la relazione:

$$V'' = V' \; n'' / n' .$$

La f.e.m. indotta nel circuito secondario V" non dipende solo dal rapporto del numero di spire, ma, essendo la f.e.m. indotta data da:

$$\text{f.e.m.} = -\frac{\Delta\phi(B)}{\Delta t} ,$$

sarà tanto maggiore quanto minore sarà il tempo di variazione del flusso magnetico.
Un condensatore consente di accumulare la carica per poi scaricarla quando la d.d.p. V è alta. Applicando una batteria di pochi volt ai morsetti, ad ogni apertura del circuito con l' interruttore, si produce una variazione del flusso, e quindi una d.d.p. V ai morsetti metallici di uscita, con grosso fattore moltiplicativo (pervenendo a circa 30000 V).
Il rocchetto di Ruhmkorff ha avuto numerose e importanti applicazioni, dalla telegrafia senza fili all'alimentazione di tubi a gas o a raggi X (Geissler, Crookes, Roengten), e al sistema di accensione del motore a scoppio.

## *TELEGRAFO SCRIVENTE DI MORSE*

Il telegrafo è uno strumento che, sfruttando fenomeni elettrici e magnetici, permette di trasmettere a distanza le parole, opportunamente «tradotte in segni».
Nel 1767 Giuseppe Bozolo aveva già proposto la possibilità di trasmettere le parole attraverso un filo conduttore, mediante un alfabeto basato sulle scintille fatte scoccare, da lontano, da un generatore elettrico.
Nel 1837 il fisico veneziano Luigi Magrini, e nel 1838 Wheatstone in Inghilterra, e Morse in America, costruirono i primi telegrafi elettromagnetici.

### *DESCRIZIONE*

Ogni sistema di telegrafia consta essenzialmente di un generatore, della linea di trasmissione, del manipolatore e del ricevitore di segnali. Per generatori si usano le coppie Daniell o gli accumulatori; possono essere anche usate delle dinamo. La linea è il reoforo che trasmette la corrente fra le due stazioni. Le linee aree del telegrafo erano di filo di ferro galvanizzato; le linee telegrafiche possono essere anche sotterranee o sottomarine. Il manipolatore (ved. figura 1) consiste in una leva metallica oscillante sul perno A, col quale comunica il filo di linea L. Una molla *f* la tiene abbassata sul contatto *b,* che comunica con il ricevitore locale R; e quindi il manipolatore viene disposto per ricevere il messaggio. Premendo sul bottone K, l'estremità a si solleva escludendo così dalla linea il ricevitore, mentre l'altra estremità *c,* abbassandosi, stabilisce il contatto con *d,* che comunica con il polo positivo della pila P; la corrente allora si trasmette nel filo di linea, L, e richiude il circuito dall'altra parte. Il ricevitore di Morse (ved. figura 2) nella sua parte essenziale è una leva orizzontale AB; la sua estremità A porta un'àncora di ferro dolce che è attratta da un'elettrocalamita, collegata con la linea, ogni volta che passa la corrente. L'altra estremità B viene allora a premere contro un nastrino di carta, mosso da un roteggio ad orologeria, e vi traccia delle linee e dei punti a seconda della durata dell'attrazione.

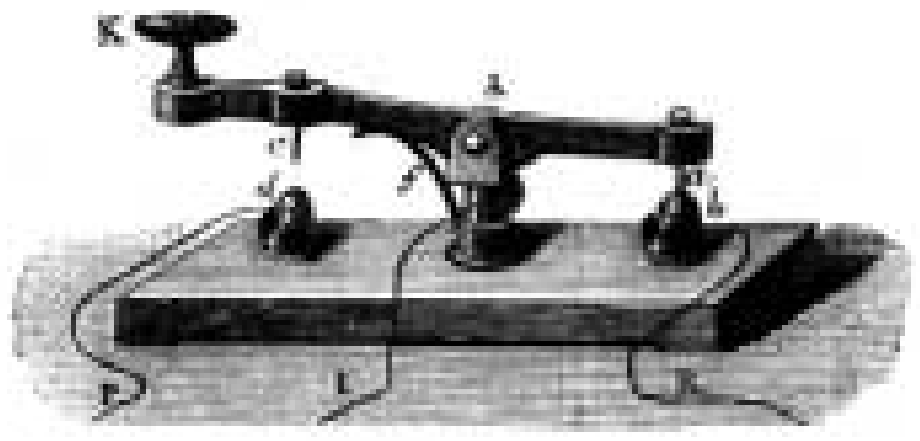

Figura 1

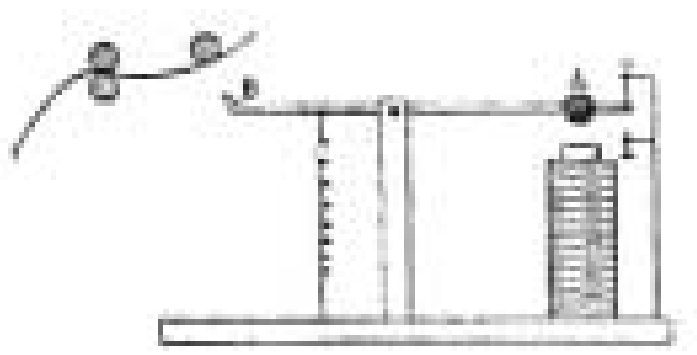

Figura 2

### *FUNZIONAMENTO*

L'insieme dell'apparecchio viene rappresentato in figura 3. La cassa contiene il roteggio di orologeria che mette in moto il cilindro a; questo, con un altro cilindro b, costituisce una specie di laminatoio, che fa scorrere il nastro di carta avvolto sulla ruota. Quando l'elettro-calamita E viene eccitata, essa attira l'ancora A vincendo la forza antagonista della molla r. L'ancora si abbassa senza venire però in contatto col nucleo, essendo impedita dalla vite a scrupolo, G; è necessario impedire il contatto, perché altrimenti la magnetizzazione residua ostacolerebbe il distacco al cessare della corrente magnetizzante. Abbassandosi l'ancora, l'estremo a sinistra, *m*, della leva si innalza, e spinge il nastro di carta YY' che si svolge dal tamburo R contro una rotellina, tinta di inchiostro sul suo contorno. Quando si abbassa la leva del manipolatore, a seconda che si trasmetta la corrente per tempo maggiore o minore, verranno tracciate lineette più o meno lunghe: cioè un messaggio trasmesso con l'alfabeto Morse, che può essere poi facilmente decifrato.

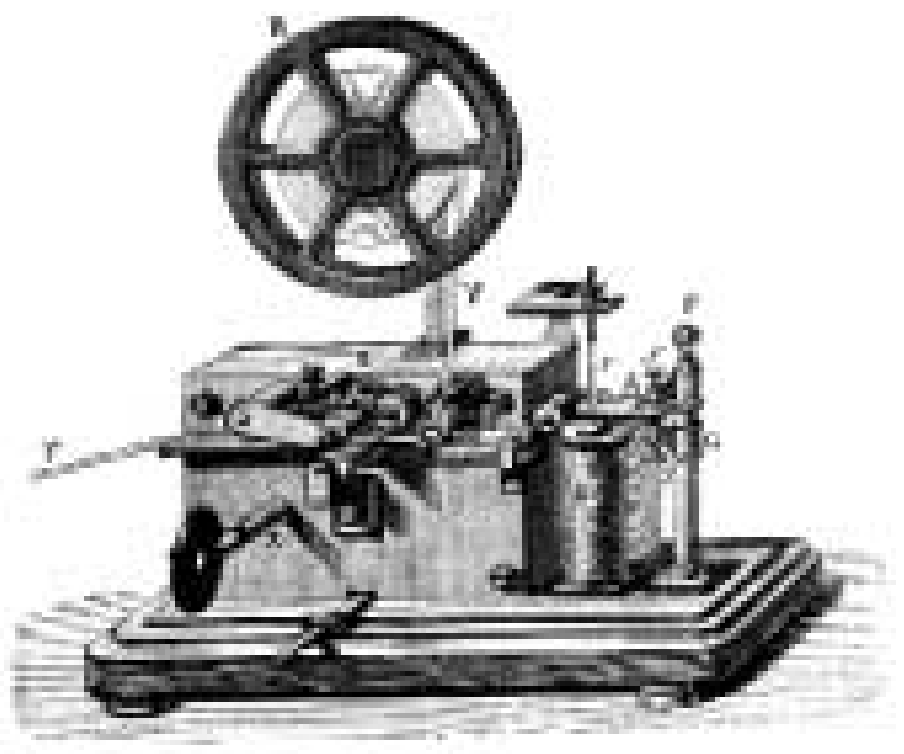

Figura 3

***OSSERVAZIONI***

Inizialmente per chiudere il circuito fra due stazioni telegrafiche si utilizzavano due fili, ma poi Steinheil eliminò il filo di ritorno, chiudendo il circuito con «la terra». Il polo negativo della batteria è collegato a terra con una lastra di rame; allo stesso modo nella stazione ricevente il circuito è collegato a terra con una seconda lamina di rame. Alle stazioni erano necessariamente annessi tre strumenti: la «bussola telegrafica», che serviva a verificare la direzione della corrente e la sua intensità, evidenziando guasti o possibili interruzioni; il commutatore, che serviva a cambiare il verso della corrente; e infine lo scaricatore, formato da due lastre metalliche fornite di punte, che servivano per togliere dal circuito le notevoli cariche provenienti dall'atmosfera.

## *MACCHINA DI WHIMSHURST (1893)*

### *MACCHINE ELETTROSTATICHE: PREMESSA*

Per ottenere elettricita statica furono inventate le macchine elettrostatiche a strofinio. La prima macchina elettrostatica fu probabilmente costruita da Otto von Guericke. Tale macchina era molto semplice e consisteva in un globo di zolfo che, mentre girava per mezzo di un manubrio, strisciava contro la mano.

### *DESCRIZIONE*

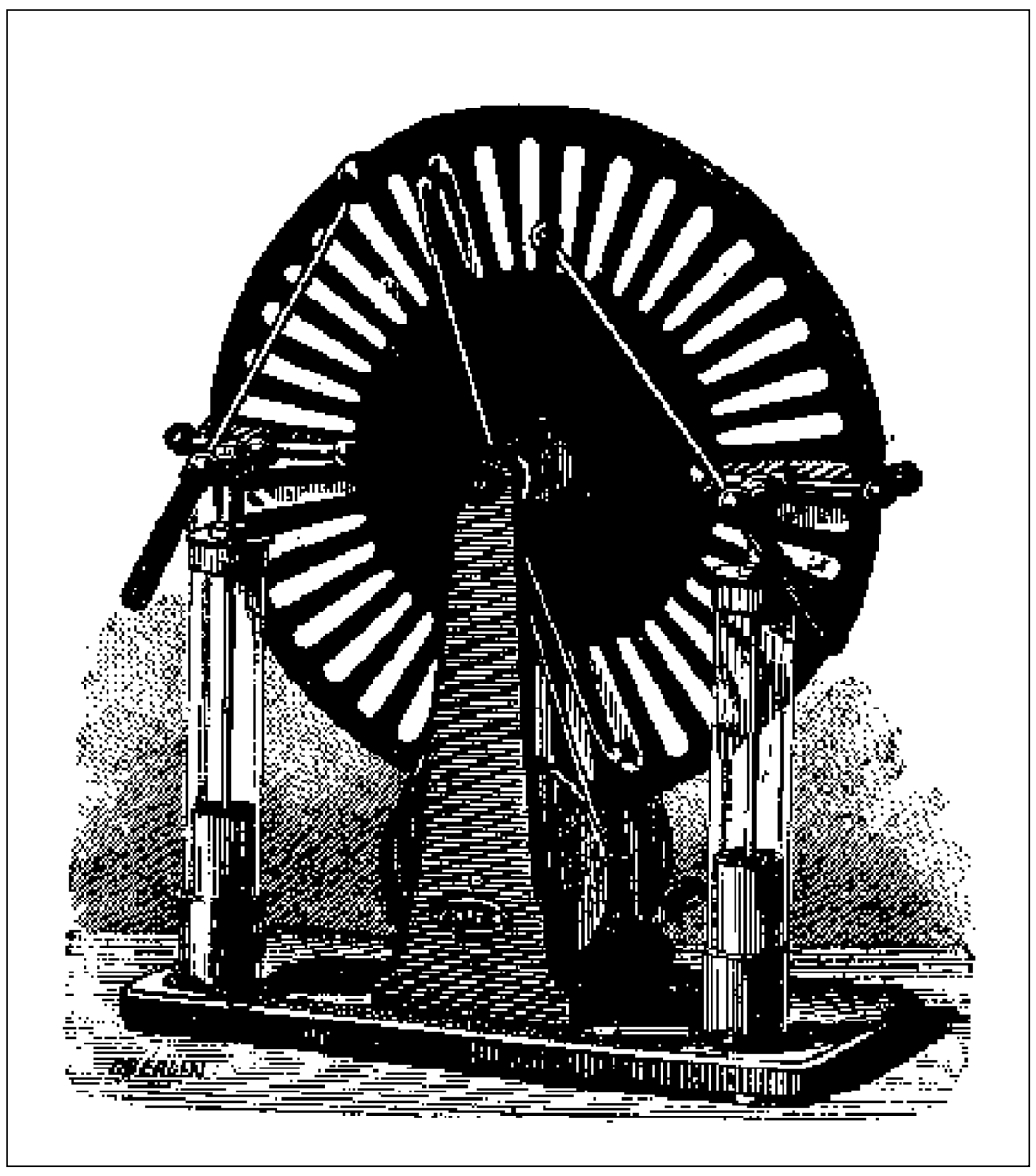

La macchina consta di dischi eguali, di ebanite o di vetro ricoperto di gomma lacca, paralleli e vicini, girevoli sullo stesso asse in senso contrario, e muniti sulle facce esterne di strisce o vettori di stagnola, radiali ed equidistanti. I dischi sono a contatto, in un piano orizzontale, con due collettori A e B della elettricità di induzione, piegati ad U, armati di punte (pettini) e comunicanti coi poli della macchina. Essi sono costruiti mediante due aste metalliche terminanti con due sferette C e D, la cui distanza può essere variata a volontà (spinterometro).
Vi sono poi due bracci conduttori, disposti diametralmente ai dischi e fissati al centro, inclinati di 45° e 60° sulla linea dei pettini (i quali poggiano leggermente sulle strisce di stagnola mediante pennellini che, per strofinio, danno elettrizzazzione alle superfici di stagnola).

### *FUNZIONAMENTO*

La macchina di Whimshurst è la macchina più atta per ottenere elettricità ad alto potenziale, mediante strofinio ed induzione.
Durante la rotazione dei dischi, le strisce radiali, che vengono debolmente elettrizzate per strofinio, inducono sui pettini con i settori di stagnola grandi quantità di cariche elettriche, le quali, tramite i conduttori, creano tra le sferette dello spinterometro una notevole d.d.p. (20000 volt).
Vengono accoppiati alla macchina condensatori cilindrici che servono ad aumentare la capacità della macchina. Aumentando il numero dei dischi si possono raggiungere d.d.p. di 250000 volt e scintille di 40 cm.
Le macchine elettrostatiche di Whimshurst sono sorgenti di elettricità ad alto potenziale, sì, ma con piccole intensità di corrente.

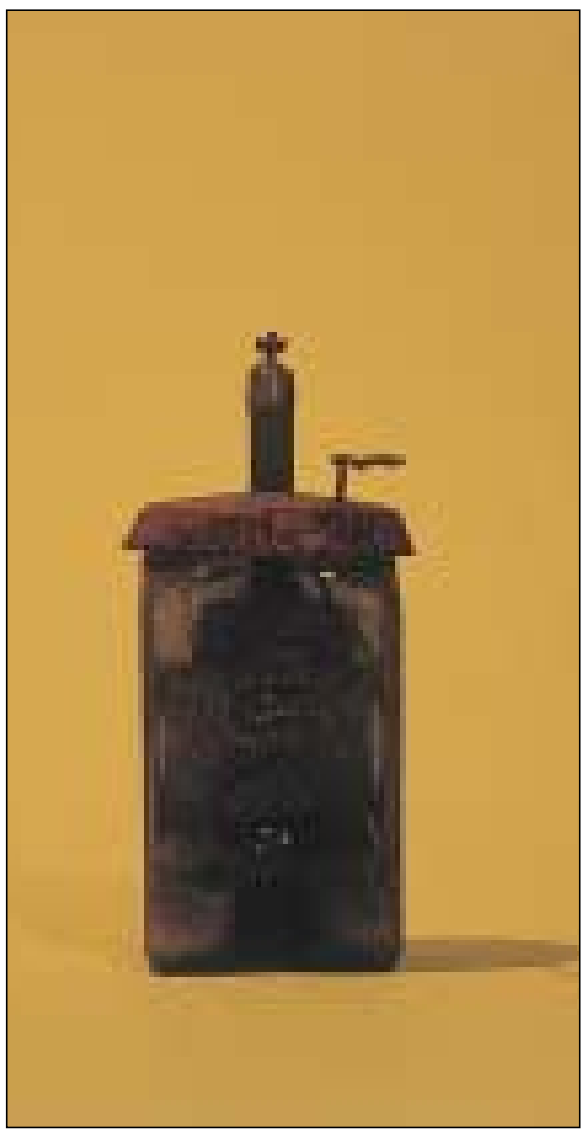

*Foto 24 - Pila Leclanché.*

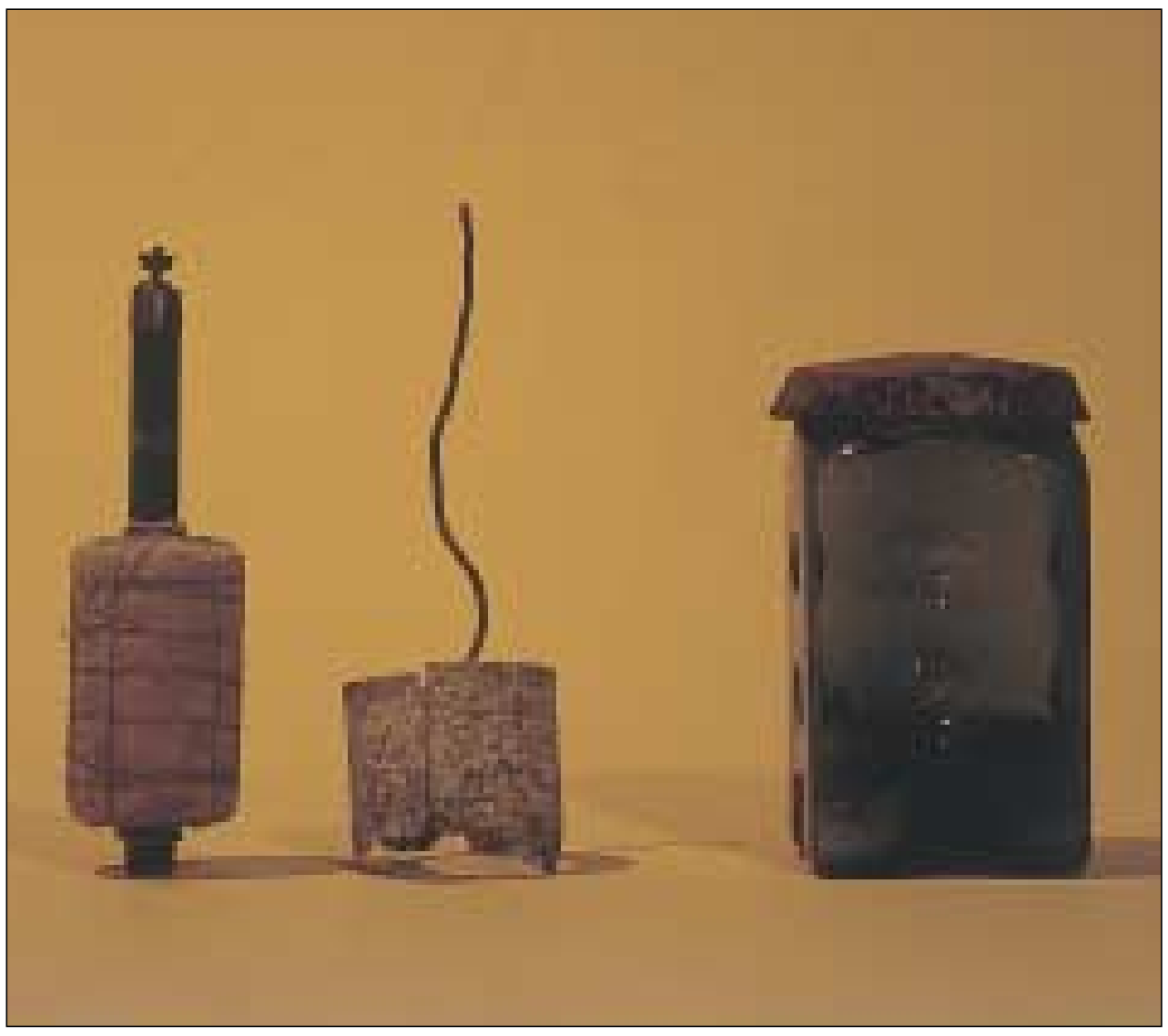

*Foto 25 - Pila Leclanché scomposta.*

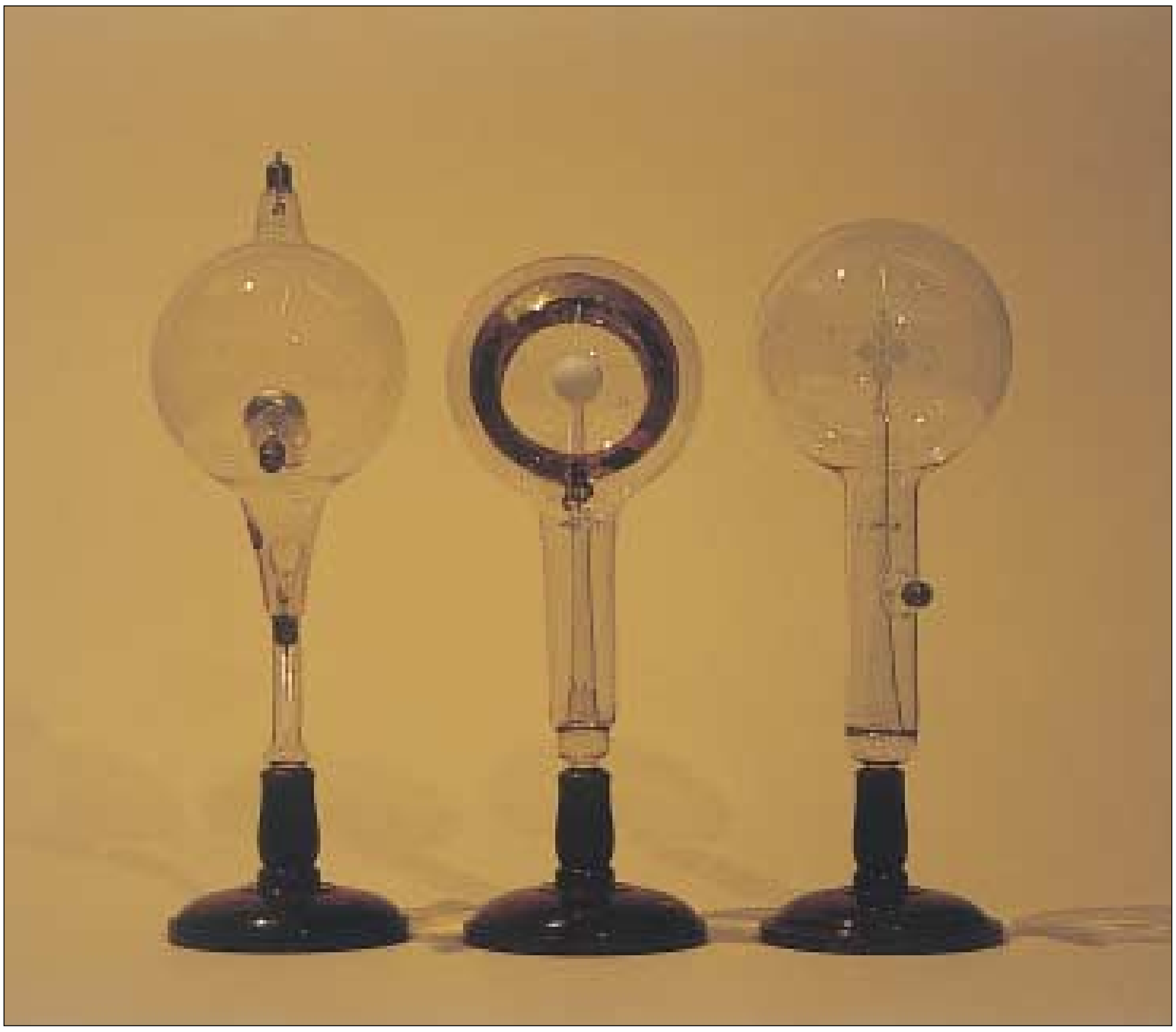

*Foto 26 - Tubi di Crookes.*

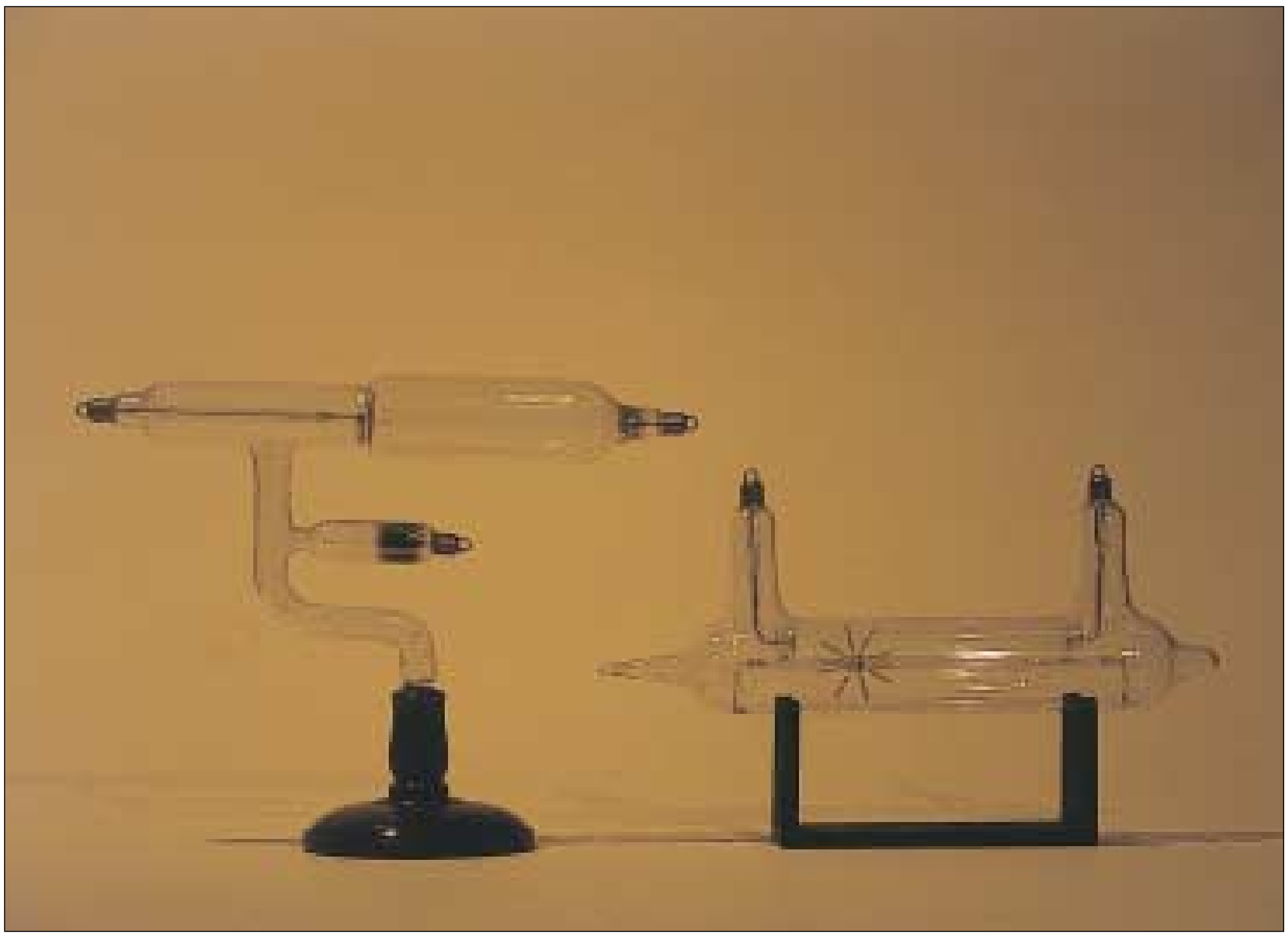

*Foto 27 - Tubi di Crookes.*

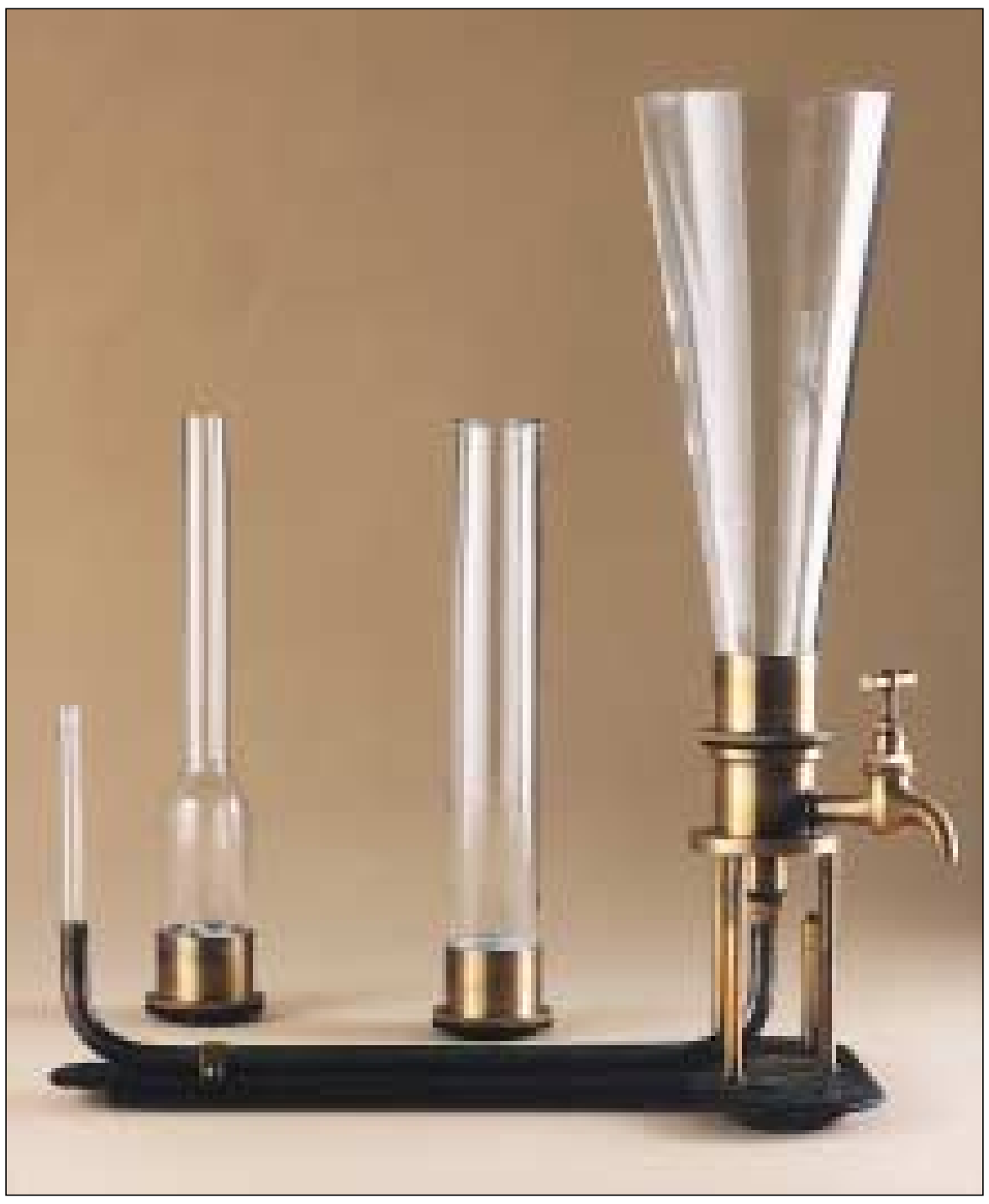

*Foto 28 - Apparecchio di Pellat.*

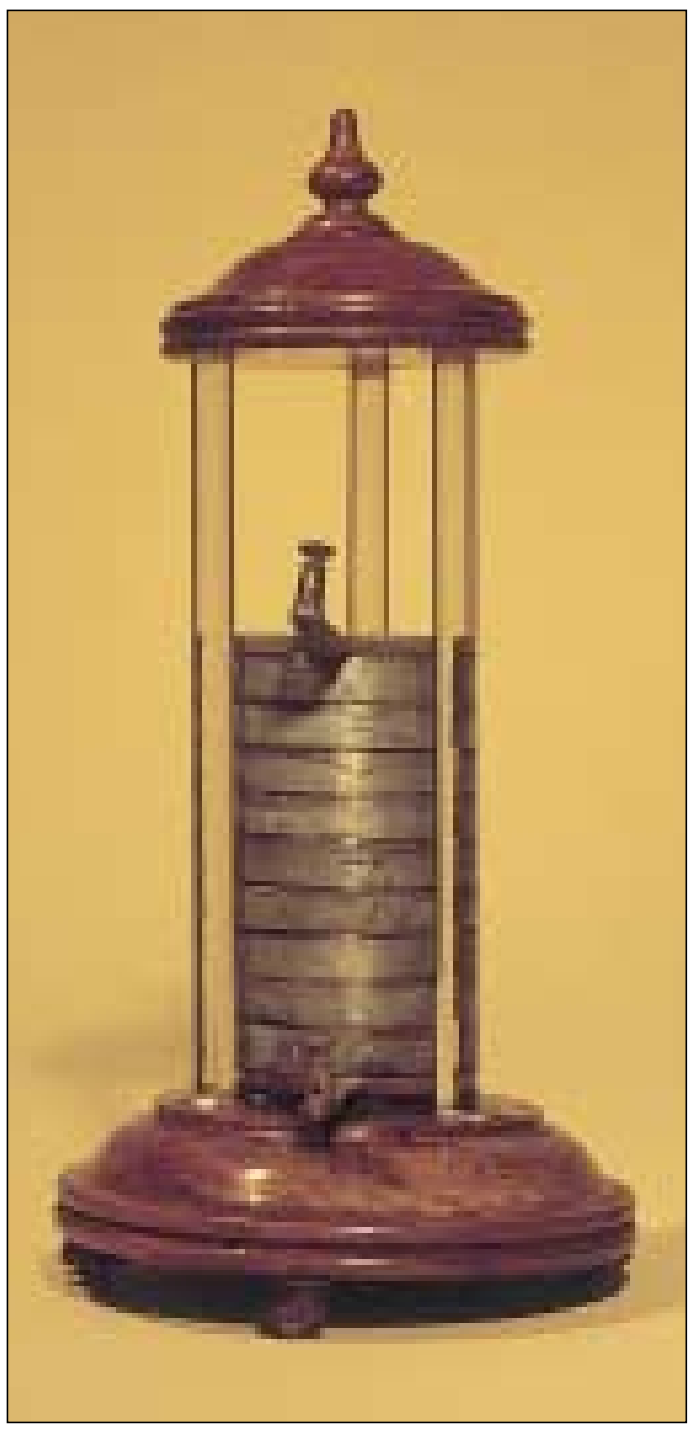

*Foto 29 - Pila di Volta.*

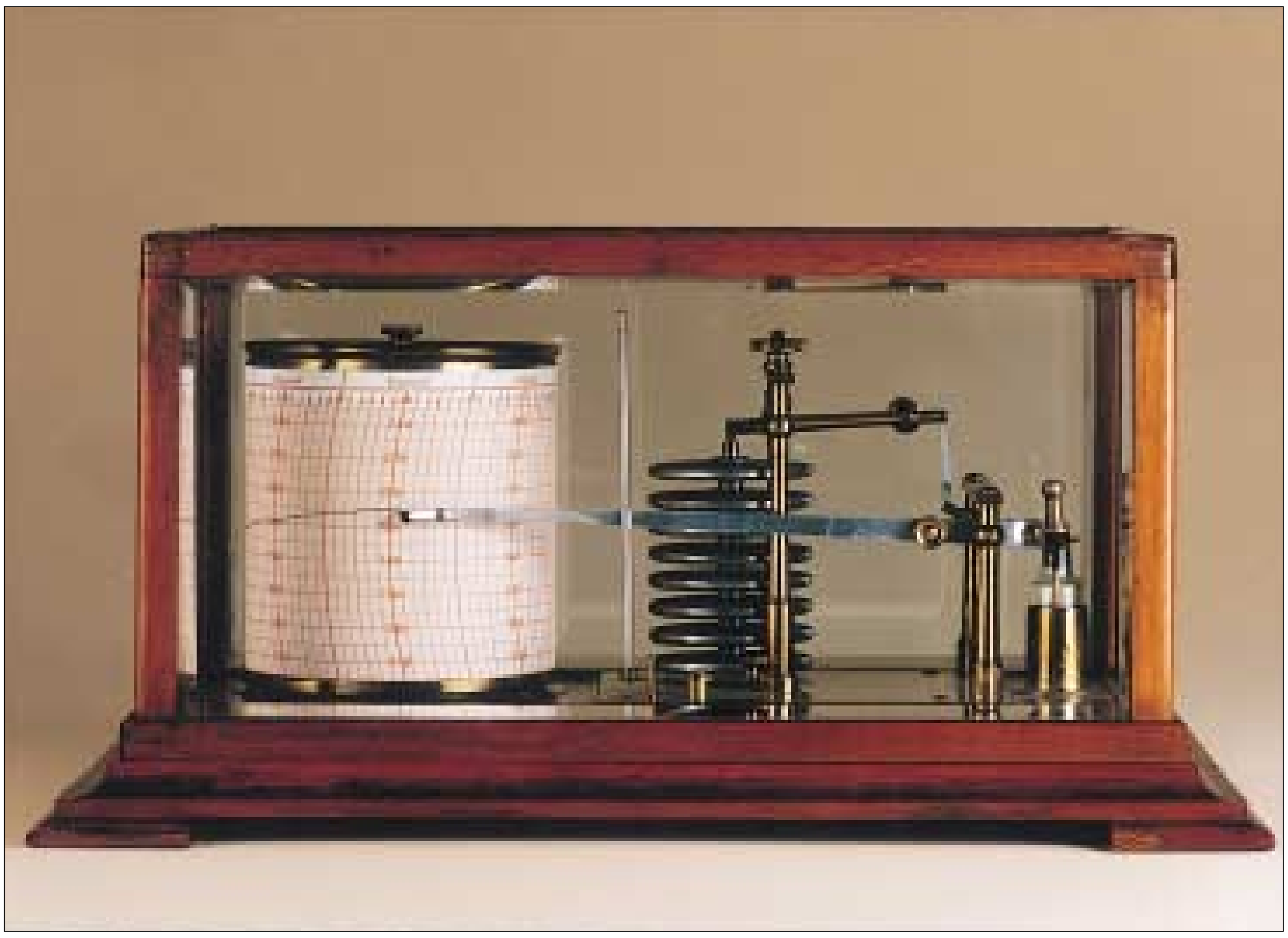

*Foto 30 - Barografo registratore Richard.*

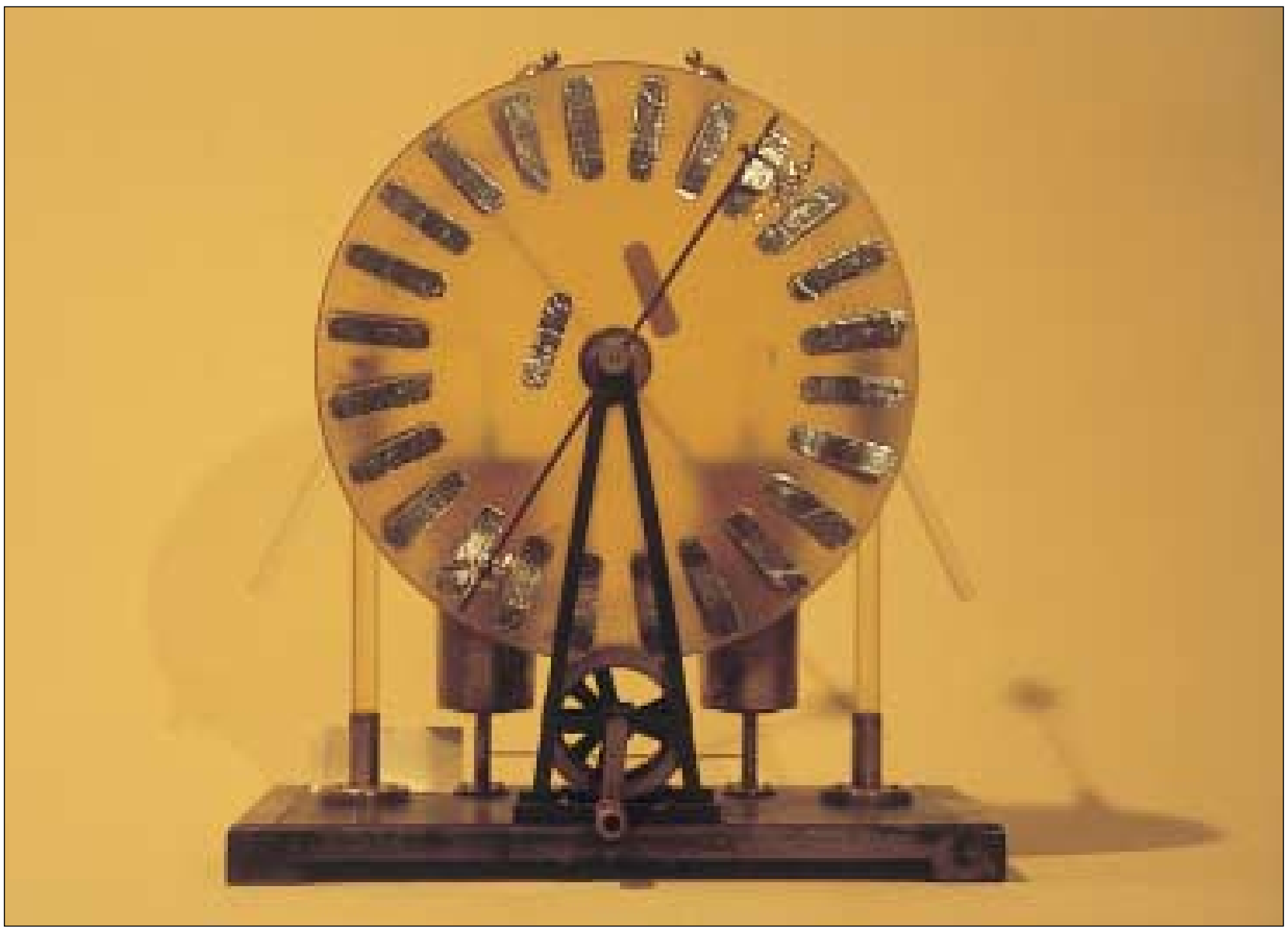

*Foto 32 - Macchina di Whimshurst.*

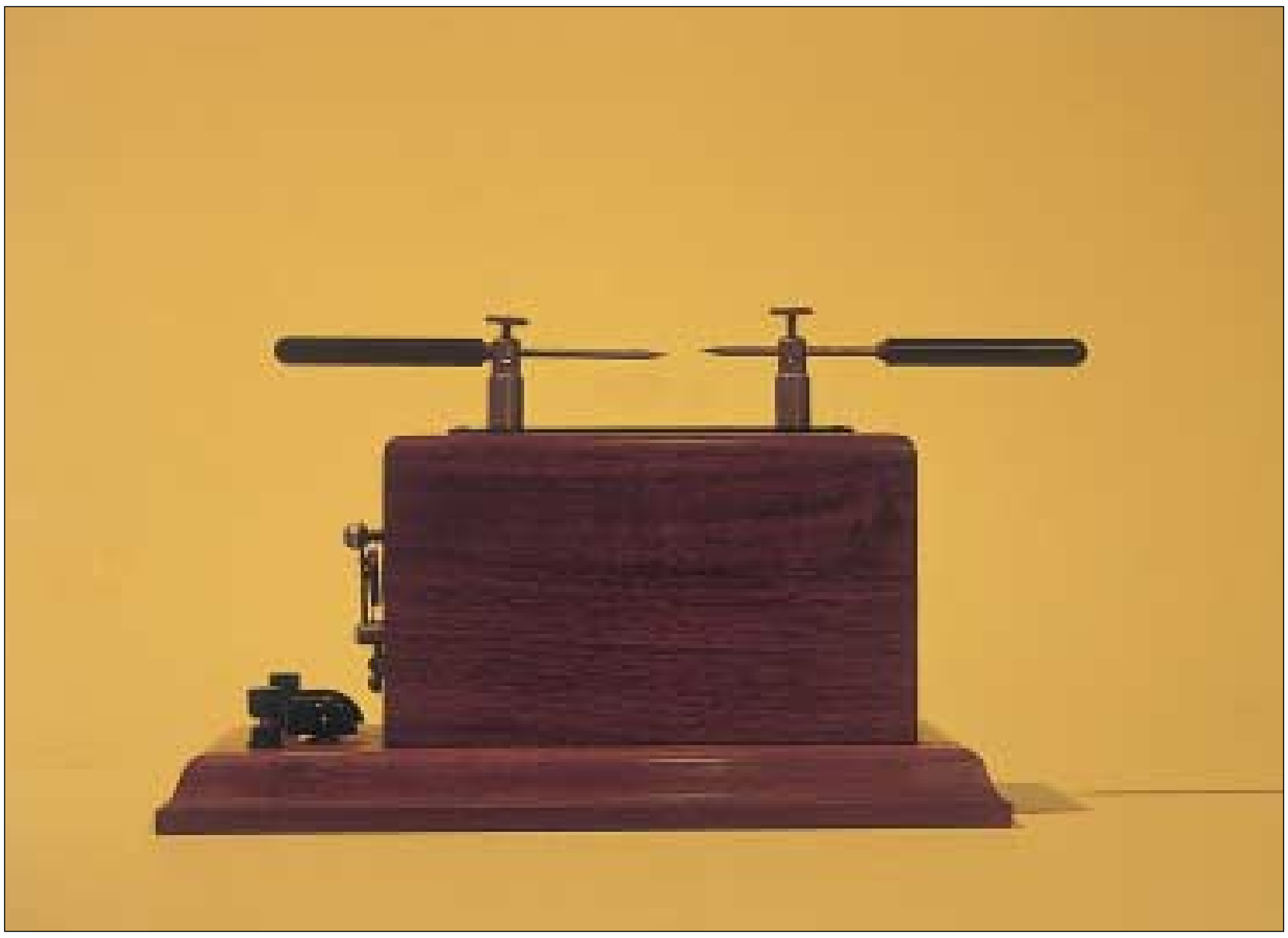

*Foto 31 - Rocchetto di Ruhmkorff.*

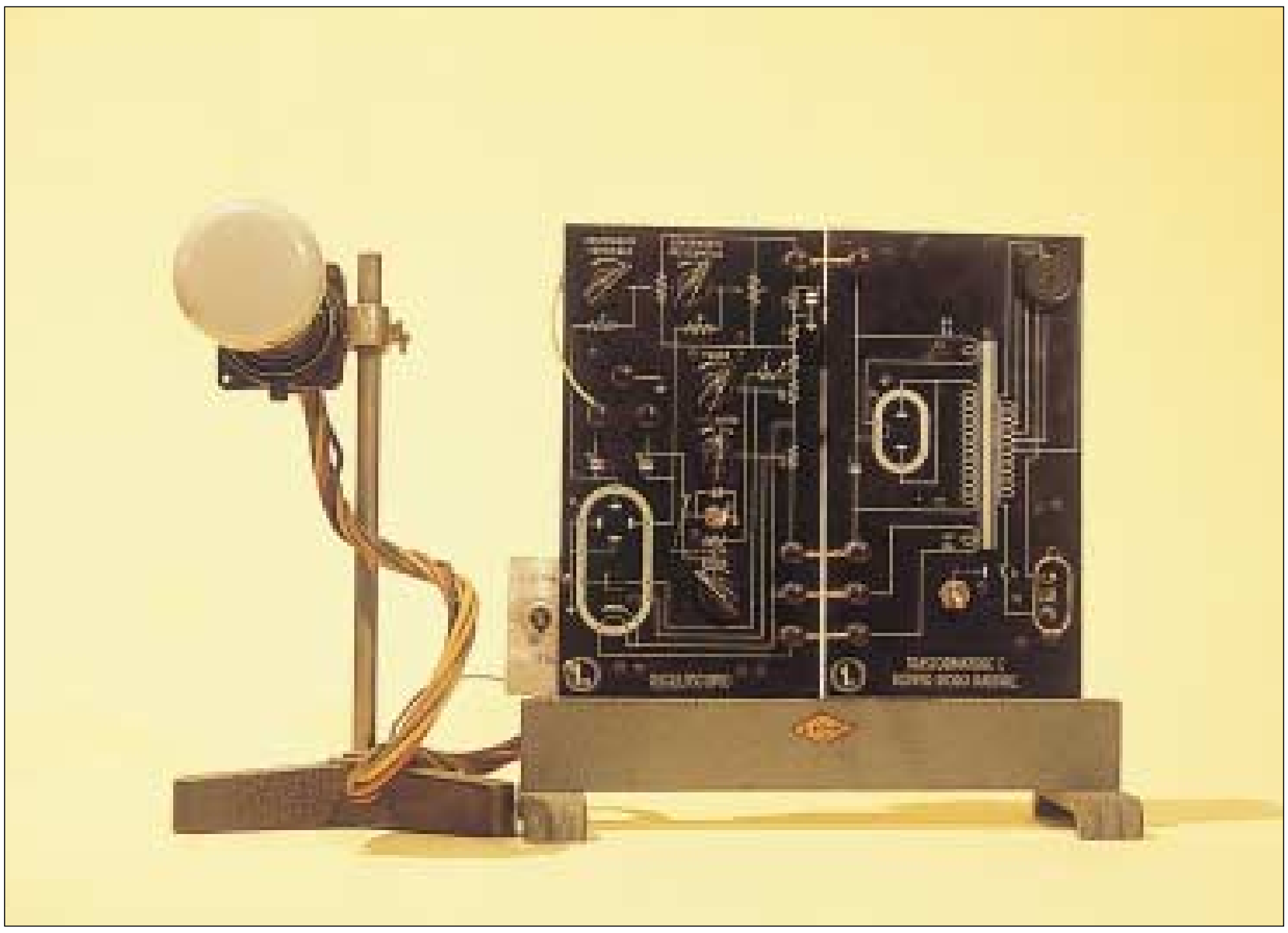

*Foto 35 - Modelli di valvole: Triodo; Pentodo.*

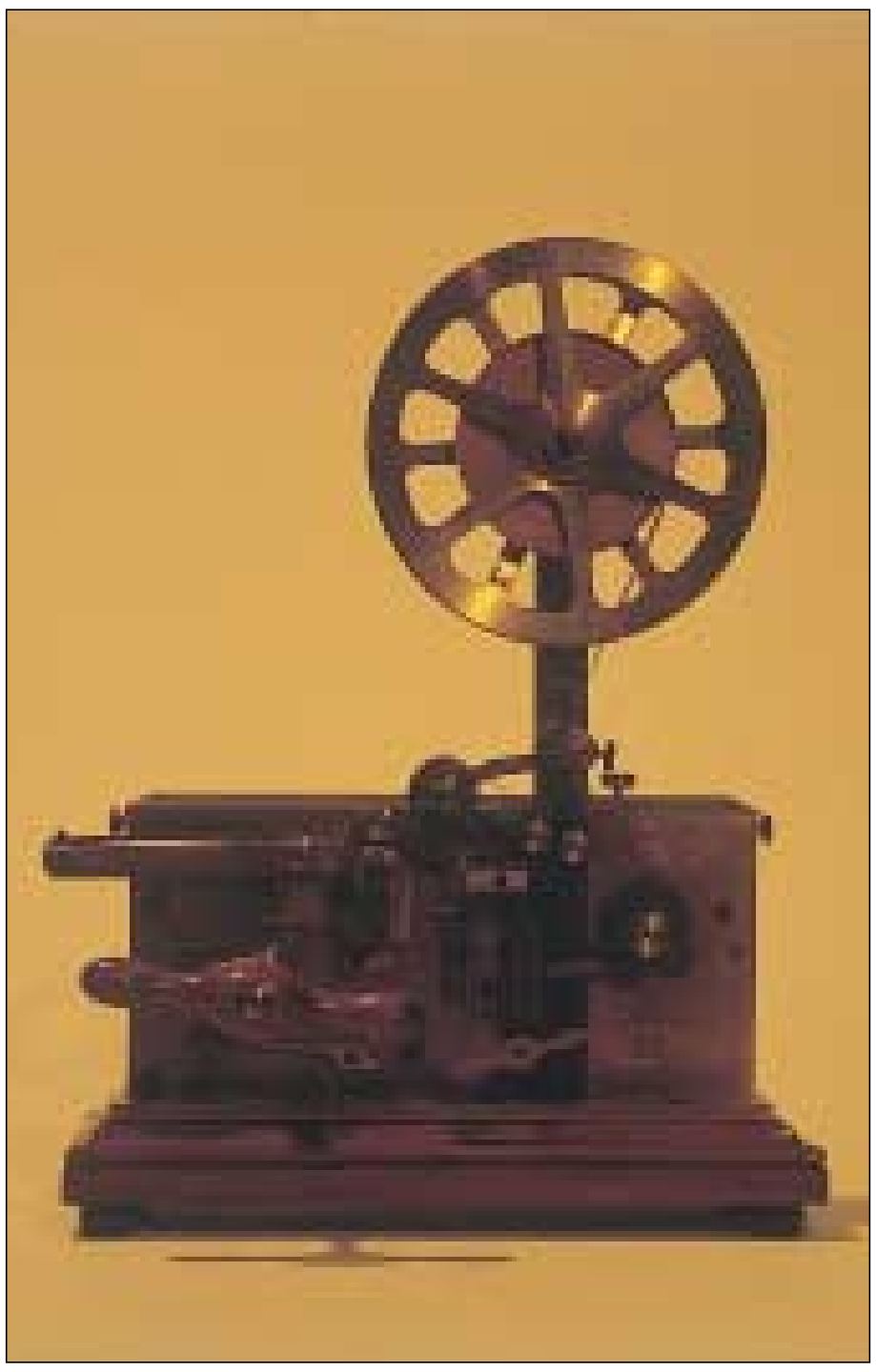

*Foto 33 - Telegrafo di Morse.*

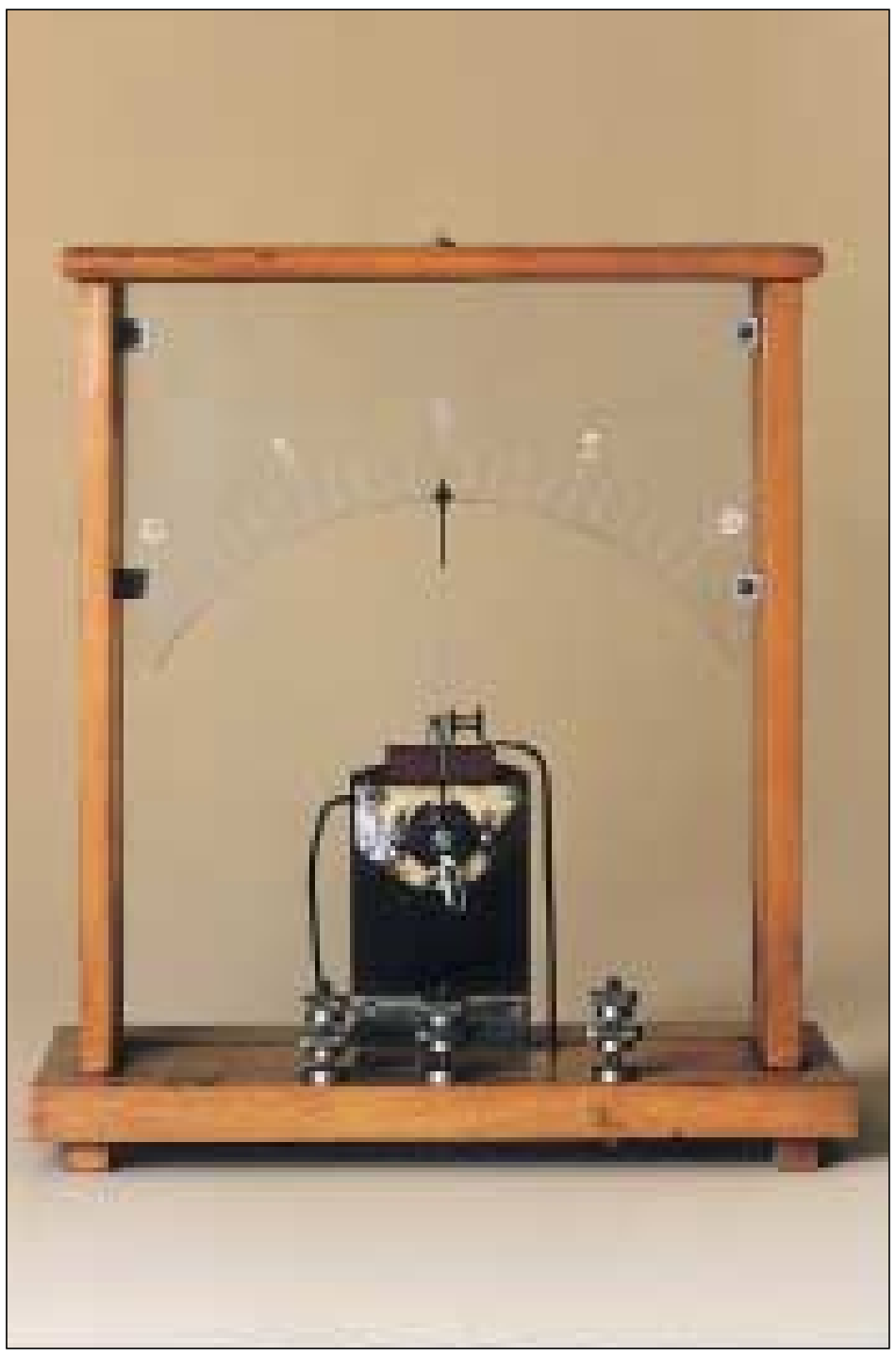

*Foto 34 - Galvanometro dimostrativo.*

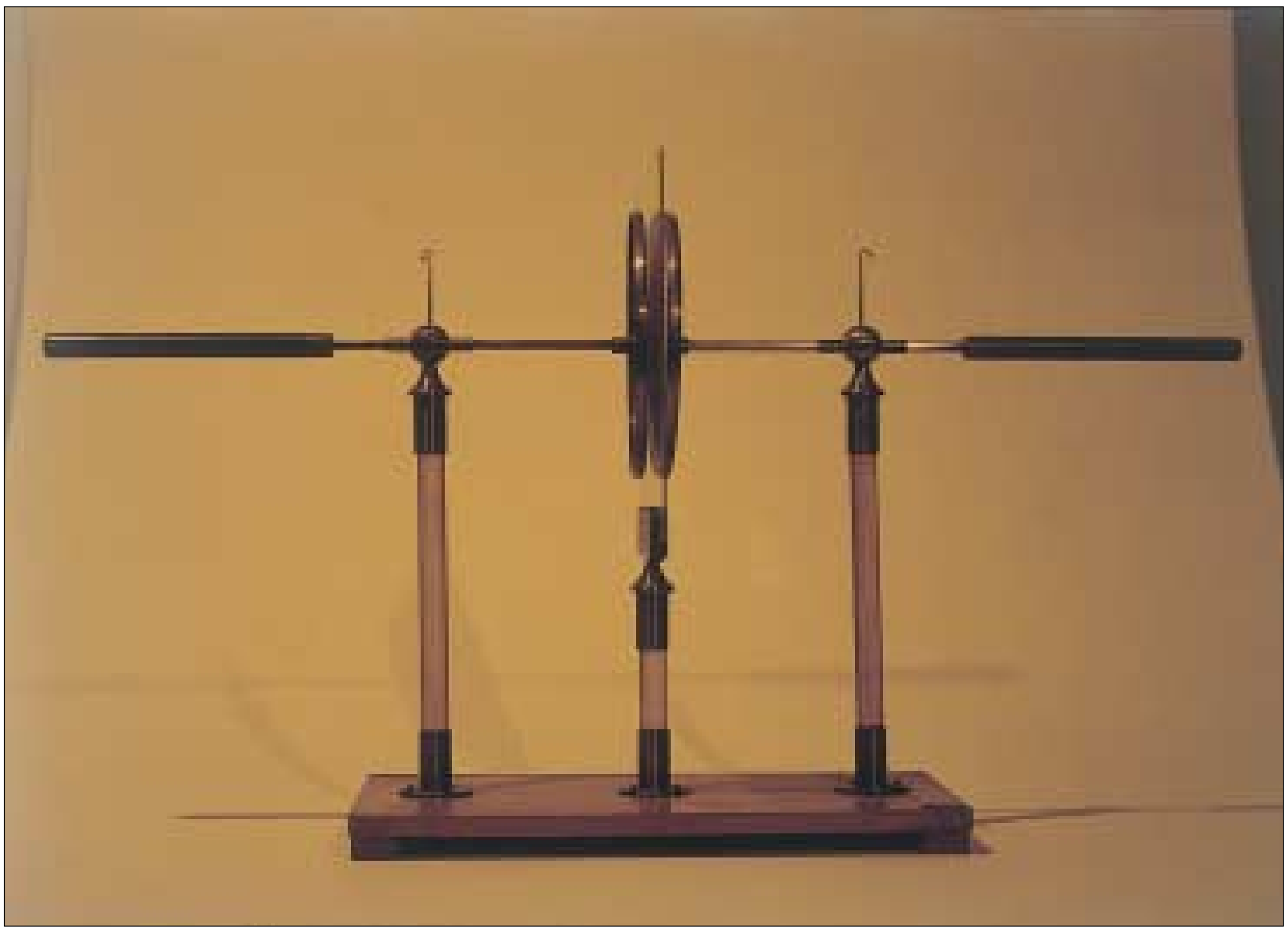

*Foto 36 - Condensatore di Epino.*

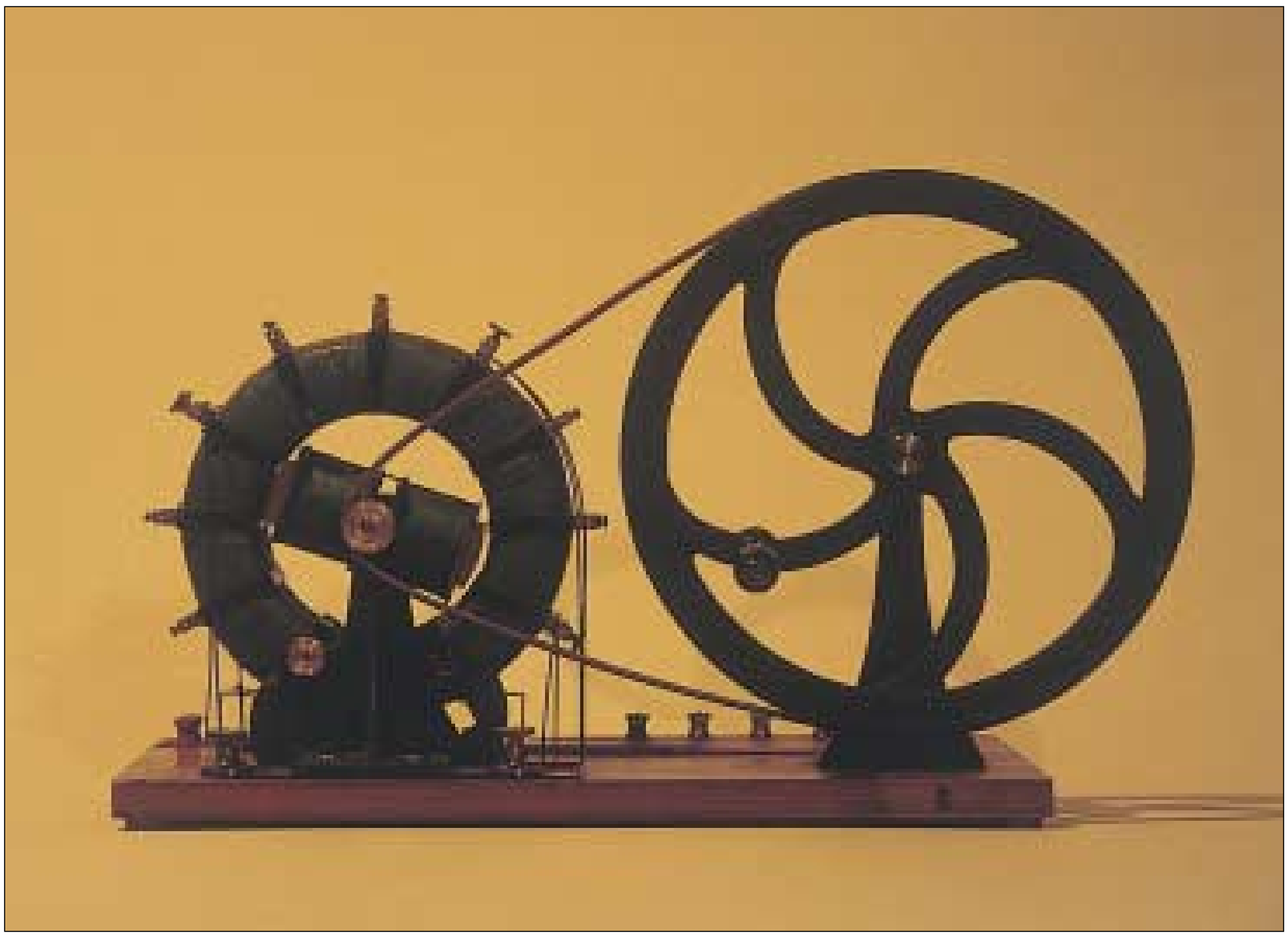

***Foto 37 - Anello di Pacinotti.***

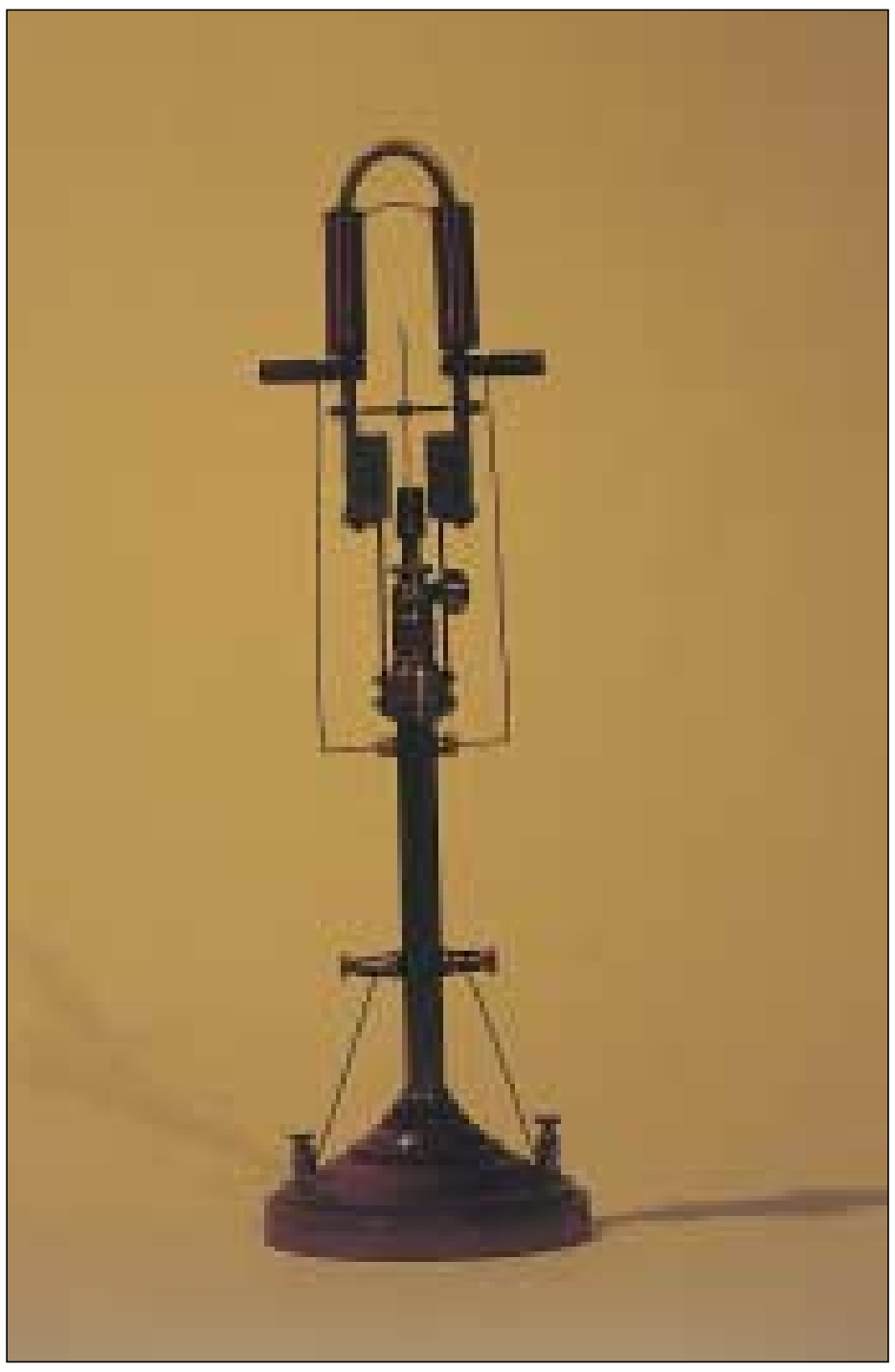

***Foto 38 - Ruota di Barlow.***

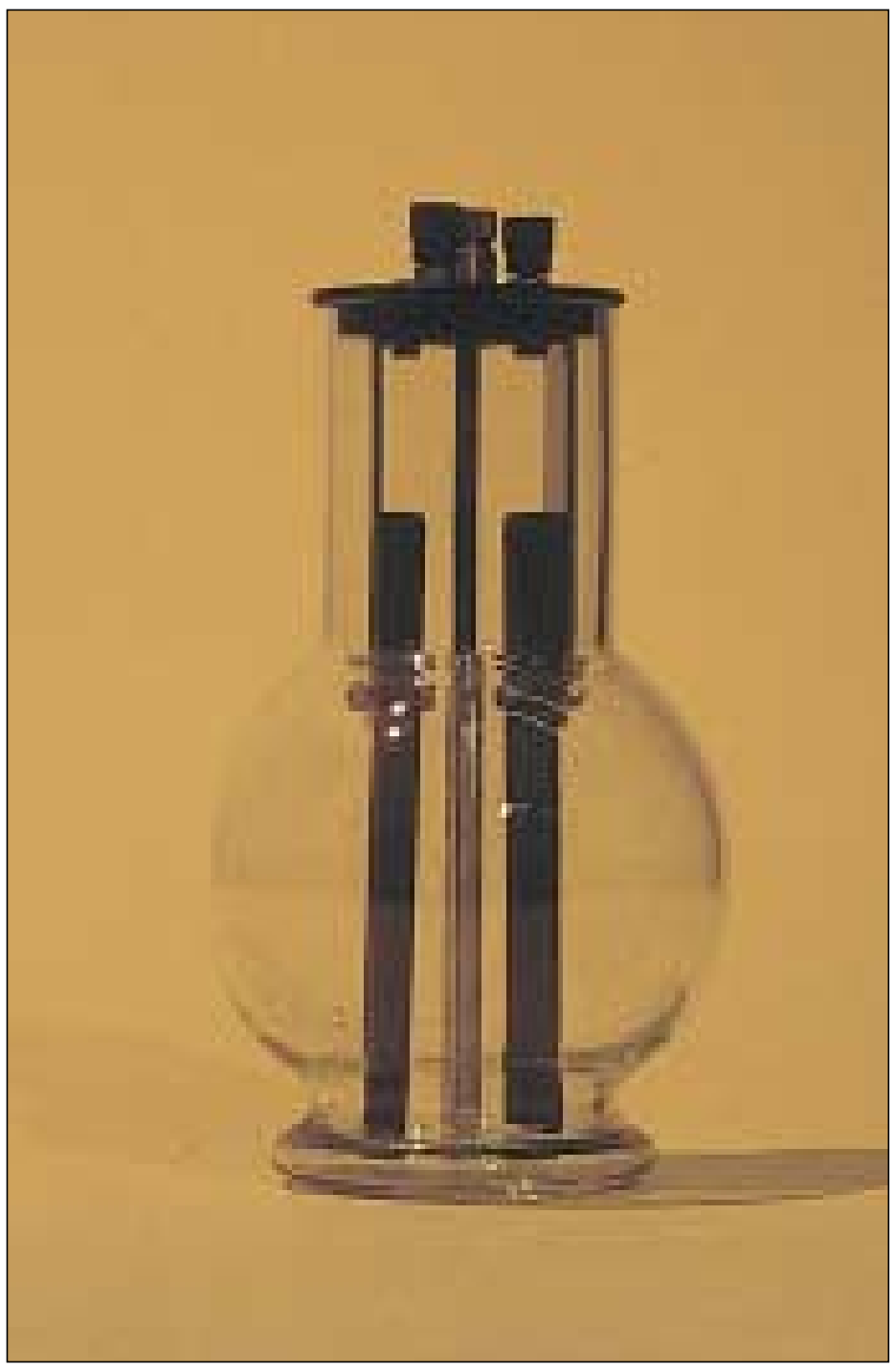

***Foto 39 - Pila di Grenet***

# 4. «SCHEDE» DEGLI STRUMENTI PIU` SIGNIFICATIVI DEL «QUARENGHI» e relative fotografie

Gli strumenti storici dell'Istituto per Geometri «G. Quarenghi» sono essenzialmente di tipo topografico, poiché quelli di Fisica sono rimasti tutti in dotazione all'Istituto «Vittorio Emanuele».
Gli strumenti dell'Ottocento sono otto e sono rappresentati nelle prime delle schede seguenti. Quelli della prima metà del Novecento (fino al 1950/60) sono diciotto; vari di essi presenti in più di un esemplare (per totali 39 pezzi).

## *RIGHE CON DIOTTRE A PINNULE (visori a traguardi)*

18mo - 19mo secolo / metallo / dimensioni varie

Componenti delle «tabulae» da rilievo, servivano per la collimazione dei punti e successivo tracciamento della direzione di mira. Combinate con le catene di misura, consentivano il rilievo di dettaglio per coordinate polari, mentre, con osservazioni di direzioni angolari da più stazioni, sui medesimi punti, fornivano per intersezione le posizioni planimetriche dei punti collimati.
Strumenti semplici e di largo impiego fino all'avvento dei goniometri a cannocchiale, permettevano la creazione di mappe disegnate direttamente sul terreno.

Come parti visibili sulla riga più piccola, per consentire anche le determinazioni altimetriche, si accessoriavano la pinnula per la collimazione con un cursore scorrevole in altezza, e la parte «oculare» con riferimenti a differenti altezze. La livella a bolla d'aria completava lo strumento garantendo che la linea di mira sia riferita all'orizzontale.

Le diottre di maggiore grandezza sono a traguardi reciproci, in modo che ciascuna pinnula possa essere utilizzata sia come «oculare» sia come «obiettivo».
Sono strumenti di scarsa precisione nel puntamento, tanto per i fenomeni di diffrazione alle aperture quanto per la difficoltà dell'occhio di mettere a fuoco contemporaneamente le fessure delle pinnule e l'oggetto collimato.

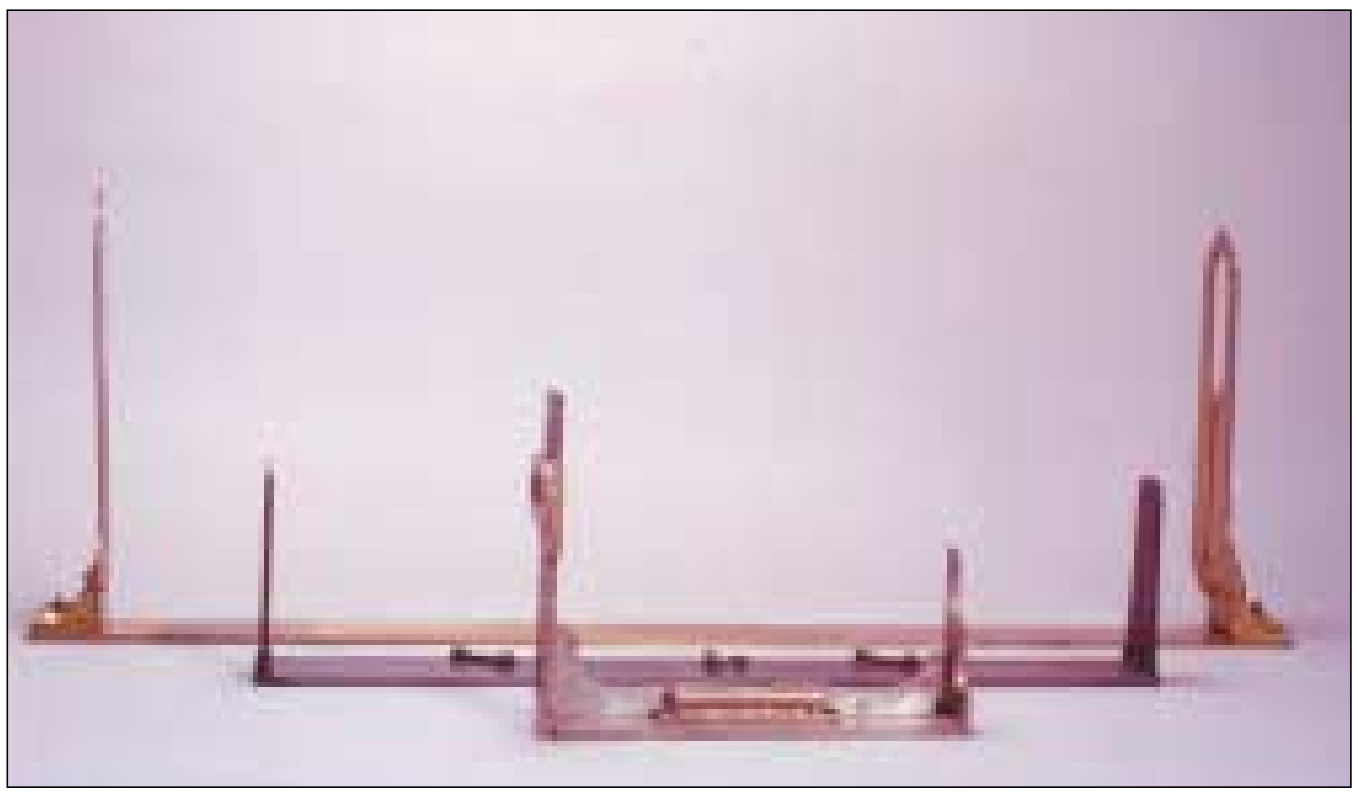

## *LIVELLO AD ACQUA*

’800 / ottone / lunghezza 95 cm, altezza 29 cm

E` tra i più semplici strumenti in grado di fornire l’allineamento orizzontale; il principio di funzionamento è quello dei vasi comunicanti, e l’operatore, traguardando i due peli liberi del liquido contenuto nei bicchieri, fissa l’orizzontale.

Usato fin dal settecento (soprattutto in lavori inerenti opere idrauliche quali canali, acquedotti, giochi d’acqua ecc.), cadde in disuso in seguito all’introduzione dei livelli a cannocchiale.

I due bicchieri sono fissati all’estremità di un tubo rigido smontabile e contenente acqua in quantità tale da dover risalire nei vetri; in posizione centrale vi è un innesto tronco-conico, per poter fissare il livello a un treppiede o a un bastone.
L’acqua impiegata veniva colorata per meglio evidenziare le superfici che materializzavano la linea di mira.
Con battute di 20-30 m, con questo strumento si hanno errori dell’ordine di 2-3 cm nel determinare il dislivello.

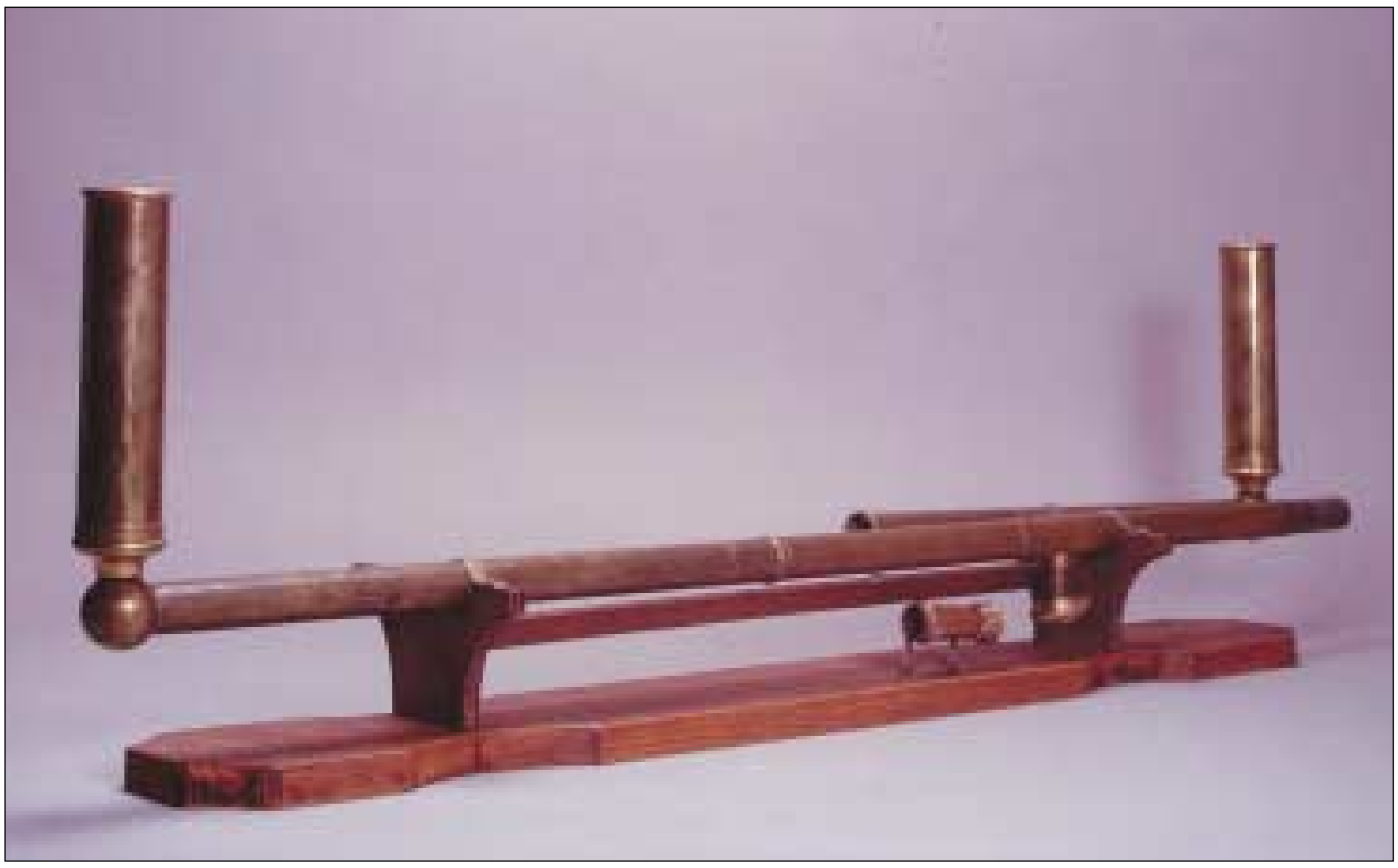

## *SEMICERCHIO TOPOGRAFICO ALTAZIMUTALE (GRAFOMETRO CON CANNOCCHIALI)*

18mo secolo / ottone / larghezza 28 cm, profondità 13 cm, altezza 25 cm

Questo strumento consente la misura di angoli sia azimutali sia zenitali, e può essere montato su treppiede. Il nostro esemplare reca la scritta «Des’cpas palais royal».
Impiegato nel rilevamento topografico, ha il semicerchio inciso con un’unica scala da 0 a 180°.
Il doppio cannocchiale consente un uso «a squadro» dello strumento.

La bussola, che rende possibile un orientamento geografico del rilievo, e il doppio cannocchiale fanno pensare ad una evoluzione, molto raffinata, del grafòmetro a pinnule. Il semicerchio è, infatti, solidale alla «alidade des stations», qui sostituita dal cannocchiale fisso inferiore, e col cannocchiale superiore mobile, che ricorda la «alidade mobille».
Il cannocchiale superiore, inoltre, ha la possibilità di ruotare verticalmente insieme con un ottavo di cerchio verticale graduato, consentendo così la misura delle distanze zenitali.
Lo strumento è dotato, infine, di una notevole livella torica collocata sulla alidada mobile; questa livella non compare sempre in strumenti peraltro molto simili, come ad es. quelli di Lennel (Parigi).

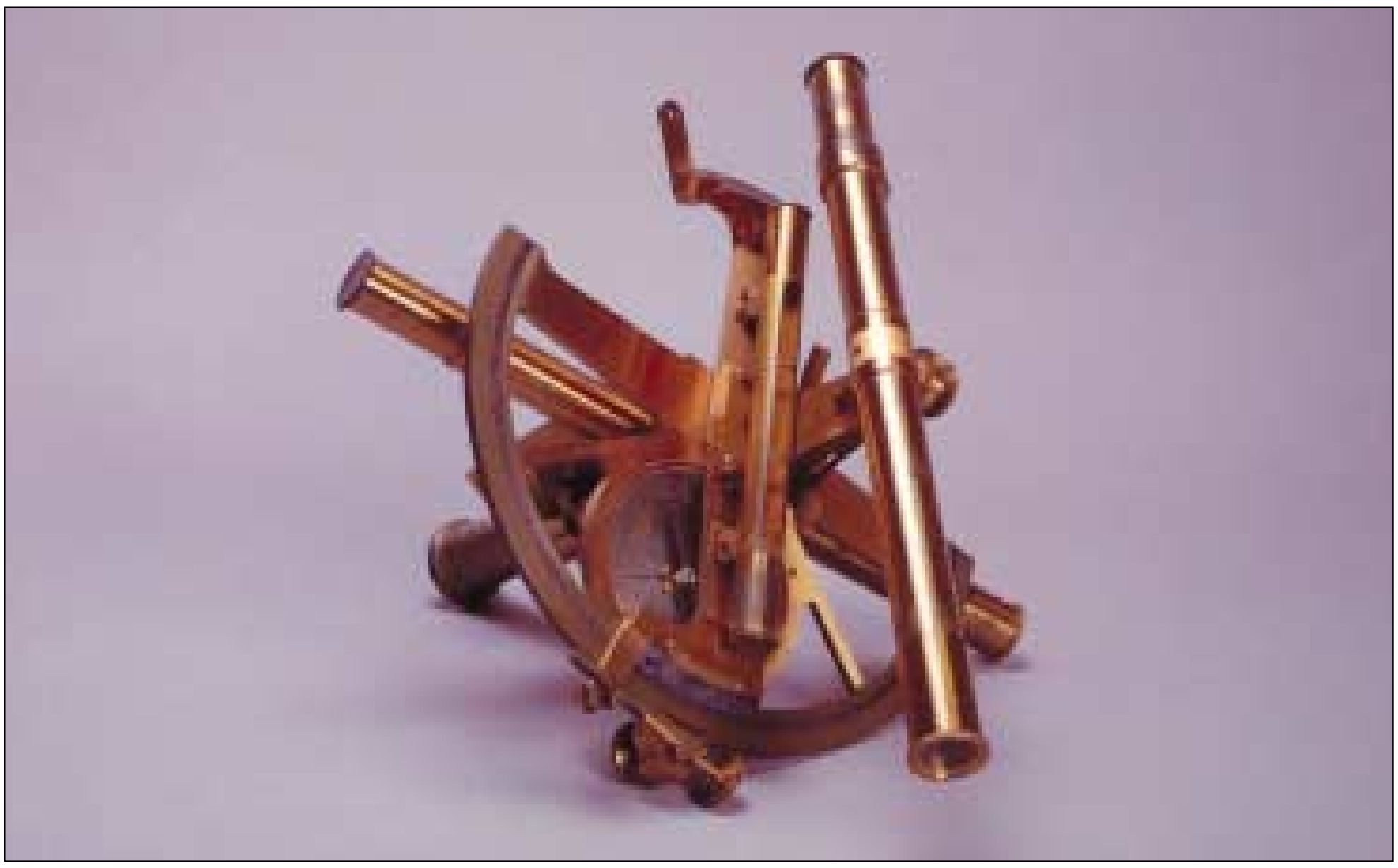

## *SQUADRO GRADUATO CON CANNOCCHIALE*

Fine ’800- primi ’900 / La Filotecnica - ing. A.Salmoiraghi

Lo strumento riporta l’incisione del rivenditore «L. Tironi – Bergamo». E’ uno squadro graduato a traguardi, munito di basamento con tre viti calanti. Il cannocchiale distanziometrico consente la misura delle distanze e, grazie alla scala laterale, la determinazione altimetrica dei punti collimati.

La livella torica montata sul cannocchiale suggerisce un uso dello strumento anche per semplici livellazioni di tipo geometrico.

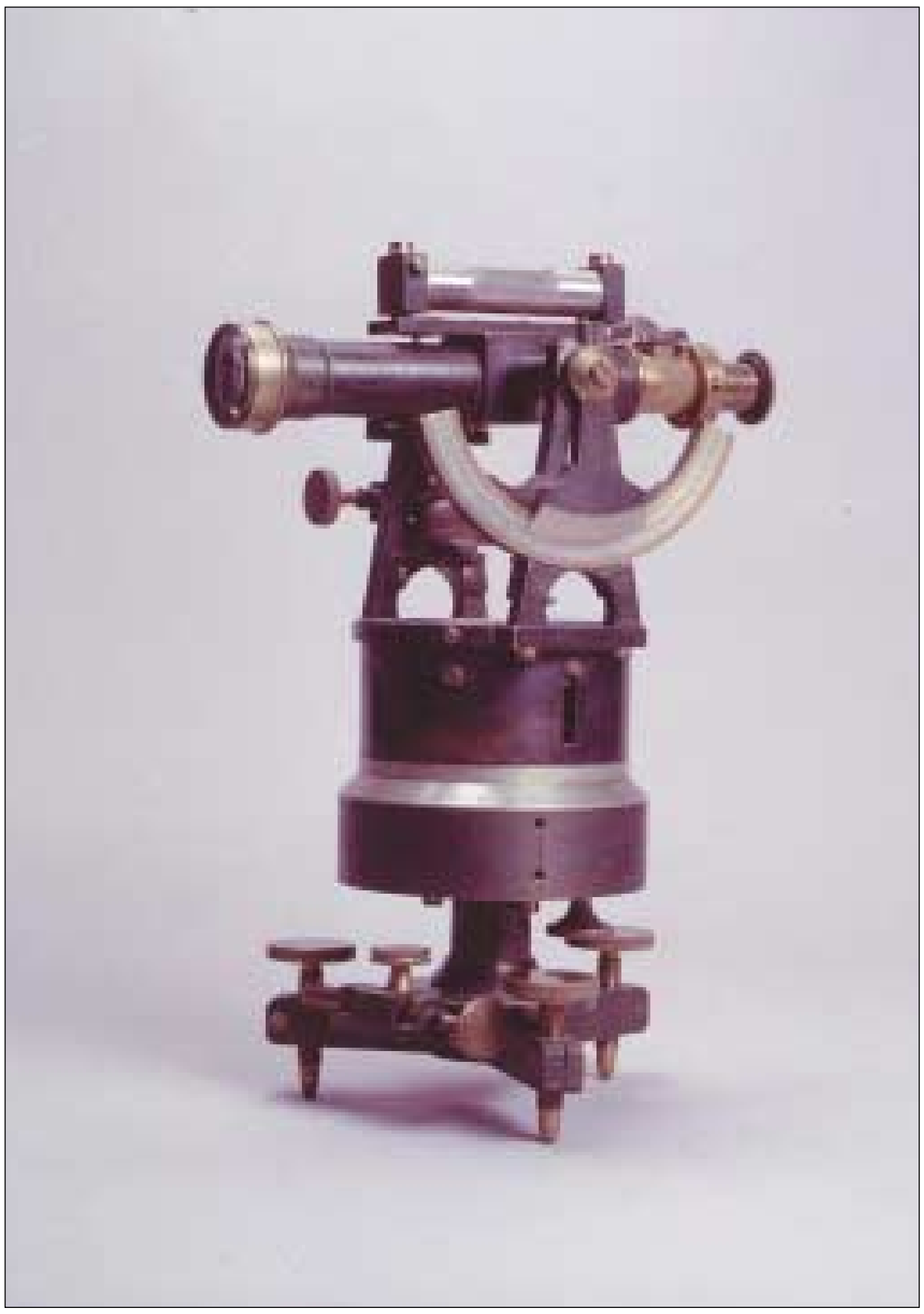

## *LIVELLO INCLINOMETRICO*
## *LA FILOTECNICA SALMOIRAGHI MOD 151*

primi ’900 / La Filotecnica - ing. A.Salmoiraghi

E’ un livello di pendenza detto anche clisigoniometro; è fornito di cerchio azimutale e bussola. L’asta graduata verticale riporta la scala delle pendenze, e il controllo dei movimenti in altezza della traversa avviene con dispositivo micrometrico.
Il cannocchiale è a lunghezza variabile e provvisto di fili distanziometrici.
Lo schema strumentale è derivato dal livello di tipo Egault; esso era impiegato per il tracciamento di ferrovie, strade e canali.

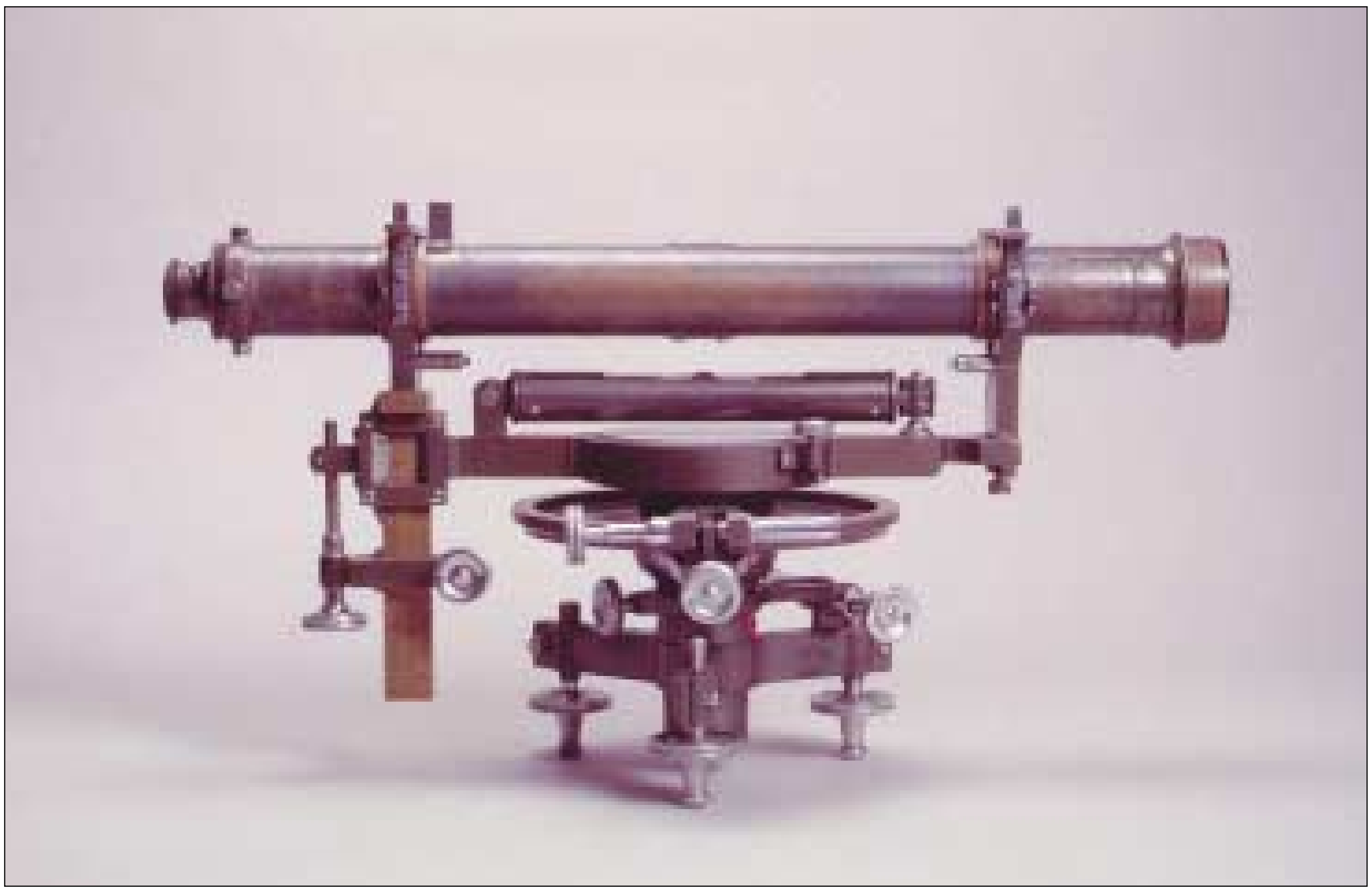

## *LIVELLO TIPO EGAULT*

Fine ’800 - primi ’900 / La Filotecnica - ing. A. Salmoiraghi

Classico livello ottocentesco con cannocchiale a lunghezza variabile e non centralmente anallattico; è munito anche di cerchio azimutale, con lettura diretta mediante microscopio a stima.
La livella è fissata alla traversa con viti laterali di rettifica.
Lo strumento veniva impiegato per livellazioni di una certa precisione, e operazioni di tracciamento sul terreno di infrastrutture e costruzioni in generale.

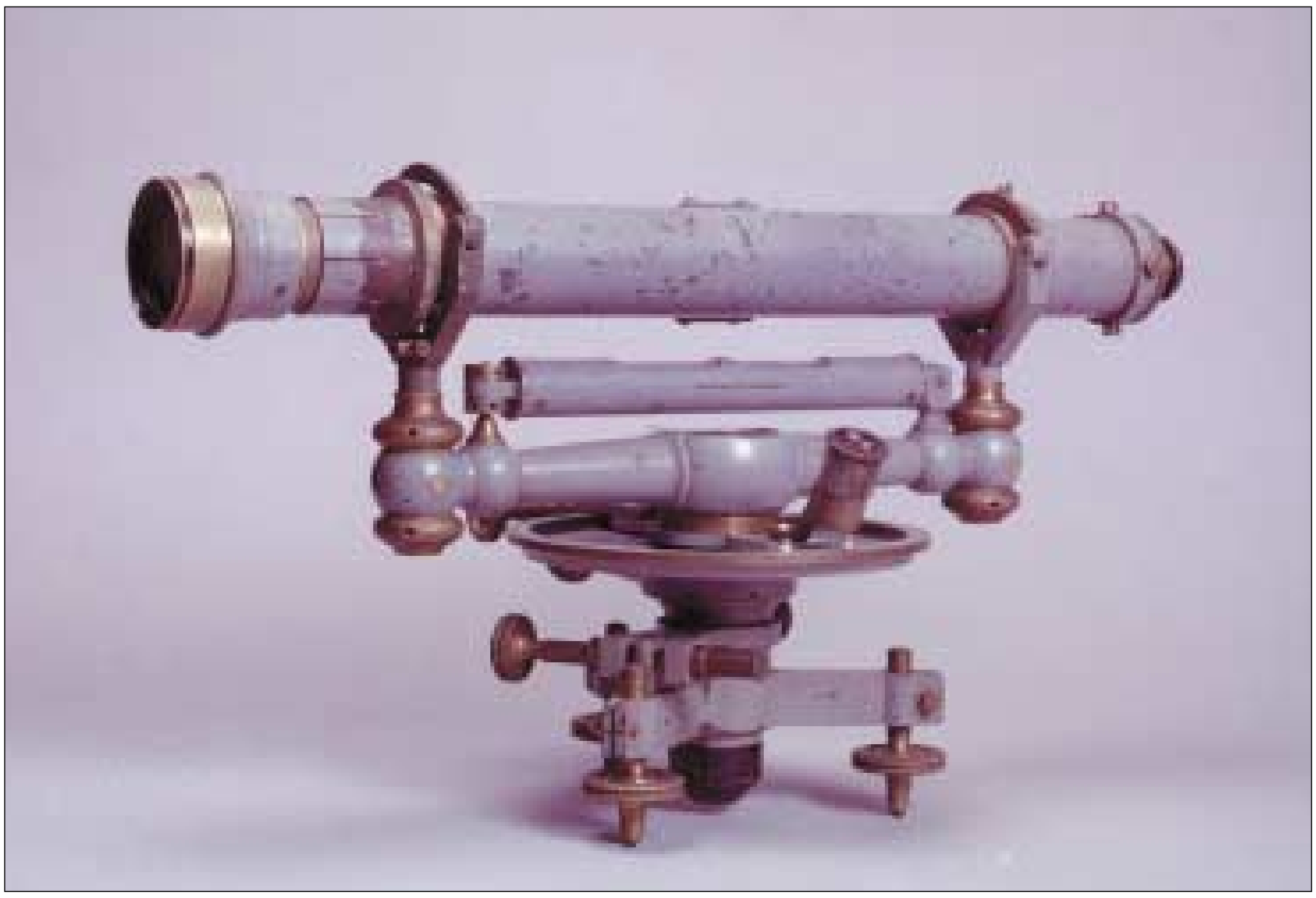

## *TACHEOMETRO SALMOIRAGHI MODELLO «CATASTO»*

fine ’800 - primi ’900 / ing. A.Salmoiraghi – Milano / larghezza 27cm, altezza 34 cm.

Goniometro per la misura di angoli azimutali e zenitali. Impiegato per lavori catastali, è noto anche come “tacheometro medio modello - n.134”, ed è caratterizzato dalla presenza di vernieri per la misura delle frazioni di graduazione dei cerchi.
Fa parte della seconda serie dei tacheometri prodotti da Salmoiraghi (la prima è quella dei tacheometri-cleps con microscopi a stima) e si allinea agli standard dei costruttori esteri dell’epoca. Il modello è tra quelli che sostituirono progressivamente i tacheometri di provenienza inglese, impiegati per la formazione del Nuovo Catasto Italiano a partire dai lavori del Catasto Modenese (1885).
Il cannocchiale è anallattico, con obiettivo da 32 mm e oculare ortoscopico con ingrandimento di circa 18 volte; su di esso compare la scritta “OMAP - Milano”. I cerchi, con divisioni su argento, misurano circa 13 cm di diametro. I vernieri di lettura sono diametralmente opposti, e permettono una precisione del primo sessagesimale. Di norma è dotato di livella torica d’alidada (mancante nel nostro esemplare), e livella mobile, da appoggiarsi su collari calibrati del cannocchiale. Sono presenti il declinatore magnetico e l’originale piombino sospeso.

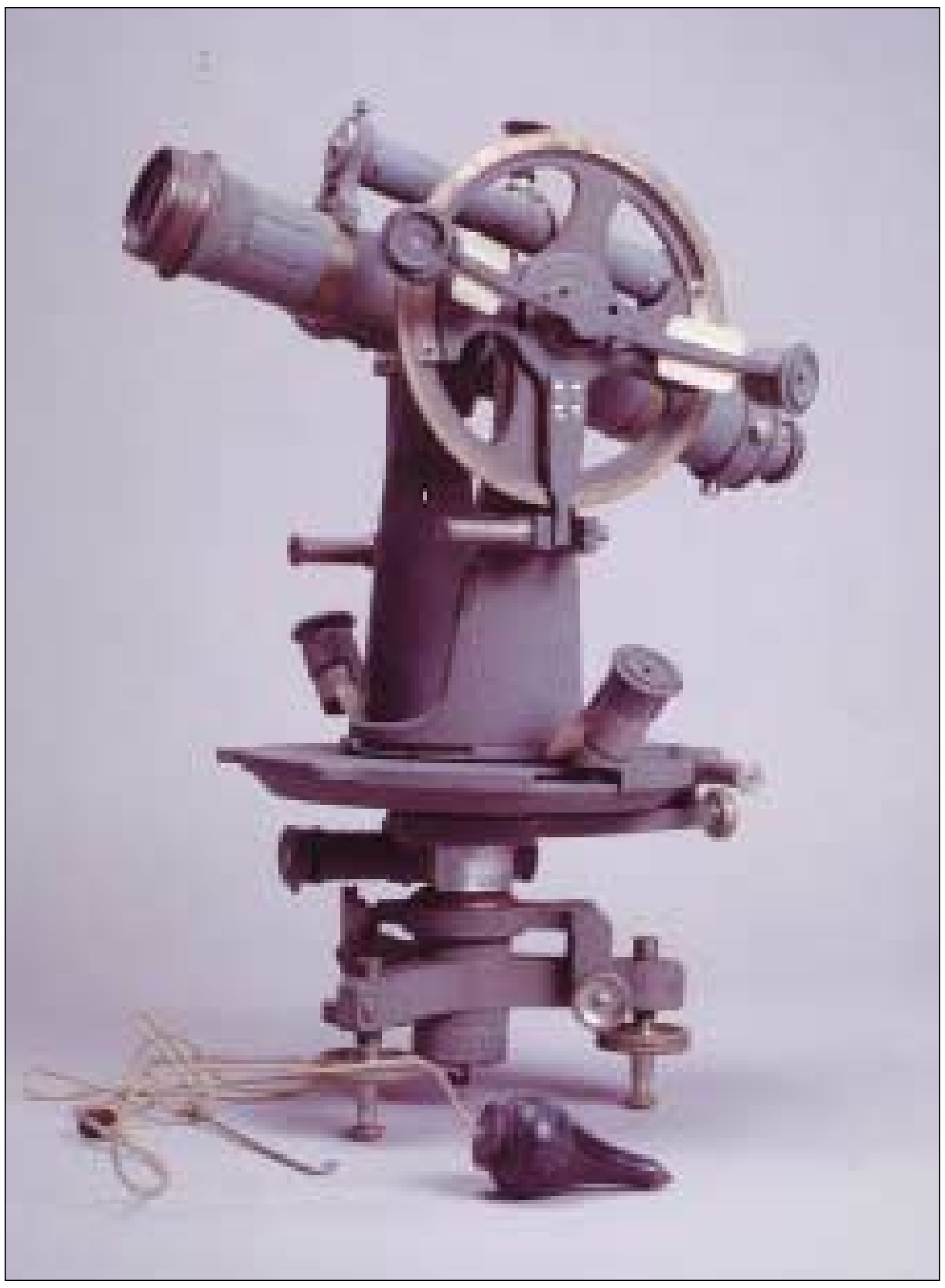

## *TACHEOMETRO A VERNIERI SAIBENE*

Primi anni ’900 / Officina Meccanica SAIBENE, Milano / lunghezza 27 cm, larghezza 16 cm, altezza 32 cm

Lo strumento è stato prodotto nelle officine Saibene di via Paullo (Milano), Casa fondata nel 1897, e costruttrice, oltre che di tacheometri, anche di livelli, squadri e flessimetri. La produzione, dopo alterne vicende, è cessata definitivamente negli anni Sessanta del XX secolo.
Il tacheometro qui rappresentato è del tipo ripetitore, con cerchi in metallo e doppi noni per le letture opposte.
Veniva impiegato per tracciamenti in cantiere, per rilievi di dettaglio e operazioni catastali di non elevata precisione.
Vistosi appaiono le lenti di ingrandimento e relativi schermetti riflettori per le letture ai cerchi. Il castello dell’alidada ha la classica forma ad A dei tacheometri europei di inizio secolo ed è di foggia semplice e leggera. I cerchi sono protetti con coperchi metallici, e visibili all’esterno attraverso finestrelle con cristallino. La notazione delle scale è centesimale.
Il cannocchiale è del tipo astronomico con apertura dell’obiettivo di 32 mm e reticolo a tre fili distanziometrici. Una livella torica è fissa sul cannocchiale. Quest’ultima, più sensibile della livella d’alidada, poteva migliorare la verticalità dell’asse di dotazione primario e servire per un impiego dello strumento “tipo livello”: per la misura di dislivelli col metodo della livellazione geometrica dal mezzo, oppure per il rilievo di piani quotati con asse di collimazione orizzontale.

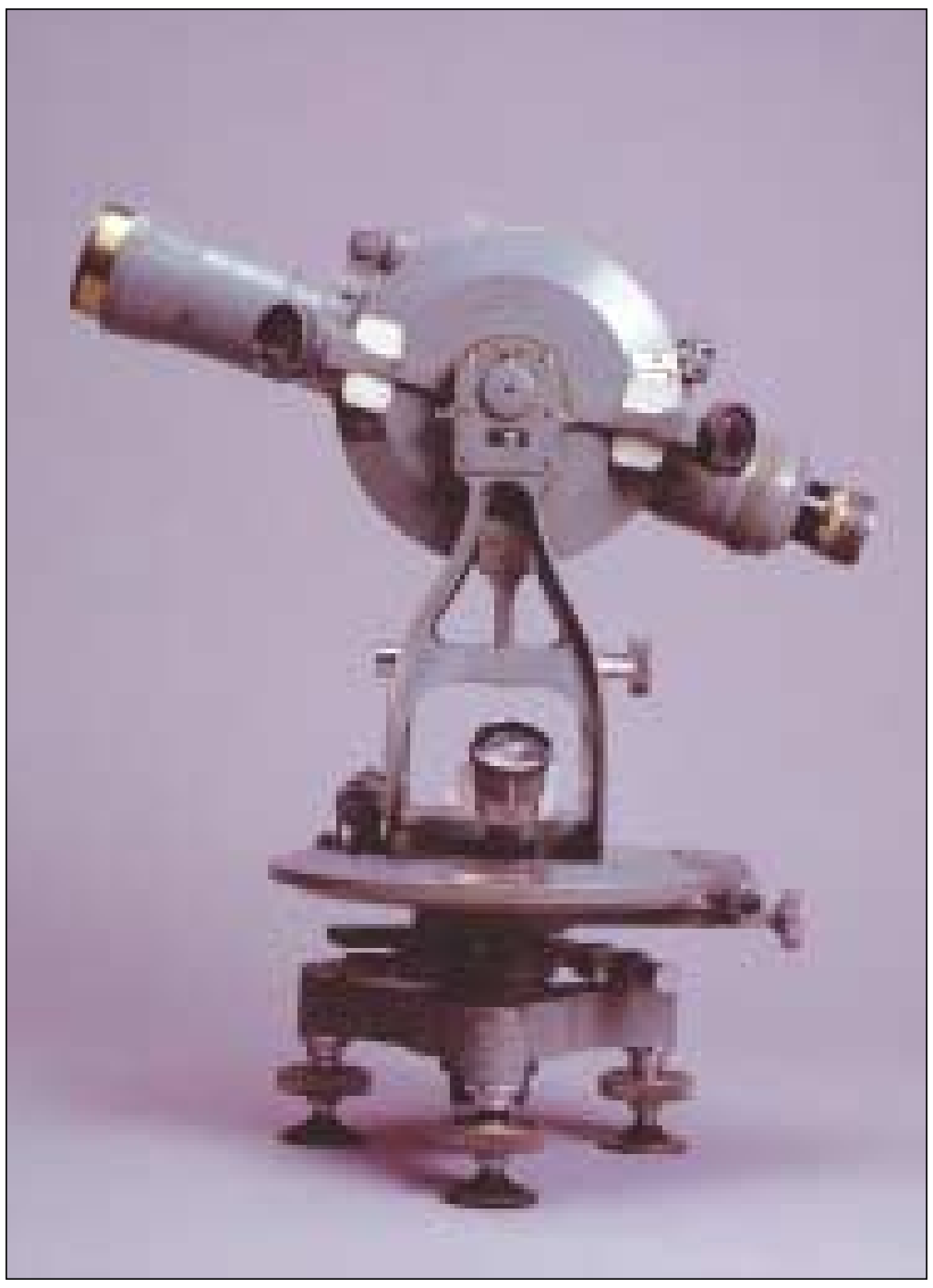

## *SESTANTE HIGGINS*

1960 / Higgins (UK)

Il sestante, goniometro a riflessione, trova scarse applicazioni nel lavoro topografico, mentre è usato in navigazione per la determinazione soprattutto della latitudine.
L'operazione che si esegue è la misura di angoli fra due direzioni, una riflessa doppiamente dagli specchi e l'altra collimata col cannocchiale dello strumento.
Il raggio mobile collegato con lo specchio rotante, infatti, individua sull'arco graduato un angolo che è la metà dell'angolo effettivo: e pertanto la graduazione è segnata in valore doppio. La lettura viene eseguita con un micrometro per migliorare la determinazione delle frazioni di divisione.
Il tipo qui rappresentatato ha una salda impugnatura, per poterlo sostenere con una mano. Il telaio riporta gli specchi e i filtri perpendicolari ad esso, e la scala graduata. E' presente il dispositivo di illuminazione per operazioni notturne.

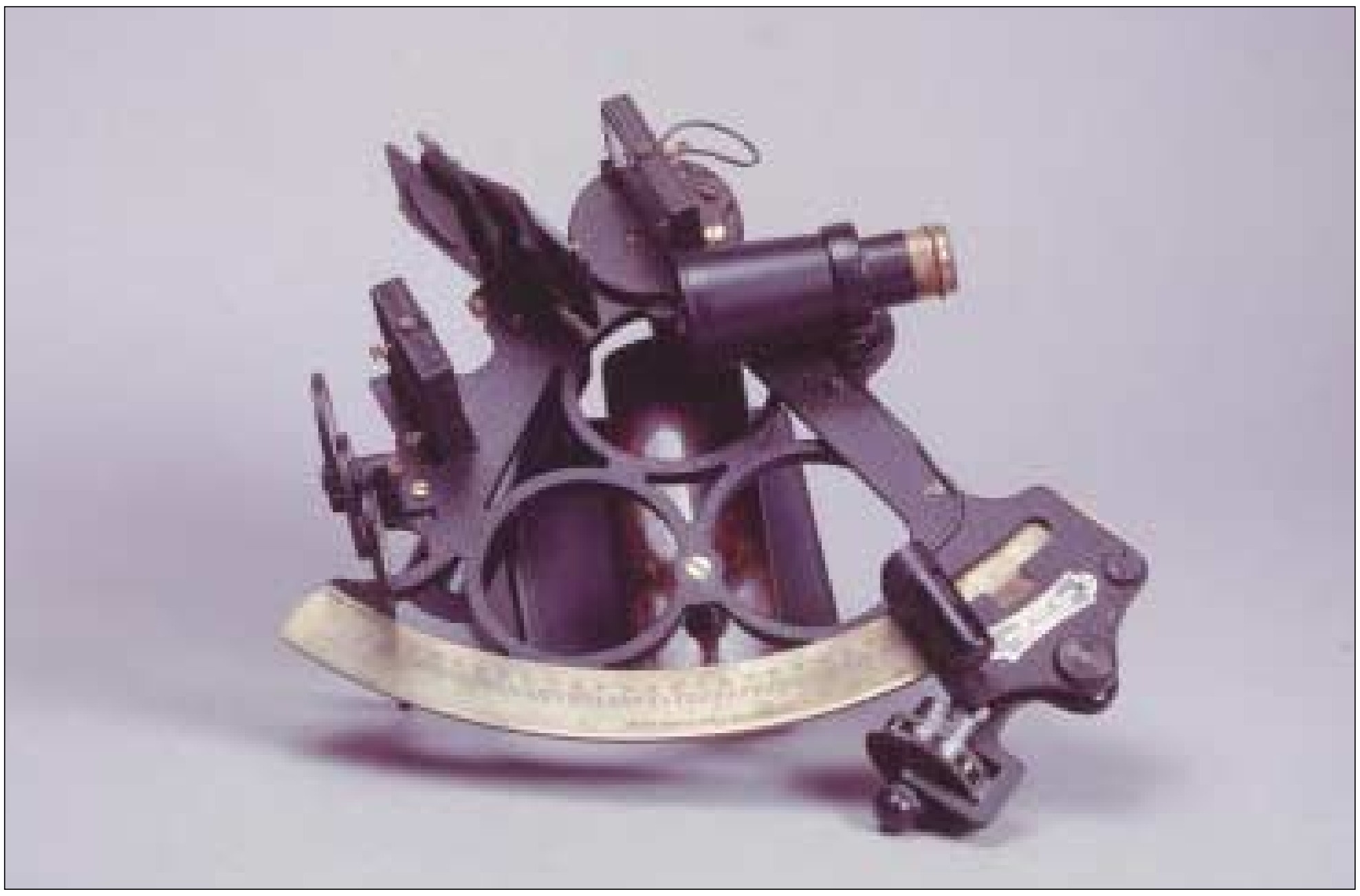

## *TAVOLETTA MONTICOLO*

1950 / Officine Galileo / lunghezza 13 cm, altezza 18 cm, larghezza 2 cm

Il nome deriva dal suo ideatore. Strumento di modesta precisione, è utile per rilevamenti speditivi mediante misure indirette di distanze e dislivelli, con la possibilità del tracciamento immediato del grafico. Il successo della soluzione si deve all'idea di condensare su una tavoletta di piccolo formato, tutti i dispositivi per un rilievo planoaltimetrico semplificato:

- un distanziometro a prismi con piccolo cannocchiale da usarsi con la tavoletta in posizione verticale;
- un ecclimetro per la determinazione della pendenza;
- una bussola per le determinazione azimutali;
- una livella e alcune scale di riduzione per completare le varie operazioni.

I valori forniti dai dispositivi per la misura della distanza, della pendenza e della direzione possono essere registrati per consentire le operazioni di calcolo e disegno. Il foglio da disegno, infatti, è fissato nel retro della tavoletta e, mediante abachi, vi si possono riportare direttamente le distanze ridotte all'orizzontale e in scala. La tavoletta Monticolo è stata usata per rilevamenti sommari in galleria, a scopo geologico, e forestale. La modesta precisione dello strumento si deve soprattutto alla bussola, quale riferimento graduato per la misura delle direzioni, alle misure con l'ecclimetro e al metodo essenzialmente grafico per il riporto delle quantità misurate.

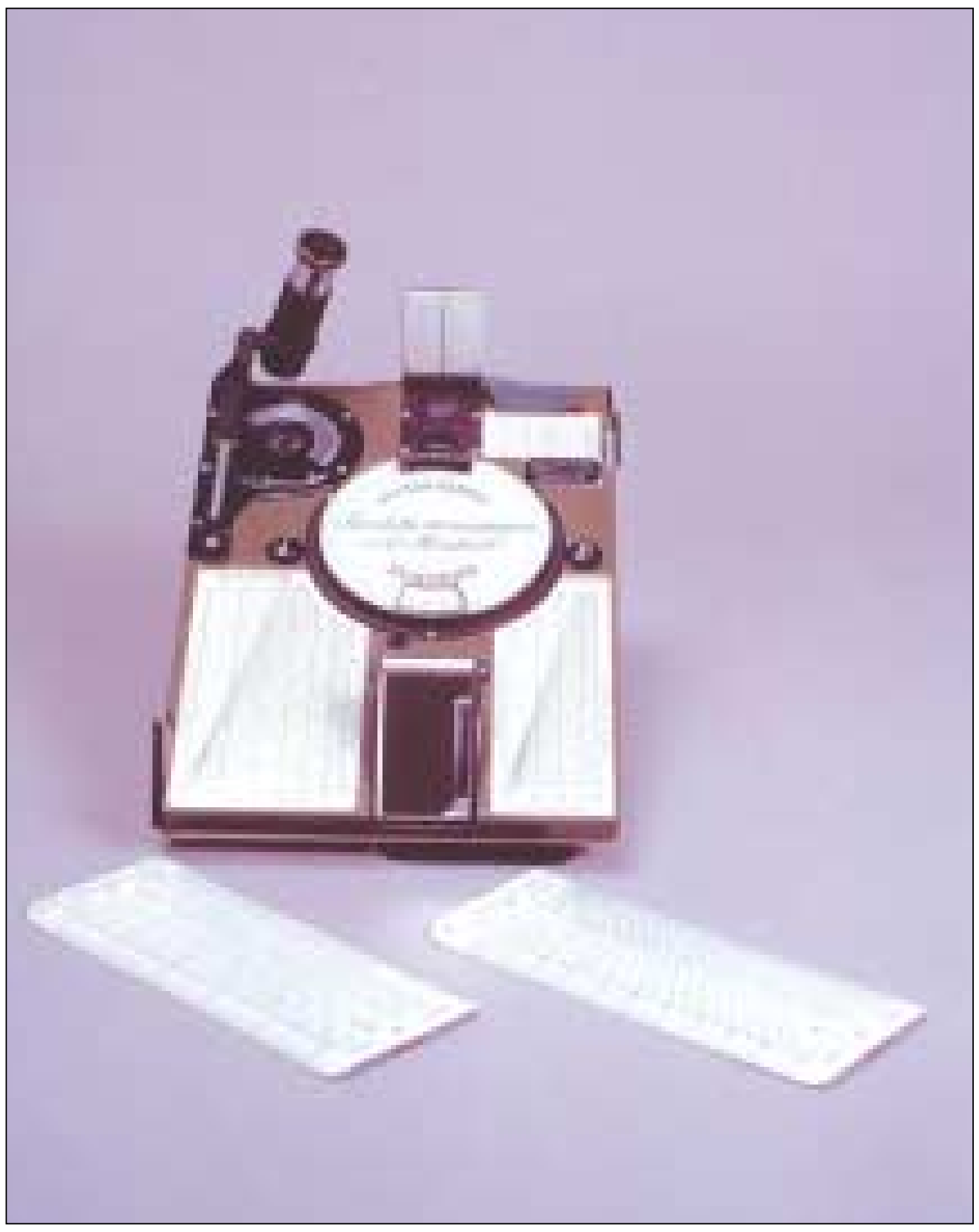

## *AUTOLIVELLO SALMOIRAGHI DI MEDIA PRECISIONE*

1960/Filotecnica Salmoiraghi, Milano / lunghezza 18.5 cm, larghezza 18.5 cm, altezza 32 cm

Livello automatico, in grado cioè di fornire l'allineamento orizzontale automaticamente, tramite un dispositivo a sospensione pendolare che mantiene la linea di mira sempre perpendicolare alla verticale. Le operazioni di misura risultano quindi meno laboriose e più brevi in termini di tempo.
Le oscillazioni del compensatore pendolare sono smorzate ad aria, e le operazioni di verifica e rettifica sono semplici ed efficaci grazie alla originalità delle soluzioni costruttive adottate dalla Casa costruttrice.

Il soprannome dato allo strumento dagli operatori del settore era "Paperino", e per la curiosa forma e per distinguerlo dal "fratello maggiore", livello automatico di grande precisione (e dimensioni), denominato invece "Paperone".

E' uno strumento semplice e compatto, impiegato in tutte le operazioni di livellazione di media precisione, come ad esempio il tracciamento di strade, canali, e lavori d'ingegneria in genere.
Il cannocchiale ha una apertura utile dell'obiettivo di 30 mm, è a immagine diritta e nel campo dell'oculare è visibile l'immagine della livella sferica: è quindi possibile controllare l'assetto dello strumento e il funzionamento del compensatore durante le letture alle stadie.
Nello strumento è presente un cerchio azimutale in vetro, con lettura a stima ai due primi centesimali.
L'errore medio chilometrico è di ± 3mm nella misura del dislivello, con livellazioni in andata e ritorno, riducibile con l'impiego del micrometro a lamina piano-parallela fornito a parte.

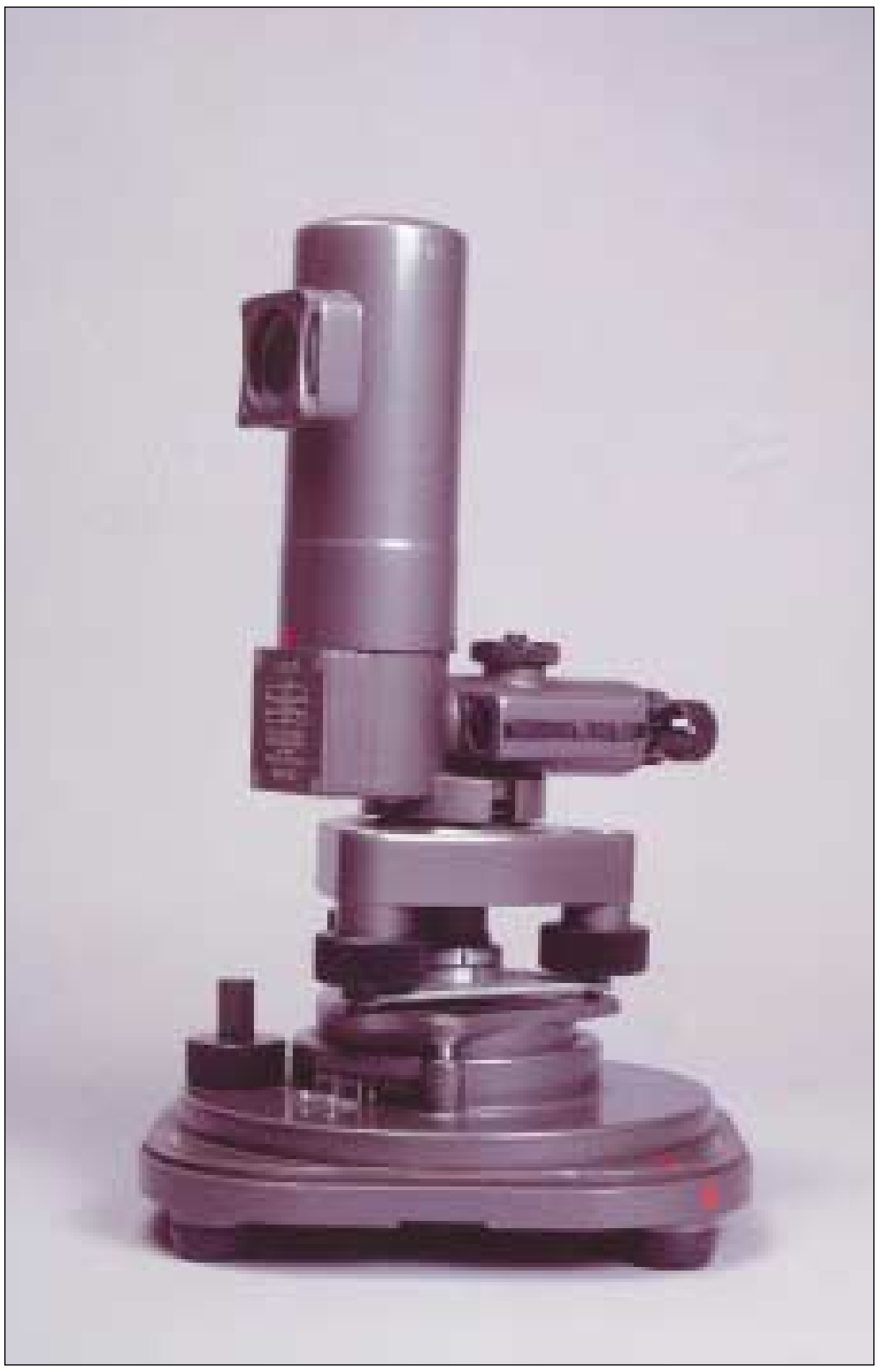

## *LIVELLO DI ALTA PRECISIONE VEB CARL ZEISS JENA "Ni 004"*

1950 / Carl Zeiss, Jena / lunghezza 46 cm, larghezza 16 cm, altezza 22 cm

Lo strumento è stato prodotto nelle officine Zeiss di Jena (Turingia), l'antica Casa fondata nel 1846 da Carl Zeiss (1816-1888), meccanico della locale università.
Alla fine della Seconda Guerra Mondiale la Turingia fu inclusa nella zona di influenza sovietica e l'impianto originale delle officine di Jena, o meglio, quello che restava dalle distruzioni della guerra, fu nazionalizzato: e riprese la produzione di strumentazioni per geodesia e fotogrammetria, oltre che per l'ottica industriale. Prima dell'occupazione sovietica della zona , le truppe americane trasferirono gran parte del gruppo dirigente nel settore occidentale, fondando nel 1946 un nuovo stabilimento a Oberkochen nel Baden-Wurttemberg. Dopo la riunificazione tedesca si è proceduto all'accorpamento delle due aziende, con varie conseguenze negative per gli stabilimenti storici di Jena.

Lo strumento è un livello con vite di elevazione di grande precisione, progettato e costruito per le determinazioni relative alle reti di livellazione del primo ordine (geodetiche), alle operazioni legate a livelli idraulici, quali canali di navigazione e vie d'acqua, a tutte le operazioni di controllo statico e di posizionamento industriale di grandi macchine.
Lo strumento è caratterizzato dalla grande accuratezza dei componenti ottico-meccanici, quali le livelle e il cannocchiale che incorpora la lamina piano-parallela a micrometro. Tutte le parti sono state protette dalle azioni dell'ambiente utilizzando materiali molto solidi, come ad esempio l'acciaio per il corpo del cannocchiale e delle livelle. Grande attenzione è stata prestata allo studio degli elementi di movimento e dei comandi.
Il cannocchiale è di grande potenza ed ha 44 ingrandimenti, elevata chiarezza e alto potere separatore, dovuti al notevole diametro dell'obiettivo (60 mm). Le livelle, molto sensibili, consentono la misura della deflessione trasversale della bolla; la livella principale, a coincidenza, ha una sensibilità di 10" su 2 mm, e compare nell'oculare del cannocchiale.

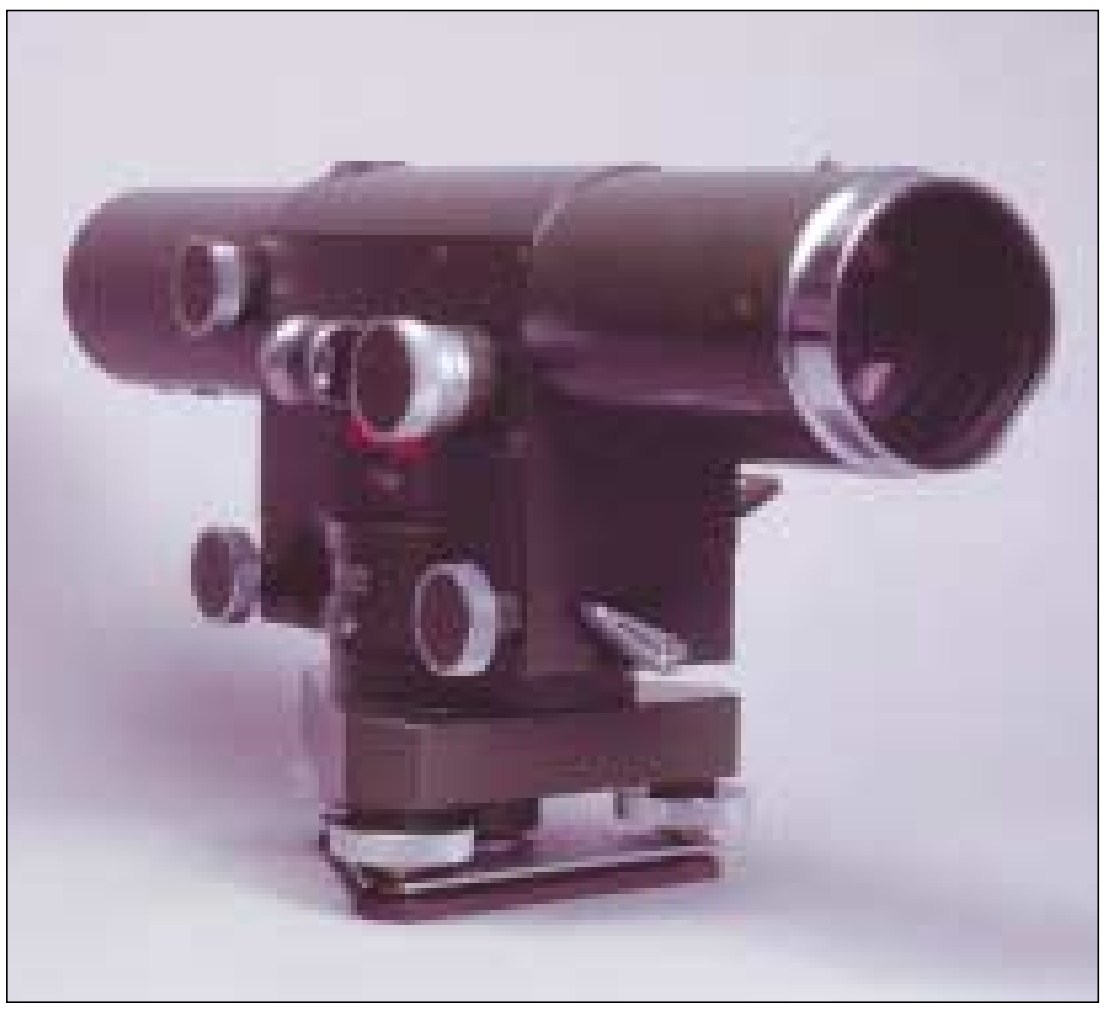

Allo strumento si accompagna una stadia di invar, semicentimetrata, che, letta a mezzo del micrometro a lamina, consente di misurare la altezza di collimazione con la precisione di 0.05 mm.
L'errore medio chilometrico nella livellazione geometrica in andata e ritorno, indicato dalla Casa costruttrice, è di ± 0.4mm.

## *RINGRAZIAMENTI;*

Come già accennato altrove, le ricerche confluite in questo piccolo volume hanno usufruito di un contributo del C.N.R. (ex-Comitato Beni Culturali); mentre la presente pubblicazione vede la luce grazie all'apposita assegnazione della PRO UNIVERSITATE BERGOMENSI (presieduta dal Cav.d.Lav. Dottor Emilio Zanetti). Si ringraziano inoltre, per la cortese e fattiva collaborazione, il Professor A. Castoldi, M. Rettore; i Professori A. Perdichizzi, M.I. Bertocchi e G. Mirandola; il Dottor R. Terranova della P.U.B.; il Dirigente Scolastico del «Lussana», Prof. C. Quarenghi; il Dirigente Scolastico del V. Emanuele, Prof. D. Frigerio; il Dirigente Scolastico del «Quarenghi», Prof. F. Fabio; il docente e tecnico di Laboratorio Sig. A. Gabuzzi.

Un caldo ringraziamento va al Sig. Stefano Castelli e ai suoi collaboratori della «Tipolito Castel», per l'appassionato e competente lavoro svolto nella produzione di questo libro.

## 5. BIBLIOGRAFIA:

# INDICE